\documentclass[12pt]{article}
\usepackage[paperwidth=21cm,paperheight=26cm, textwidth=16cm,textheight=20.6cm]{geometry}
\usepackage{amsmath,amsthm,amssymb,xspace,xcolor,hyperref}\hypersetup{colorlinks=true, linkcolor=black,citecolor=black,urlcolor=black}\usepackage{graphicx}%rewritelastpackage
\numberwithin{equation}{section}

\theoremstyle{plain}
\newtheorem{theorem}{Theorem}

\newtheorem{claim}{Claim}[section]
\newtheorem{proposition}{Proposition}
\newtheorem{lemma}{Lemma}
\newtheorem{corollary}{Corollary}
\newtheorem{definition}{Definition}
\newtheorem{question}{Question}
\newtheorem{hypothesis}{Hypothesis}

\newcommand{\APC}{\ensuremath{{\sf APC_1}}\xspace}

\newcommand{\SB}{\ensuremath{{\sf S^1_2}}\xspace}

\newcommand{\PV}{\ensuremath{{\sf PV_1}}\xspace}
\newcommand{\pv}{\ensuremath{{\sf PV}}\xspace}

\newcommand{\EF}{\ensuremath{{\sf EF}}\xspace}
\newcommand{\WF}{\ensuremath{{\sf WF}}\xspace}

\newcommand{\Ptime}{\ensuremath{{\sf P}}\xspace}
\newcommand{\Ppoly}{\ensuremath{{\sf P/poly}}\xspace}

\newcommand{\NPtime}{\ensuremath{{\sf NP}}\xspace}
\newcommand{\NP}{\ensuremath{{\sf NP}}\xspace}
\newcommand{\coNP}{\ensuremath{{\sf coNP}}\xspace}

\newcommand{\Circuit}{\ensuremath{{\sf Circuit}}\xspace}

\newcommand{\SAT}{\ensuremath{{\sf SAT}}\xspace}

\newcommand{\MCSP}{\ensuremath{{\sf MCSP}}\xspace}
\newcommand{\GCSP}{\ensuremath{{\sf GCSP}}\xspace}

\newcommand{\LB}{\ensuremath{{\sf LB}}\xspace}
\newcommand{\LBtt}{\ensuremath{{\sf LB_{tt}}}\xspace}
\newcommand{\ttable}{\ensuremath{{\sf tt}}\xspace}

\begin{document}

\title{{\bf Learning algorithms versus\\ automatability of Frege systems}}
\author{J\'an Pich \\ \small University of Oxford \and Rahul Santhanam \\ \small University of Oxford}
\date{October 2021}%rewriteoxOctober
\maketitle

\begin{abstract} We connect %two central concepts in complexity theory, 
learning algorithms and algorithms automating proof search in propositional proof systems: for every sufficiently strong, well-behaved propositional proof system $P$, %, which satisfies some basic properties, \APC-provably p-simulates Je\v{r}\'abek's system \WF, and which proves efficiently for some boolean function $h$ that $h$ is hard for circuits of subexponential size, 
we prove that the following statements are equivalent,
\begin{itemize}
\item[1.] {\bf Provable learning.} $P$ proves efficiently that p-size circuits are learnable by subexponential-size circuits over the uniform distribution with membership queries.
\item[2.] {\bf Provable automatability.} $P$ proves efficiently that $P$ is automatable by non-uniform circuits %srewrite' by nonuni circuits '
 on propositional formulas expressing p-size circuit lower bounds.
\end{itemize}
Here, $P$ is sufficiently strong and well-behaved if I.-III. holds: I. $P$ p-simulates Je\v{r}\'abek's system \WF (which strengthens the Extended Frege system \EF by a surjective weak pigeonhole principle)%rewriteoxpredch' (which..)'
; II. $P$ satisfies some basic properties of standard proof systems which p-simulate \WF; III. $P$ proves efficiently for some Boolean %rewriteoxpredch' Boolean 'anaslmedzera
 function $h$ that $h$ is hard on average for circuits of subexponential size. For example, if III. holds for $P=\WF$, then Items 1 and 2 are equivalent for $P=\WF$.%rewriteoxpredch'Items'

If there is a function $h\in\mathsf{NE}\cap \mathsf{coNE}$ which is hard on average for circuits of size $2^{n/4}$, for each sufficiently big $n$, then there is an explicit %rewriteoxpredch' which '' an explicit 'anaslmedzera
propositional proof system $P$ satisfying properties I.-III., i.e. the equivalence of Items 1 and 2 holds for $P$.%rewriteoxpredch'Items'
 %The equivalence is, however, conditional under the assumption that \WF proves efficiently some circuit lower bound.%rewriteabstract
\end{abstract}

%\tableofcontents

\section{Introduction}

Learning algorithms and automatability algorithms searching for proofs in propositional proof systems are central concepts in complexity theory, but a priori they appear rather unrelated. %rewriteoxpredch' unrelated. '
\bigskip

\noindent {\bf Learning %rewriteoxpredch'rnin'anaslmedzera
 algorithms.} In the PAC model of learning introduced by Valiant \cite{LT}, a circuit class $\mathcal{C}$ is learnable by a randomized %srewrite'a randomized '
 algorithm $L$ over the uniform distribution, up to error $\epsilon$, with confidence $\delta$ and membership queries, if for every Boolean %rewriteoxpredch' Boolean 'anaslmedzera
 function $f$ computable by a circuit from $\mathcal{C}$, when given oracle access to $f$, $L$ outputs with probability $\ge \delta$ over the uniform distribution %' over.. distribution '
 a circuit computing $f$ on $\ge (1-\epsilon)$ inputs. An important task of learning theory is to find out if standard circuit classes such as \Ppoly %rewriteoxprech' such as Ppoly'naslmedzeraanasl' we can try '
 are learnable by efficient circuits. A way to approach the question is to connect the existence of efficient learning algorithms to other standard conjectures in complexity theory. For example, we can try to prove that efficient learning of \Ppoly is equivalent to $\Ptime =\NP$ or to the non-existence of strong pseudorandom generators. In both cases one implication is known: $\Ptime=\NP$ implies efficient learning of \Ppoly (with small error and high confidence) %rewriteoxpredchanaslmedzera
 which in turn breaks pseudorandom generators. However, while some progress on the opposite implications has been made, they remain open, cf. \cite{ABX,S19}. %or, e.g., whether statements such as $\Ptime =\NP$ can be reduced to the existence of efficient algorithm learning \Ppoly (or whether the existence of efficient algorithms learning \Ppoly follows from the non-existence of one-way f's %(Equivalence to natural proofs and nonuniform prgs.)%rewritetentoadalsiparaokrem{\bf..}
\medskip

\noindent {\bf Automatability.} The notion of automatability was introduced in the work of Bonet, Pitassi and Raz \cite{BPR}. A propositional proof system $P$ is automatable if there is an algorithm $A$ such that for every tautology $\phi$, $A$ finds a $P$-proof of $\phi$ %srewrite' of phi '
 in p-time in the size of the shortest $P$-proof of $\phi$. That is, even if $P$ does not prove all tautologies efficiently, it can %rewriteoxpredch' can'anaslmedzera
 still be automatable. Establishing (non-)automatability results for concrete proof systems is one of the main tasks of proof complexity. This led to many attempts to link the notion of automatability to other standard complexity-theoretic conjectures. For example, recently Atserias and M\"uller \cite{AM} proved that automating Resolution is \NP-hard and their work has been extended to other weak proof systems, e.g. \cite{DGNPRS, Gkdnf, GKMP}. For stronger systems, it is known that automating Extended Frege system \EF, Frege or even constant-depth Frege %rewriteoxpredchmedzera,naslmedzera,medzerapredpredchadz' or'anasl'Hellman'
 would break specific cryptographic assumptions such as the security of RSA or Diffie-Hellman scheme, cf. \cite{KP,BPR,BDGMP}. It remains, however, open to obtain non-automatability of strong systems like Frege under a generic assumption such as the existence of strong pseudorandom generators, let alone to prove the equivalence between such notions. %(Feasible interpolation, not known how to base impossibility of FIP on a generic assumption).
\bigskip

In the present paper we derive a conditional equivalence between learning algorithms for p-size circuits and automatability of proof systems on tautologies encoding circuit lower bounds.%rewrite'on tautologies.. bounds.'a'derive'

\subsection{Our result}

An ideal connection between learning and automatability would say that for standard proof systems $P$, %rewrite'say that.. $P$ 'a'\sf{P/poly}'

\bigskip
\centerline{``$P$ is automatable if and only if \Ppoly is learnable efficiently%srewrite' efficiently' {\sf P/poly}
".}

\bigskip
We establish this modulo some provability conditions and a change of parameters. Additionally, we need to consider automatability only w.r.t. formulas encoding circuit lower bounds. More precisely, denote by $\ttable(f,s)$ a propositional formula which expresses that boolean function $f$ represented by its truth-table is not computable by a boolean circuit of size $s$ represented by free variables, see Section $\ref{s:formalizations}$. So $\ttable(f,s)$ is a tautology if and only if $f$ is hard for circuits of size $s$. Similarly, let $\ttable(f,s,t)$ be a formula expressing that circuits of size $s$ fail to compute $f$ on $\ge t$-fraction of inputs. In our main result (Theorem~\ref{t:informal}) we use a slightly modified notion of automatability where the automating algorithm for a proof system $P$ is non-uniform and %srewrite' is non-uniform and '
 outputs a $P$-proof of a given formula $\ttable(f,n^{O(1)})$ in p-time in the size of the shortest $P$-proof of $\ttable(f,2^{n^{o(1)}},1/2-1/2^{n^{o(1)}})$, see Section \ref{s:formalizations}%srewrite' see..daut}'
.%rewritetentoparaacontentnaslvety

\begin{theorem}[Informal, cf. Theorem \ref{t:main}]\label{t:informal} Let $P$ be a propositional proof system which \APC-provably p-simulates \WF and satisfies some basic properties. Moreover, assume that $P$ proves efficiently $\ttable(h,2^{n/4},1/2-1/2^{n/4})$ for some boolean function $h$. Then, the following statements are equivalent: 
\begin{itemize}
\item[1.] {\bf Provable learning.} $P$ proves efficiently that p-size circuits are learnable by $2^{n^{o(1)}}$-size circuits, over the uniform distribution, up to error $1/2-1/2^{n^{o(1)}}$, with membership queries and confidence $1/2^{n^{o(1)}}$.
\item[2.] {\bf Provable automatability.} $P$ proves efficiently that $P$ is automatable by non-uniform circuits %srewrite' by nonuni circuits '
 on formulas $\ttable(f,n^{O(1)})$.
\end{itemize}
%$P$-provable automatability of $P$ on formulas $\ttable(f,n^{O(1)})$ is equivalent to $P$-provable learning of p-size circuits by subexponential-size circuits over the uniform distribution with membership queries.
\end{theorem}

\WF is an elegant strengthening of \EF introduced by Je\v{r}\'abek \cite{Jwphp}, which corresponds to the theory of approximate counting \APC, a theory formalizing probabilistic p-time reasoning, see Section \ref{sba}. Concrete proof systems which \APC-provably p-simulate \WF and satisfy the basic properties from Theorem \ref{t:informal} include \WF itself or even much stronger systems such as set theory $\mathsf{ZFC}$ (if we interpret {\sf ZFC} as a suitable system for proving tautologies, see Section \ref{s:main}). The error and confidence of learning algorithms can be amplified `for free', see Section \ref{ss:npl}, but we did not make the attempts to prove that the amplification is efficiently provable already in \WF. %srewrite 'in WF. %' \APC-decent systems. %\WF-provability in Theorem \ref{t:informal} always refers to `efficient provability' in the sense that the tautology in question has p-size \WF-proof.%rewritethispara%rewriteprvavvetanextpar,ktomu'an interesting''Frege''possible''major''ZFC'' \cite{Wacc}''celufootnote2'a'ttable(..)'vsade.
\medskip

\noindent {\bf Plausibility of the assumption.} The main assumption in Theorem \ref{t:informal} is the provability of a circuit lower bound $\ttable(h,2^{n/4},1/2-1/2^{n/4})$. This assumption has an interesting status. Razborov's conjecture about hardness of Nisan-Wigderson generators implies a conditional hardness of formulas $\ttable(h,n^{O(1)})$ for Frege (for every $h$), cf. \cite{Rkdnf}, and it is possible to consider extensions of the conjecture to all standard proof systems, even set theory $\mathsf{ZFC}$. On the other hand, all major circuit lower bounds for weak circuit classes and explicit boolean functions are known to be effciently provable in $\EF$ %rewriteoxpredch' EF 'anaslmedzera
\footnote{This has not been verified for lower bounds obtained via the algorithmic method of Williams \cite{Wacc}.}, cf. \cite{Rup,MP}. If we believe that explicit circuit lower bounds such as $\ttable(h,2^{n/4},1/2-1/2^{n/4})$, for some $h\in\mathsf{EXP}$, %rewriteox'h'vpredch\ttableapredch', for some h in EXP, '
 are true, it is also perfectly plausible %rewriteoxpredchanaslmedzera
 that they are efficiently provable in a standard proof system such as $\mathsf{ZFC}$ %rewriteoxpredchanaslmedzera
\footnote{Efficient provability of $\ttable(h,2^{n/4},1/2-1/2^{n/4})$ in {\sf ZFC}, for some $h\in\mathsf{EXP}$, %rewriteox'h'vpredch\ttableapredch' in ZFC, for some h in EXP, '
 would follow from the standard provability of this lower bound in {\sf ZFC}.} or \EF. Notably, if \EF proves efficiently $\ttable(h,2^{n/4})$ for some boolean function $h$, then \EF simulates \WF, cf. \cite[Lemma 19.5.4]{Kpc}. %srewrite')'vpredchttformulea' simulates WF, cf. []. %'anasled2vety
If there is a p-time algorithm which given a string of length $2^n$ (specifying the size of $\ttable(h,2^{n/4})$) %rewriteoxpredch(..)anaslmedzera
 generates an \EF-proof of $\ttable(h,2^{n/4})$, then \EF is p-equivalent to \WF. To see that, combine Lemma \ref{singlehard} with the fact (proved in \cite{Jwphp}) that \APC proves the reflection principle for \WF.
\bigskip%rewriteodbigskippodalsiusubsekciu

As a corollary of Theorem \ref{t:informal} we show that, under a standard hardness assumption, there is an explicit %rewriteoxpredch' an explicit 'anaslmedzera
 proof system $P$ for which the equivalence holds. This, follows, essentially, by `hard-wiring' tautologies $\ttable(h,2^{n/4},1/2-1/2^{n/4})$ to \WF.

\begin{corollary}[cf. Corollary \ref{c:main}]\label{c:informal}
Assume there is a $\mathsf{NE}\cap \mathsf{coNE}$-function $h_n:\{0,1\}^n\mapsto \{0,1\}$ such that for each sufficiently big $n$, $h_n$ is not $(1/2+1/2^{n/4})$-approximable by $2^{n/4}$-size circuits.%rewriteoxpredch'.'amedzerazanaslfootnote
\footnote{A circuit $C$ with $n$ inputs $\gamma$-approximates function $f:\{0,1\}^n\mapsto \{0,1\}$ if $\Pr_{x\in \{0,1\}^n}[C(x)=f(x)]\ge \gamma$.} Then there is a proof system $P$ (which can be described explicitly given the definition of $h_n$) %rewriteoxpredch' (..) 'anaslmedzera
 such that Items 1 and 2 from Theorem \ref{t:informal} are equivalent.  %$P$-provable automatability of $P$ on formulas $\ttable(f,n^{O(1)})$ is equivalent to $P$-provable learning of p-size circuits by subexponential-size circuits over the uniform distribution with membership queries.
\end{corollary}

The proof of Theorem \ref{t:informal} reveals also a conditional proof complexity collapse, which we discuss in Section \ref{s:main}.%rewriteoxpredch' which we discuss.. main}.'

\begin{corollary}[cf. Corollary \ref{c:collapse}]\label{c2:informal}
Let $P, P_0$ be propositional proof systems which \APC-provably p-simulate \WF and satisfy some basic properties. Moreover, assume that systems $P, P_0$ prove %srewrite' systems P,P0 prove '
 efficiently $\ttable(h_n,2^{n/4},1/2-1/2^{n/4})$ for some boolean function $h_n$. Then, Item~1 implies Item 2:
\begin{itemize}
\item[1.] {\bf $P$-provable automatability.} $P$ proves efficiently that $P$ is automatable by non-uniform circuits %srewrite' by nonuni circuits '
 on formulas $\ttable(f,n^{O(1)})$. % For each $k,c\ge 1$, there is a constant $K'$ and $2^{K'n}$-size circuits $B$ such that $P$ proves efficiently $\mathsf{aut}_{P}(B, \Phi,2^{cn})$, where $\Phi$ is the set of pairs $\langle \ttable(f,2^{Kn^{\gamma}},1/2-1/2^{Kn^{\gamma}}), \ttable(f,n^k)\rangle$ for all boolean functions $f$ with $n$ inputs.
\item[2.] {\bf $P_0$-provable proof search.} There are p-size circuits $B$ %srewrite ' B '
 such that $P_0$ proves efficiently that circuits $B$ %srewrite' B '
 (given just $\ttable(f,n^{O(1)})$) generate $P$-proofs of $\ttable(f,n^{O(1)})$ or $2^{n^{o(1)}}$-size circuits $(1/2+1/2^{n^{o(1)}})$-approximating $f$. %For each $k\ge 1$, there is a constant $K'$ and $2^{K'n}$-size circuits $B$ such that $P_0$ proves efficiently ``$B$ (given just $\ttable(f,n^k)$) outputs a $P$-proof of $\ttable(f,n^k)$ or $B$ outputs a $2^{Kn^{\gamma}}$-size circuit $(1/2+1/2^{Kn^{\gamma}})$-approximating $f$.". %rewritethisthm
%\item[1'.] There is a subexponential-size circuit $A$ such that \WF proves efficiently that $A$ learns circuits of size $n^{\Omega(l)}$.
%\item[2'.] There are p-size circuits $B$ such that \WF proves efficiently $\mathsf{aut}_{\WF}(B,\Phi,2^{O(n)})$ where $\Phi$ is the set of formulas $\ttable(f,n^l)$ for all boolean functions $f$ with $n$ inputs.
\end{itemize}\end{corollary}

\subsection{Outline of the proof} 

Our starting point for the derivation of Theorem \ref{t:informal} is a relation between natural proofs and automatability which goes back to a work of Razborov and Kraj\'{i}\v{c}ek. Razborov \cite{Rba,Rpairs} proved that certain theories of bounded arithmetic cannot prove explicit circuit lower bounds assuming strong pseudorandom generators exist. Kraj\'{i}\v{c}ek \cite{Kint,Kfor} developed the concept of feasible interpolation (a weaker version of automatability, cf. \cite{Kpc}) and reformulated Razborov's unprovability result in this language, see \cite[Section 17.9]{Kpc} for more historical remarks. %rewriteoxvpredchvete' Krajicek developed ''reformulated'', See .. remarks.'
%This reformulation appears also in \cite{Rpairs} \textcolor{red}{(doublecheck the history with Krajicek and Razborov)}.%rewritevtomtoparaprve'Rpairs''Kint''Kpc'%vnaslvete:'Kint''Automatable proof.. EF,'' useful'

\begin{theorem}[Razborov-Kraj\'{i}\v{c}ek \cite{Rba,Rpairs,Kint} - informal version]\label{t:razbkra} Let $P$ be a proof system which simulates \EF. If $P$ is automatable and $P$ proves %rewriteoxpredch' Let P.. proves 'snaslmedzerou
 efficiently $\ttable(h,n^{O(1)})$ for some %rewriteoxpredch' some 'snaslmedzerou
 function $h$, then %rewriteoxpredch', then 'snaslmedzerou
 there are \Ppoly-natural proofs useful against \Ppoly.
\end{theorem}

The second crucial ingredient %rewriteoxpredch' ingredient '
 we will use is a result of Carmosino, Impagliazzo, Kabanets and Kolokolova, who showed that natural proofs can be turned into learning algorithms \cite{CIKK}. This allows us to conclude the following.

\begin{theorem}[Informal, cf. Theorem \ref{t:rkcikk}]\label{t:intro2} Let $P$ be a proof system simulating \EF. If $P$ proves %rewriteoxpredch' Let P.. proves 'snaslmedzerou
 efficiently $\ttable(h,n^{O(1)})$ for some function $h$, then %rewriteoxpredch' then 'smedzerami
 automatability of $P$ implies the existence of subexponential-size circuits learning p-size circuits over the uniform distribution, with membership queries.%rewrite'eo'v'cf. Theorem','t:rkcikk','For proof.. prove',dalejvnaslvete'of Theorem \ref{t:intro2}'a't:orkcikk'vnaslthm
\end{theorem}
%rewriteoxnaslmedzeravetaa' Here a pps is optimal.. .'

Theorem \ref{t:intro2} directly implies that if strong pseudorandom generators exist and \EF proves efficiently $\ttable(h,n^{O(1)})$ for some $h$, then \EF is automatable if and only if there are subexponential-size circuits learning p-size circuits over the uniform distribution, with membership queries. The disadvantage of this observation is that, unlike in Theorem \ref{t:informal}, its assumptions are known to imply that both sides of the desired equivalence are false.

We note that the proof of Theorem \ref{t:intro2} can be used to show also that optimal and automatable proof systems imply learning algorithms. Here, a propositional proof system $P$ is optimal, if for each propositional proof system $R$, an $R$-proof $\pi$ of $\phi$, implies the existence of a $poly(|\pi|)$-size $P$-proof of $\phi$.

\begin{theorem}[Optimality and automatability implies learning, cf. Theorem \ref{t:orkcikk}] If there is an optimal proof system which is automatable, then there are subexponential-size circuits infinitely often learning p-size circuits over the uniform distribution.%rewritevtejtovete','a' inifinitely often'
\end{theorem}

In fact, it is possible to prove, unconditionally, that there is some propositional proof system $P$ such that automatability of $P$ is equivalent to the existence of subexponential-size circuits inifnitely often learning \Ppoly over the uniform distribution, cf. Theorem~\ref{t:unconditional}. The proof is, however, non-constructive so (unlike in Corollary \ref{c:informal}) we do not know which system $P$ satisfies the equivalence.%rewriteoxpredch2vetysmedzeramiaskips
\medskip%rewriteoxthismedskipanaslemptyriadok

\noindent {\bf The entrance of metamathematics.} Unfortunately, it is unclear how to derive the opposite implication in Theorem \ref{t:intro2}. We do not know how to automate, say, \EF assuming just the existence of efficient learning algorithms. In order to get the reverse, we need to assume that an efficient learning algorithm is provably correct in a proof system $P$, which p-simulates \WF. For simplicity, %rewriteoxpredch' simplicity, 'anaslmedzera
 let $P=\WF$. If we assume that \WF proves efficiently for some small circuits that they can learn p-size circuits, we can show that there are small circuits such that \WF proves efficiently that these circuits automate \WF on formulas $\ttable(f,n^{O(1)})$. In more detail, we first formalize in \APC the implication that \WF-provable learning yields automatability of \WF on $\ttable(f,n^{O(1)})$ - if a learning circuit $A$ does not find a small circuit for a given function $f$, the automating circuit uses \WF-proof of the correctness of $A$ to produce a short \WF-proof of $\ttable(f,n^{O(1)})$. %srewrite' tt(n^{O(1)})-if .. f,n^O(1)). 'smedzeramia'ttable(..)'nakoncipara
 Then, we translate the \APC-proof to \WF and conclude that \WF proves that \WF-provable learning implies automatability of \WF. This allows us to show that if we have \WF-provable learning, then \WF is \WF-provably automatable on $\ttable(f,n^{O(1)})$.%rewritevtomtopara'in a proof system P .. P=WF.'

It is important that assuming \WF-provable learning, we are able to derive \WF-provable automatability of \WF, and not just automatability of \WF. This makes it possible to obtain %rewrite' makesitpossitoobtain '' and establish.. equivalence'amedzerazaparagrafom
 the opposite direction and establish the desired equivalence: If we know that \WF proves that \WF is automatable, we can formalize the proof of Theorem \ref{t:intro2} in \WF and conclude the existence of \WF-provable learning algorithms. 

One could expect that \WF-provable learning would yield just automatability of \WF and \WF-provability of \WF-provable learning would be needed to get \WF-provable automatability of \WF. It might be thus surprising that already the first metamathematical level achieves the desired `fixpoint'.
\bigskip

\noindent {\bf Benefits of bounded arithmetic.} The proof of Theorem \ref{t:informal} relies heavily on formalizations. Among other things we need to formalize the result of Carmosino, Impagliazzo, Kabanets and Kolokolova in \APC %rewriteoxpredchanaslmedzera
\footnote{We will actually formalize `CIKK' just conditionally, in order to avoid the formalization of Bertrand's postulate.}, %, including Goldreich-Levin amplification, %rewritevtomtoriadku', %'skip;footnote:`CIKK'
and use an elaborated way of expressing complex statements about metacomplexity by propositional formulas: existential quantifiers often need to be witnessed before translating them to propositional setting. The framework of bounded arithmetic allows us to deal with these complications in an elegant way: we often reason in bounded arithmetic, possibly using statements of higher quantifier complexity, and only subsequently translate the outcomes to propositional logic, if the resulting (proved) statement has \coNP form. Notably, already propositional formulas expressing probabilities in the definition of learning algorithms require more advanced tools - the probabilities are encoded using suitable Nisan-Wigderson generators which come out of the notion of approximate counting in \APC, cf. Section \ref{ss:lform}.  %rewriteoxpredchvetasmedzerami
%rewritemedzeranasledsubsectiontitleafootnoteabove

\subsection{Related results}

Learning algorithms and automatability have been linked already in the work of Alekhnovich, Braverman, Feldman, Klivans and Pitassi \cite{ABFKP}, who showed an informal connection between learning of weak circuit classes and automatability of some weak systems such as tree-like Resolution. As already mentioned, Atserias and M\"uller \cite{AM} proved that automating Resolution is \NP-hard and their work has been extended to other weak proof systems, see e.g. \cite{DGNPRS, Gkdnf, GKMP}. A direct consequence of these results is that efficient algorithms automating the respective proof systems can be used to learn efficiently classes like \Ppoly. %rewriteoxpredch' some weak systems.. Ppoly.'anaslveta
A major difference between these results and ours is that for our results to apply, the proof system needs to be sufficiently strong, while for the other results, the proof system needs to be weak (in the sense that lower bounds for the system are already known).

\subsection{Open problems}

\noindent {\bf Unconditional equivalence between learning and automatability.} Is it possible to avoid the assumption on the provability of a circuit lower bound in Theorem \ref{t:informal} and establish an unconditional equivalence between learning and automatability?

\medskip

\noindent {\bf Complexity theory from the perspective of metamathematics.} %rewriteprveCv{bf}medzeraprednaslvetou,'possible'vnej,druhuvetuporesolvesmedzeroupred
Our results demonstrate that in the context of metamathematics it is possible to establish some complexity-theoretic connections which we are not able to establish otherwise. We exploit the metamathematical nature of the notion of automatability: efficient $P$-provability of the correctness of an algorithm implies efficient $P$-provability of automatability of $P$. Is it possible to take advantage of metamathematics in other contexts and resolve other important open problems in this setting? For example, could we get a version of the desired equivalence between the existence of efficient learning algorithms and the non-existence of cryptographic pseudorandom generators, cf. \cite{OS,S19,Plc}? The question of basing cryptography on a worst-case assumption such as $\Ptime\ne\NP$ could be addressed in this setting by showing that if a sufficiently strong proof system $P$ proves efficiently that there is no strong pseudorandom generator\footnote{The formalization of this statement would assume the existence of a p-size circuit which for any p-size circuit defining a potential pseudorandom generator outputs its distinguisher.}, then $P$ is p-bounded.%rewritefootnotevpredchvete,ktomu'a sufficiently strong'a2x$P$

\medskip%rewritemedskiparestofsubsection

\noindent {\bf Circuit lower bound tautologies.} How essential are %rewriteoxpredchanadchmedzera
 circuit lower bound tautologies %rewriteoxpredch' tautologies 'anaslmedzera
 in our results? Consider fundamental questions of proof complexity (p-boundness, %rewriteoxpredch'boundness, 'anaslmedzera
 optimality, automatability) w.r.t. formulas $\ttable(f,s)$. Do they coincide with the original ones? Are %rewriteoxpredch' Are 'anaslmedzera
 formulas $\ttable(f,s)$ the hardest ones, do they admit optimal proof systems, or can we turn automatability on formulas $\ttable(f,s)$ into automatability on all formulas? %(If WF proves 'lear->p=np' then tt hardest taus for WF.)
\medskip

\noindent {\bf Proof complexity magnification.} %The proof of Theorem \ref{t:main} reveals an interesting magnification phenomenon. Assuming \WF proves a sufficiently strong circuit lower bound, \WF-provable automatability implies \EF provable algorithms generating \WF-proofs. In other words, \EF lower bound for formulas encoding efficient proof-search algorithms imply \WF lower bounds either for tautologies expressing circuit lower bounds or for formulas encoding automatability of \WF. Can we get even \EF lbs for tt implying \WF lower bounds for tt (by employing the implication: tt lbs imply lbs for lear)? \textcolor{red}{(Would work if we could get lear with confidence 1 from lear with confidence 2/3)}
Is it possible to obtain the collapse from Corollary~\ref{c2:informal} for formulas expressing standard conjectures? For example, is it possible to show that the %srewrite'the '
 hardness of $\ttable(\SAT,n^{O(1)})$ for some proof system $P_0$, implies hardness of $\ttable(\SAT,n^{O(1)})$ for stronger proof systems? An instance of such a magnification phenomenon appeared in \cite{MP} (with $P_0=$ constant-depth Frege, $P=$ Frege and a different formalization %srewrite' n '
 of $\ttable(\SAT,n^{O(1)})$ in $P$).

\section{Preliminaries}

\subsection{Natural proofs and learning algorithms}\label{ss:npl}%rewritelabela%vnaslpara

$[n]$ denotes $\{1,\dots,n\}$. $\Circuit[s]$ denotes fan-in two Boolean circuits of size at most $s$. The size of a circuit is the number of gates. %A function $f:\{0,1\}^n\mapsto \{0,1\}$ is $\gamma$-approximated by a circuit $C$, if $\Pr_x[C(x)=f(x)]\ge\gamma$.

\begin{definition}[Natural property \cite{RR}]\label{d:natpr} Let $m=2^n$ and $s,d:\mathbb{N} \mapsto \mathbb{N}$. A sequence of circuits $\{C_{m}\}^{\infty}_{m=1}$ is a $\Circuit[s(m)]$-natural property useful against $\Circuit[d(n)]$ if
\begin{itemize}
\item[1.] {\em Constructivity.} $C_{m}$ has $m$ inputs and size $s(m)$,
\item[2.] {\em Largeness.} $\Pr_x[C_{m}(x)=1]\ge 1/m^{O(1)}$,
\item[3.] {\em Usefulness.} For each sufficiently big $m$, $C_{m}(x)=1$ implies that $x$ is a truth-table of a function on $n$ inputs which is not computable by circuits of size $d(n)$.
\end{itemize}
\end{definition}

%rewritetherestofthesection also move Def of Lear before boosting?

%\begin{definition}[Pseudorandom generator]\label{d:prg}
%A function $g:\{0,1\}^n\mapsto\{0,1\}^{n+1}$ computable by p-size circuits is a pseudorandom generator safe against circuits of size $s(n)$, if for each circuit $D$ of size $s(n)$, $$\left|\Pr_{y\in\{0,1\}^{n+1}}[D(y)=1]-\Pr_{x\in\{0,1\}^n}[D(g(x))=1]\right|<\frac{1}{s(n)}.$$
%\end{definition}

\begin{definition}[PAC learning]\label{d:lear}
%  Let $C_n$ be a concept class over the set $X_n$ and let $\cC = \cup_{n \in \N} C_n$ and $\cX = \cup_{n \in \N} X_n$. We say that
A circuit class $\mathcal{C}$ is learnable over the uniform distribution %rewriteoxpredch' distribution 'anaslmedzera
 by a circuit class $\mathcal{D}$ up to error $\epsilon$ with confidence $\delta$, if there are randomized oracle circuits $L^f$ from $\mathcal{D}$ such that for every Boolean function $f:\{0,1\}^n\mapsto\{0,1\}$ computable by a circuit from $\mathcal{C}$, when given oracle access to $f$, input $1^n$ and the internal randomness $w \in \{0,1\}^*$, $L^f$ outputs the description of a circuit satisfying 
  \begin{equation*}
    \Pr_w [ L^f(1^n, w) \text{ } (1-\epsilon) \text{-approximates } f ] \geq \delta.
  \end{equation*}
%  and runs in time polynomial in $n, 1/\epsilon, 1/\delta$. 

\noindent $L^f$ uses non-adaptive membership queries if the set of queries which $L^f$ makes to the oracle does not depend on the answers to previous queries. $L^f$ uses random examples if the set of queries which $L^f$ makes to the oracle is chosen uniformly at random. %If $\delta=1$, we omit mentioning the confidence parameter.
\end{definition}

In this paper, PAC learning always refers to learning over the uniform distribution. While, a priori, learning over the uniform distribution might not reflect real-world scenarios very well (and on the opposite end, learning over all distributions is perhaps overly restrictive), as far as we can tell it is possible that PAC learning of p-size circuits over the uniform distribution implies PAC learning of p-size circuits over all p-samplable distributions. %Note that if we can lear p-size circuits over the uniform distribution we can learn p-size circuits even if the queries to f are drawn from a distribution samplable by a p-size circuit R. This is because instead of learning f over the distribution given by R we can learn f\circ R over the unifrom distribution - in such case the quality of the output of the learner is measured with respect to the distribution given by R?
\bigskip

\noindent {\bf Boosting confidence and reducing error.} The confidence of the learner can be efficiently boosted in a standard way. Suppose an $s$-size circuit $L^f$ learns $f$ up to error $\epsilon$ with confidence $\delta$. We can then run $L^f$ $k$ times, test the output of $L^f$ from every run with $m$ new random queries and output the most accurate one. By Hoeffding's inequality, $m$ random queries fail to estimate the error $\epsilon$ of an output of $L^f$ up to $\gamma$ with probability at most $2/e^{2\gamma^2m}$. %P[|Xm-EXm|\ge gamma m]\le 2/e^{2gamma^2m}; EXm=mEX1=m\sum_x Pr[hypothesis correct on x=X1]=mPr[hypothesis correct], Xm=estimated correctness of hypothesis
Therefore the resulting circuit of size $poly(s,m,k)$ learns $f$ up to error $\epsilon+\gamma$ with confidence at least $1-2k/e^{2\gamma^2m}-(1-\delta)^k\ge 1-2k/e^{2\gamma^2m}-e^{-k\delta}$. If we are trying to learn small circuits  we can get even confidence 1 by fixing internal randomness of learner nonuniformly without losing much on the running time or the error of the output. It is also possible to reduce the error up to which $L^f$ learns $f$ without a significant blowup in the running time and confidence. If we want to learn $f$ with a better error, we first learn an amplified version of $f$, $Amp(f)$. Employing direct product theorems and Goldreich-Levin reconstruction algorithm, Carmosino et. al. \cite[Lemma 3.5]{CIKK} showed that for each $0<\epsilon,\gamma<1$ it is possible to map a Boolean function $f$ with $n$ inputs to a Boolean function $Amp(f)$ with $poly(n,1/\epsilon,\log(1/\gamma))$ inputs so that $Amp(f)\in \Ppoly^f$ and there is a probabilistic $poly(|C|,n,1/\epsilon,1/\gamma)$-time machine which given a circuit $C$ $(1/2+\gamma)$-approximating $Amp(f)$ and an oracle access to $f$ outputs with high probability a circuit $(1-\epsilon)$-approximating $f$. %We thus typically ignore the optimisation of the confidence and error parameter in the rest of the paper.
We can thus often ignore the optimisation of the confidence and error parameter. Note, however, that the error reduction of Carmosino et al. requires membership queries.
\bigskip

\noindent{\bf Natural proofs vs learning algorithms.} %rewritethisparasmedzeramiapredchadzajucibigskip
Natural proofs are actually equivalent to efficient learning algorithms with suitable parameters. In this paper we need just one implication.

\begin{theorem}[Carmosino-Impagliazzo-Kabanets-Kolokolova \cite{CIKK}]\label{t:cikkcore}
Let $R$ be a $\Ppoly$-natural property useful against $\Circuit[n^k]$ for $k \geq 1$. Then, for each $\gamma\in (0,1)$, $\Circuit[n^{k\gamma/a}]$ is learnable by $\Circuit[2^{O(n^{\gamma})}]$ over the uniform distribution with non-adaptive membership queries, confidence 1, up to error $1/n^{k\gamma/a}$, where %$k = \frac{d\gamma}{a}$ and 
$a$ is an absolute constant.
\end{theorem}

\subsection{Bounded arithmetic and propositional logic}\label{sba}

Theories of bounded arithmetic capture various levels of feasible reasoning and present a uniform counterpart to propositional proof systems.

The first theory of bounded arithmetic formalizing p-time reasoning was introduced by Cook \cite{Cpv} as an equational theory \pv. We work with its first-order conservative extension \PV from \cite{KPT}. The language of \PV, denoted \pv as well, consists of symbols for all p-time algorithms given by Cobham's characterization of p-time functions, cf. \cite{Cptime}. A \pv-formula is a first-order formula in the language \pv. $\Sigma^b_0$ (=$\Pi^b_0$) denotes \pv-formulas with only sharply bounded quantifiers $\exists x, x\leq |t|$, $\forall x, x\leq |t|$, where %$t$ is a term not containing $x$ and 
$|t|$ is ``the length of the binary representation of $t$". Inductively, $\Sigma^b_{i+1}$ resp. $\Pi^b_{i+1}$ is the closure of $\Pi^b_i$ resp. $\Sigma^b_i$ under positive Boolean combinations, sharply bounded quantifiers, and bounded quantifiers $\exists x, x\le t$ resp. $\forall x, x\le t$. %rewriteodtialpotial predchadzajucu vetu s medzerami nahraduzjucu: For $i\geq 0$, $\Sigma^b_{i+1}$ and $\Pi^b_{i+1}$ are defined inductively by counting the number of alternations of bounded quantifiers $\exists x, x\leq t$, $\forall x, x\leq t$, ignoring the sharply bounded ones 
Predicates definable by $\Sigma^b_{i}$ resp. $\Pi^b_{i}$ formulas are in the $\Sigma^p_{i}$ resp. $\Pi^p_{i}$ level of the polynomial hierarchy, and vice versa. %The hierarchy of $\Sigma^b_i(PV)$- and $\Pi^b_i(PV)$-formulas is defined similarly to $\Sigma^b_i$ and $\Pi^b_i$ (in first-order logic with equality) but in the language of $PV$. 
\PV is known to prove $\Sigma^b_0(\pv)$-induction: $$A(0)\wedge \forall x\ (A(x)\rightarrow A(x+1))\rightarrow \forall x A(x),$$ for $\Sigma^b_0$-formulas $A$, cf. Kraj\'{i}\v{c}ek \cite{Kba}. %(Every p-time function can be straightforwardly defined as a well-behaved \pv-function, so we can interpret provability in \PV as capturing the idea of what can be demonstrated when our reasoning is restricted to manipulation of p-time objects.)%rewrite':'and','predapodisplayedflasvtomtoadalsompara

%Buss \cite{Bba} gave an elegant formalization of p-time reasoning as a theory \SB in the language $L:=\{0,S,+,\cdot, =,\leq, \lfloor \frac{x}{2}\rfloor, |x|,\#\}\subseteq\pv$ where $x\# y=2^{|x|\cdot |y|}$. \SB consists of a finite set of axioms capturing basic properties of symbols in $L$ and a polynomial induction: $$A(0)\wedge \forall x, (A(\lfloor x/2 \rfloor)\rightarrow A(x))\rightarrow \forall x A(x)$$ for $\Sigma^b_1(L)$-formulas $A$ (defined as $\Sigma^b_1$ but in the language $L$). 
Buss \cite{Bba} introduced the theory \SB extending \PV with the $\Sigma^b_1$-length induction: $$A(0)\wedge \forall x<|a|, (A(x)\rightarrow A(x+1))\rightarrow A(|a|),$$%rewriteoxpredch' A(|a|)'anaslmedzera
 for $A\in\Sigma^b_1$. \SB proves the sharply bounded collection scheme $BB(\Sigma^b_1)$: $$\forall i<|a|\ \exists x<a, A(i,x)\rightarrow \exists w\ \forall i<|a|, A(i,[w]_i),$$ for $A\in\Sigma^b_1$ ($[w]_i$ is the $i$th element of the sequence coded by $w$), which is unprovable in \PV under a cryptographic assumption, cf. \cite{CTcol}. On the other hand, \SB is $\forall\Sigma^b_1$-conservative over \PV. This is a consequence of Buss's witnessing theorem stating that $\SB\vdash \exists y, A(x,y)$ for $A\in\Sigma^b_1$ implies $\PV\vdash A(x,f(x))$ for some \pv-function $f$. %When proving a $\Sigma^b_2$ formula in \SB we are free to use the sharply bounded collection scheme for $A\in\Sigma^b_2$, denoted $BB(\Sigma^b_2)$, because $\SB+BB(\Sigma^b_2)$ is $\forall\Sigma^b_2$-conservative over \SB, cf. \cite{Rcon}.%rewrite'$\Sigma^b_1$-length induction:'and'%'predposlednouvetou

Following a work by Kraj\'{i}\v{c}ek \cite{Kwphp}, Je\v{r}\'{a}bek \cite{Jwphp,Jphd,Japx} systematically developed a theory \APC capturing probabilistic p-time reasoning by means of approximate counting.\footnote{Kraj\'{i}\v{c}ek \cite{Kwphp} introduced a theory $BT$ defined as $\SB+dWPHP(\pv)$ and proposed it as a theory for probabilistic p-time reasoning.} The theory \APC is defined as $\PV+dWPHP(\pv)$ where $dWPHP(\pv)$ stands for the dual (surjective) pigeonhole principle for \pv-functions, i.e. for the set of all formulas $$x>0\rightarrow\exists v<x(|y|+1)\forall u<x|y|,\ f(u)\neq v,$$ where $f$ is a \pv-function %rewriteoxpredchanaslmedzera
 which might involve other parameters not explicitly shown. We devote Section \ref{approximatec} to a more detailed description of the machinery of approximate counting in \APC.%rewritevtomtopara'Following.. systematically 'celufootnotes'.'prednou,dalejvdisplayedfla','a',which might.. not explicitly shown.'
\bigskip

Any $\Pi^b_1$-formula $\phi$ provable in \PV can be expressed as a sequence of tautologies $||\phi||_n$ with proofs in the Extended Frege system \EF which are constructible in p-time (given a string of the length $n$), cf. \cite{Cpv}. Similarly, $\Pi^b_1$-formulas provable in \APC translate to tautologies with p-time constructible proofs in \WF, an extension of \EF introduced by Je\v{r}\'{a}bek \cite{Jwphp}. We describe the translation and system \WF in more detail below. %, see Definition \ref{d:wf}. We will work with the presentation of the translation from \cite{Jwphp}. %In the same way \V resp. \VNC correspond to \AC-\F resp. \F. (see wphp.pdf)%rewritezvysoksubsekcieod'Jwphp'ktomu'\phi'a'||phi||_n'vpredchpara

As it is often easier to present a proof in a theory of bounded arithmetic than in the corresponding propositional system, bounded arithmetic functions, so to speak, as a uniform language for propositional logic.
\smallskip

We refer to Kraj\'i\v{c}ek \cite{Kpc} for basic notions in proof complexity.

\begin{definition}[\WF (WPHP Frege), cf. Je\v{r}\'abek \cite{Jwphp}]\label{d:wf} Let $L$ be a finite and complete language for propositional logic, i.e. $L$ consists of finitely many boolean connectives of constant arity such that each boolean function of every arity can be expressed %srewrite'expressed '
by an $L$-formula, and let $\mathcal{R}$ be a finite, sound and implicationally complete set of Frege rules (in the langauge $L$). A \WF-proof of a ($L$-)circuit $A$ is a sequence of circuits $A_0,\dots,A_k$ such that $A_k=A$, and each $A_i$ is derived from some $A_{j_1},\dots, A_{j_{\ell}}$, $j_1,\dots,j_{\ell}<i$ by a Frege rule from $\mathcal{R}$, or it is similar to some $A_j$, $j<i$, or it is the dWPHP axiom, $$\bigvee_{\ell=1}^m (r_{\ell}\ne C_{i,\ell}(D_{i,1},\dots,D_{i,n})),$$ where $n<m$ and $r_{\ell}$ are pairwise distinct variables which do not occur in circuits $A$, $C_{i,\ell'}$, or $A_j$ for $j<i$, but may occur in circuits $D_{i,1},\dots, D_{i,n}$.
\end{definition}

The similarity rule in Definition \ref{d:wf} is verified by a specific p-time algorithm %rewriteoxpredchanaslmedzera
 which checks that circuits $A_i$ and $A_j$ can be `unfolded' to the same (possible huge) formula, cf. \cite[Lemma 2.2.]{Jwphp}. Intuitively, the $\mathsf{NLOG}$ ($\subseteq \Ptime$) algorithm recognizes if two circuits are not similar by guessing a partial path through them, going from the output to the inputs, where on at least one instruction the circuits disagree. As defined \WF depends on the choice of Frege rules and language $L$, but for each choice the resulting systems are p-equivalent, so we can identify them. The dWPHP axiom refers to `dual weak pigeonhole principle' postulating the existence of an element $r_1,\dots,r_m$ %rewriteoxpredch' r_m'anaslmedzera
outside the range of a p-size map $C_{i,1},\dots,C_{i,m}:\{0,1\}^n\mapsto \{0,1\}^m$. %srewrite' m}'toistevdalsejvete
The dWPHP axiom comes with a specification of circuits $C_{i,1},\dots,C_{i,m}, D_{i,1},\dots,D_{i,n}$ so that we can recognize the axiom efficiently.
\smallskip

The translation of a $\Pi^b_1$ formula $\phi$ into a sequence of propositional formulas $||\phi||_{\overline{n}}$ works as follows. For each $\pv$-function $f(x_1,\dots,x_k)$ and numbers $n_1,\dots,n_k$ we have a p-size circuit $C_f$ computing the restriction $f:2^{n_1}\times \dots\times 2^{n_k}\mapsto 2^{b(n_1,\dots,n_k)}$, where $b$ is a suitable `bounding' polynomial for $f$. The formula $||f||_{\overline{n}}(p,q,r)$ expresses that $C_f$ outputs $r$ on input $p$, with $q$ being the auxiliary variables corresponding to the nodes of $C_f$.  The formula $||\phi(x)||_{\overline{n}}(p,q)$ is defined as $||\phi'(x)||_{\overline{n}}(p,q)$, where $\phi'$ is the negation normal form of $\phi$, i.e. negations in $\phi'$ are only in front of atomic formulas. The formula $||\phi'(x)||_{\overline{n}}(p,q)$ %rewriteoxpredch'The formula phi(x)..  q) 'naslmedzeraanasl'$\vee,'
 is defined inductively in a straightforward way so that $||\dots||$ commutes with $\vee,\wedge$. The atoms $p$ correspond to variables $x$, atoms $q$ correspond to the universally quantified variables of $\phi$ and to the outputs and auxiliary variables of circuits $C_f$ for functions $f$ appearing in $\phi$. Sharply bounded quantifiers are replaced by polynomially big conjuctions resp. disjunctions. For the atomic formulas we have, \begin{align*}||f(x)=g(x)||_{\overline{n}}:= & ||f(x)||_{\overline{n}}(p,q,r)\wedge ||g(x)||_{\overline{n}}(p,q',r')\rightarrow \bigwedge_i r_i=r'_i,\\
||\neg f(x)=g(x)||_{\overline{n}}:= & ||f(x)||_{\overline{n}}(p,q,r)\wedge ||g(x)||_{\overline{n}}(p,q',r')\rightarrow \neg\bigwedge_i r_i=r'_i,\\
||f(x)\le g(x)||_{\overline{n}}:= & ||f(x)||_{\overline{n}}(p,q,r)\wedge ||g(x)||_{\overline{n}}(p,q',r')\rightarrow \bigwedge_i (r_i\wedge \bigwedge_{j>i}(r_j=r_j')\rightarrow r'_i),\\
||\neg f(x)\le g(x)||_{\overline{n}}:= & ||f(x)||_{\overline{n}}(p,q,r)\wedge ||g(x)||_{\overline{n}}(p,q',r')\rightarrow \neg \bigwedge_i(r_i\wedge\bigwedge_{j>i}(r_j=r_j')\rightarrow r'_i).\end{align*}

\subsection{Approximate counting}\label{approximatec}
%rewriteodtialpotial next sentence a 2x (tiez v %) if s medzerami after i.e. in the 3rd paragraf
In order to prove our results % formalize notions such as PAC learning in \WF 
we will need to use Je\v{r}\'{a}bek's theory of approximate counting. This section recalls the properties of \APC we will need.
\medskip

%The dual (or surjective) pigeonhole principle for $f$, written as $dWPHP(f)$, is the universal closure of the formula $$x>0\rightarrow\exists v<x(|y|+1)\forall u<x|y|,\ f(u)\neq v.$$ For a set of functions $\Gamma$, $dWPHP(\Gamma):=\{dWPHP(f)\ |\ f\in\Gamma\}$.

%The theory \APC is defined as $\PV+dWPHP(\pv)$ where $dWPHP(\pv)$ stands for the dual (surjective) pigeonhole principle for \pv-functions, i.e. the set of all formulas $$x>0\rightarrow\exists v<x(|y|+1)\forall u<x|y|,\ f(u)\neq v$$ where $f$ is a \pv-functions.
%\medskip

By a definable set we mean a collection of numbers satisfying some formula, possibly with parameters. When a number $a$ is used in a context which asks for a set it is assumed to represent the integer interval $[0,a)$, e.g. $X\subseteq a$ means that all elements of set $X$ are less than $a$. If $X\subseteq a$, $Y\subseteq b$, then $X\times Y:=\{bx+y\mid x\in X, y\in Y\}\subseteq ab$ and $X\dot{\cup} Y:=X\cup\{y+a\mid y\in Y\}\subseteq a+b$. Rational numbers are assumed to be represented by pairs of integers in the natural way. We use the notation $x\in Log\leftrightarrow \exists y,\ x=|y|$ and $x\in LogLog\leftrightarrow \exists y,\ x=||y||$.%rewriteposlveta,medzeraprednouavnaslvete'Let '
\medskip

Let $C: 2^n\rightarrow 2^m$ be a circuit and $X\subseteq 2^n, Y\subseteq 2^m$ definable sets. %We write $C:X\rightarrow Y$ if $C[X]\subseteq Y$, i.e. $\forall x\in X\ C(x)\in Y$.  We write $C:X\hookrightarrow Y$ if $C[X]\subseteq Y$ and the function computed by $C$ is injective on $X$. 
We write $C:X\twoheadrightarrow Y$ if $\forall y\in Y\exists x\in X,\ C(x)=y$.
Je\v{r}\'{a}bek \cite{Japx} gives the following definitions in \APC (but they can be considered in weaker theories as well).

%\begin{theorem}[Je\v{r}\'{a}bek \cite{Japx}] (in $APC_1$) Let $C:2^n\rightarrow 2$ be a boolean circuit and $\epsilon^{-1}\in Log$. Denote $X:=\{x< 2^n| C(x)=1\}$. 

%There exists $s\leq 2^n, v\leq poly(n\epsilon^{-1}|C|)$, and circuits $G_i, H_i; i=0,1,$ of size $poly(n\epsilon^{-1}|C|)$ such that $$G_0: v(s+\epsilon 2^n) \twoheadrightarrow v\times X \ \ \ \ \ \ H_0: v\times X\hookrightarrow v(s+\epsilon 2^n)$$ $$G_1: v\times (X \dot{\cup} \epsilon 2^n ) \twoheadrightarrow vs \ \ \ \ \ \ H_1: vs \hookrightarrow v\times (X\dot{\cup} \epsilon 2^n)$$ and such that $$G_i\circ H_i=id$$ on their respective domains. 
%\end{theorem}

\begin{definition} Let $X,Y\subseteq 2^n$ be definable sets, and $\epsilon\leq 1$. The size of $X$ is approximately less than the size of $Y$ with error $\epsilon$, written as $X\preceq_{\epsilon} Y$, if there exists a circuit $C$, and $v\neq 0$ such that $$C: v\times (Y\dot{\cup}\epsilon 2^n)\twoheadrightarrow v\times X.$$ $X\approx_{\epsilon}Y$ stands for $X\preceq_{\epsilon} Y$ and $Y\preceq_{\epsilon} X$.
\end{definition}

Since a number $s$ is identified with the interval $[0,s)$, $X\preceq_{\epsilon} s$ means that the size of $X$ is at most $s$ with error $\epsilon$.

%\begin{definition} Let $X\subseteq 2^{|t|}$ be a definable set and $0\leq \epsilon, p\leq 1$. We define $$\Pr_{x<t}[x\in X]\preceq_{\epsilon}p\ \ \ \text{iff}\ \ \ X\cap t\preceq_{\epsilon} pt$$ and similarly for $\approx_\epsilon$.
%\end{definition}%rewriteodtialpotial tuto poznamku: Moritz's extended definition follows from this one in case $Z\subseteq 2^{|t|+|s|}$ where $t$,$s$ are powers of 2 (and this is how we use it in Razborov-Smolensky's aproximation), otherwise we will always have $Z\subseteq ts\subseteq 2^{|ts|}$ (and the coding of $<x,y>\in Z$ can be assumed to be z=xs+y), so the extra definition is not needed.%rewritevtomtopara2x'%'predbeginaend

The definition of $X\preceq_{\epsilon} Y$ is an unbounded $\exists \Pi^b_2$-formula even if $X,Y$ are defined by circuits so it cannot be used freely in bounded induction. Je\v{r}\'{a}bek \cite{Japx} solved this problem by working in ${\sf HARD}^A$, a conservative extension of \APC, defined as a relativized theory $\PV(\alpha)+dWPHP(\pv(\alpha))$ extended with axioms postulating that $\alpha(x)$ is a truth-table of a function on $||x||$ variables hard on average for circuits of size $2^{||x||/4}$, see Section \ref{ss:lform}. In ${\sf HARD}^A$ there is a $\PV(\alpha)$ function $Size$ approximating the size of any set $X\subseteq 2^n$ defined by a circuit $C$ so that $X\approx_{\epsilon} Size(C,2^n,2^{\epsilon^{-1}})$ for $\epsilon^{-1}\in Log$, cf. \cite[Lemma 2.14]{Japx}. %rewriteoxpredch' cf. ..Japx}. 'anaslmedzera
 If $X\cap t\subseteq 2^{|t|}$ is defined by a circuit $C$ and $\epsilon^{-1}\in Log$, we can define %rewriteoxpredch' can define 'anaslmedzera 
 $$\Pr_{x<t}[x\in X]_{\epsilon}:=\frac{1}{t} Size(C,2^{|t|},2^{\epsilon^{-1}}).$$%rewritevposlvete'\cap t\subseteq 2^|t|'avtomtopara', see Section \ref{ss:lform}.'

\def\relatth{
\begin{definition}[in \PV] Let $f:2^k\rightarrow 2$ be a truth-table of a Boolean function with $k$ inputs ($f$ is encoded as a string of $2^k$ bits, hence $k\in LogLog$). We say that $f$ is (worst-case) $\epsilon$-hard, written as $Hard_{\epsilon}(f)$ if no circuit $C$ of size $2^{\epsilon k}$ computes $f$. The function $f$ is average-case $\epsilon$-hard, written as $Hard^A_{\epsilon}(f)$, if for no circuit $C$ of size $\leq 2^{\epsilon k}$: $$|\{u<2^k|C(u)=f(u)\}|\geq (1/2+2^{-\epsilon k})2^k.$$
\end{definition}

\begin{proposition}[Je\v{r}\'{a}bek \cite{Jdual}] For every constant $\epsilon <1/3$ there exists a constant $c$ such that \APC proves: for every $k\in LogLog$ such that $k\geq c$, there exist average-case $\epsilon$-hard functions $f:2^k\rightarrow 2$.  
\end{proposition}

\PV can be relativized to $\PV(\alpha)$. The new function symbol $\alpha$ is then allowed in the inductive clauses for introduction of new function symbols. % in definition \ref{newp}. 
This means that the language of $\PV(\alpha)$, denoted also $\pv(\alpha)$, contains symbols for all p-time oracle algorithms.%rewriteJwphpvtitulkenasldef

\begin{definition}[Je\v{r}\'{a}bek \cite{Jwphp}] The theory ${\sf HARD}^A$ is an extension of the theory $\PV(\alpha)+dWPHP(\pv(\alpha))$ by the axioms
\smallskip

1. $\alpha(x)$ is a truth-table of a Boolean function in $||x||$ variables,
 
2. $x\geq c\rightarrow Hard^A_{1/4}(\alpha(x))$,

3. $||x||=||y||\rightarrow \alpha(x)=\alpha(y)$,
\smallskip

\noindent where $c$ is the constant from the previous proposition. 
\end{definition}

\begin{theorem}[Je\v{r}\'{a}bek \cite{Jdual,Japx}]\label{sec} ${\sf HARD}^A$ is a conservative extension of \APC. Moreover, there is a $\pv(\alpha)$-function $Size$ such that ${\sf HARD}^A$ proves: if $X\subseteq 2^n$ is definable by a circuit $C$, then $$X\approx_{\epsilon} Size(C,2^n,e)$$ where $\epsilon=|e|^{-1}$
\end{theorem} 

We will abuse the notation and write $Size(X,\epsilon)$ instead of $Size(C,2^n,e)$. 

\begin{definition}[in \APC] If $X\subseteq 2^{|t|}$ is defined by a circuit $C$ and $\epsilon^{-1}\in Log$, we put $$\Pr_{x<t}[x\in X]_{\epsilon}:=\frac{1}{t} Size(C,2^{|t|},\epsilon).$$ % Size(X\cap t,\epsilon).$$
\end{definition}
}

The presented definitions of approximate counting are well-behaved:

\begin{proposition}[Je\v{r}\'{a}bek \cite{Japx}](in \PV)\label{lem} Let $X,X',Y,Y',Z\subseteq 2^n$ and $W,W'\subseteq 2^m$ be definable sets, and $\epsilon, \delta<1$. Then

$i)\ \ X\subseteq Y\Rightarrow X\preceq_{0} Y$,

$ii)\ \ X\preceq_{\epsilon} Y \wedge Y\preceq_{\delta} Z\Rightarrow X\preceq_{\epsilon+\delta} Z$,

$iii)\ \ X\preceq_\epsilon X'\wedge W\preceq_{\delta}W'\Rightarrow X\times W\preceq_{\epsilon+\delta+\epsilon\delta} X'\times W'$.

$iv)\ \ X\preceq_\epsilon X'\wedge Y\preceq_{\delta}Y'\ and\ X',Y'\ are\ separable\ by\ a\ circuit,\ then\ X\cup Y\preceq_{\epsilon+\delta}X'\cup Y'$.

\end{proposition}

\begin{proposition}[Je\v{r}\'{a}bek \cite{Japx}](in \APC)\label{really} 

\noindent 1.\ \ Let $X,Y\subseteq 2^n$ be definable by circuits, $s,t,u\leq 2^n$, $\epsilon, \delta,\theta, \gamma < 1, \gamma^{-1}\in Log $. Then

i)\ \ $X\preceq_{\gamma} Y$ or $Y\preceq_{\gamma} X$,

ii)\ \ $s\preceq_\epsilon X\preceq_\delta t\Rightarrow s<t+(\epsilon+\delta+\gamma)2^n$,

iii)\ \ $X\preceq_{\epsilon} Y\Rightarrow 2^n\backslash Y\preceq_{\epsilon+\gamma} 2^n\backslash X $,

iv)\ \ $X\approx_{\epsilon} s\wedge Y\approx_{\delta} t\wedge X\cap Y\approx_{\theta} u\Rightarrow X\cup Y\approx_{\epsilon+\delta+\theta+\gamma} s+t-u$.

\medskip

\noindent 2. (Disjoint union) Let $X_i\subseteq 2^n$, $i<m$ be defined by a sequence of circuits and $\epsilon,\delta\leq 1$, $\delta^{-1}\in Log$. If $X_i\preceq_\epsilon s_i$ for every $i<m$, then $\bigcup_{i<m} (X_i\times \{i\})\preceq_{\epsilon+\delta} \sum_{i<m} s_i$.

\medskip

\noindent 3. (Averaging) Let $X\subseteq 2^n\times 2^m$ and $Y\subseteq 2^m$ be definable by circuits, $Y\preceq_{\epsilon} t$ and $X_y\preceq_{\delta} s$ for every $y\in Y$, where $X_y:=\{x| \left<x,y\right>\in X\}$. Then for any $\gamma^{-1}\in Log$, $$X\cap (2^n\times Y) \preceq_{\epsilon+\delta+\epsilon\delta+\gamma} st.$$

\def\chernoff{\noindent 3. (Chernoff's bound) Let $X\subseteq 2^n, m\in Log, 0\leq \epsilon,\delta,p\leq 1$ and $X\succeq_{\epsilon} p2^n$. Then $$\{w\in (2^n)^m|\ |\{i<m|w_i\in X\}|\leq m(p-\delta)\}\preceq_0 c4^{m(c\epsilon-\delta^2)}2^{nm}$$ for some constant $c$, where $w$ is treated as a sequence of $m$ numbers less than $2^n$ and $w_i$ is its $i$-th member.}
\end{proposition}

%Consequently:

%\begin{proposition}(in $APC_1$)\label{Si} Let $X,Y\subseteq 2^n, a\leq 2^n$ be definable by a circuit and $\epsilon^{-1}\in Log$. Then

%$Size(X\times a,\epsilon)\approx_{2\epsilon} Size(X,\epsilon)\cdot a$

%$Size(X\times Y,\epsilon)\approx_{3e} Size(X,\epsilon) Size(Y,\epsilon)$ 
%\end{proposition}
%rewritetatomedzera

%We conclude this section with an observation that 
When proving $\Sigma^b_1$ statements in \APC we can afford to work in $\SB+dWPHP(\pv)+BB(\Sigma^b_2)$ and, in fact, assuming the existence of a single hard function in \PV gives us the full power of \APC. Here, $BB(\Sigma^b_2)$ is defined as $BB(\Sigma^b_1)$ but with $A\in\Sigma^b_2$. %rewriteoxpredchvetasmedzerami%rewritevnasllemme'$\pv$-functions $g,h$'%srewrite'where f(y).. y>|f|%rewrite', and C_h.. f,x.)'

\begin{lemma}[\cite{MP}]\label{singlehard} Suppose $\SB+dWPHP(\pv)+BB(\Sigma^b_2)\vdash \exists y A(x,y)$ for $A\in\Sigma^b_1$. Then, for every $\epsilon<1$, there is $k$ and $\pv$-functions $g,h$ such that \PV proves $$|f|\ge |x|^k\wedge \exists y, |y|=||f||, C_h(y)\ne f(y)\rightarrow A(x,g(x,f))$$ where $f(y)$ is the $y$th bit of $f$, $f(y)=0$ for $y>|f|$, and $C_h$ is a circuit of size $\leq 2^{\epsilon ||f||}$ generated by $h$ on $f,x$. Moreover, $\APC\vdash\exists y A(x,y)$.
\end{lemma}

\def\prfofMPlemma{
\proof By \cite[Corollary 4.12]{Jdual}, $\SB+dWPHP(\pv)+BB(\Sigma^b_2)\vdash\exists y A(x,y)$ implies $\SB+dWPHP(\pv)\vdash\exists y A(x,y)$. Then, following Thapen's proof of \cite[Theorem 4.2]{Tstr} (cf. also \cite[Proposition 1.14]{Jdual}), there is $\ell$ and $h\in\pv$ such that \SB proves $$(\forall v\leq 2^{8|x|^\ell}\exists u\leq 2^{4|x|^\ell},\ h(u)=v)\vee \exists y A(x,y).$$ By Buss's witnessing theorem it now suffices to show that for every $\epsilon<1$ there is $k$ such that \SB proves \begin{multline*}(\forall v\leq 2^{8|x|^\ell}\exists u\leq 2^{4|x|^\ell},\ h(u)=v)\rightarrow\\ 
(|f|\ge |x|^k\rightarrow \exists\text{ circuit }C\text{ of size} \leq 2^{\epsilon||f||}\ \forall y, |y|=||f||, C(y)=f(y)).\end{multline*} Argue in \SB. The surjection $h:2^m\rightarrow 2^{2m}$, where $m=4|x|^\ell$, is computed by a circuit of size $m^{\ell'}$ for a standard $\ell'$. Following Je\v{r}\'{a}bek's \SB-proof of \cite[Proposition  3.5]{Jdual}, this implies that every (number) $f$ viewed as a truth-table of length $|f|$ is computed by a size $O(m|m|+m^{\ell'}|\lceil |f|/m \rceil|)$ circuit with $||f||$ inputs. % (w.l.o.g. $|x|^l$ is a power of two)
For sufficiently large $k$, $|f|\ge |x|^k$ implies that this size is $\le 2^{\epsilon||f||}$.%rewriteodtialpotial poslednu bodku, multline above and in Lemma 4.1: h in \pv, displayed fla a 'where Ch is... h on f,x.  '

The ``moreover'' part is a consequence of $\APC\vdash \forall n\in LogLog\ \exists f:2^n\rightarrow 2, \LBtt(f,2^{n/4})$, cf. \cite[Corollary 3.3]{Jdual}. 
 \qed}

\def\theroadnottaken{
The counting (and Size function in HARD^A) are built on NW generator, and consequently on the existence of a function hard on average. In order to count it is possible to use a single NW generator, see Corollary in Dai. 
We can push the assumption further. Jerabek formalized hardness amplification, i.e. delta in Thm 2.7 [apx paper] can be arbitrary, not necessarilly 1/4. All in all, for any epsilon and n, PV>Proposition 4.2 assuming the existence of a function hard in worst-case.}

\def\inverselemma{
%rewritedefamedzerabove
\bigskip

Lemma \ref{singlehard} allows us to use the $BB(\Sigma^b_2)$ collection scheme for proving $\Sigma^b_1$-statements in \APC. Unfortunately, when collecting circuits witnessing $\preceq_\epsilon$ predicates given by $\exists\Pi^b_2$ formulas the $BB(\Sigma^b_2)$ collection is a priori not sufficient. To overcome this complication the quantifier complexity of $\preceq_\epsilon$ can be pushed down to $\Sigma^b_2$ because the circuits counting sizes of sets in \APC are invertible.

\begin{lemma}[\cite{MP}]\label{invert}(in \APC) Let $X\subseteq 2^n$ be defined by a circuit and $\epsilon^{-1}\in Log$. Suppose $X\preceq_\epsilon s$. Then, $X\preceq_\epsilon s+3\epsilon2^n$ is expressible by a provable $\Sigma^b_2$ formula.
\end{lemma}

%\proof By \cite[Theorem 2.7]{Japx}, there exists $t$ such that $X\approx_\epsilon t$ is witnessed by invertible circuits of size $poly(n\epsilon^{-1}S)$ where $S$ is the size of the circuit defining $X$. Applying Proposition \ref{really} 1.$ii$) we get $t<s+3\epsilon2^n$. \qed
}
%rewritemedzeratatoa}

\def\pvine{
\subsection{Standard inequalities in \PV}\label{spv}

For a \pv-function symbol $f$ and $n\in Log$, in \PV we can define inductively $\sum_{i=0}^n f(i)$. Similarly, we can define iterated products, factorials, and binomial coefficients. It is easy to see that, by induction, \PV proves: $n\in Log\rightarrow \sum_{i=0}^n {n\choose i}=2^n$. %(VTC0+IMUL needed.)

\begin{proposition}[Stirling's bound, cf. Je\v{r}\'{a}bek \cite{Jdual}]\label{stirling} There is a $c>1$ such that \PV proves: $$0<k<n\in Log\rightarrow \frac{1}{c}{n \choose k}<\frac{n^n}{k^k(n-k)^{n-k}}\left(\left\lfloor\sqrt{\frac{k(n-k)}{n}}\right\rfloor+1\right)^{-1}<c {n \choose k}.$$
\end{proposition}

\begin{proposition}\label{concentration} For each $\epsilon>0$ there is an $n_0$ such that \PV proves: $$n_0<n\in Log\rightarrow \sum_{i=0}^{\lfloor n/2+n^{1/3}\rfloor} {n\choose i}<\left(\frac{1}{2}+\epsilon\right)2^n.$$
\end{proposition}

\proof $\sum_{i=0}^{\lfloor n/2\rfloor-1} {n\choose i}=\frac{1}{2}\left(\sum_{i=0}^{\lfloor n/2\rfloor-1} {n\choose i}+\sum_{i=0}^{\lfloor n/2\rfloor-1} {n\choose n-i}\right)<2^{n-1}$ and by Stirling's bound, for some constant $c>1$, $$\sum_{i=\lfloor n/2\rfloor}^{\lfloor n/2+n^{1/3}\rfloor} {n\choose i}<(n^{1/3}+1){n\choose \lfloor n/2\rfloor}<2^n4c\left(\frac{n^{1/3}}{\lfloor n^{1/2}/2\rfloor}+\frac{1}{\lfloor n^{1/2}/2\rfloor}\right)$$ where to verify the last inequality for odd $n$ we used also the provability of $a,b\in Log$, $b>0\rightarrow (1+a/b)\leq 4^{a/b}$ shown in \cite[Stirling's bound, Claim 1]{Jdual}. \qed

\begin{proposition}\label{expbound} \PV proves: $$a,b\in Log,\ b>a+1\rightarrow (b-a)^b\leq b^b/2^a.$$ Note that the conclusion implies $(1-a/b)\leq 2^{-a/b}$.
\end{proposition}

\proof Proceed as in the proof of Claim 2 in the proof of Stirling's bound \cite{Jdual} but instead of Claim 1 use the inequality $b^b\leq  (b+1)^b/2$. \qed}

\section{Formalizing complexity-theoretic statements}\label{s:formalizations}

\subsection{Circuit lower bounds} An `almost everywhere' formulation of a circuit lower bound for circuits of size $s$ and a function $f$ says that for every sufficiently big $n$, for each circuit $C$ with $n$ inputs and size $s$, there exists an input $y$ on which the circuit $C$ fails to compute $f(y)$.%rewrite'analmosteverywhere'a'foreach'above

If $f:\{0,1\}^n\rightarrow \{0,1\}$ is an \NPtime function and $s=n^k$ for a constant $k$, this can be written down as a $\forall\Sigma^b_2$ formula $\LB(f,n^k)$, $$\forall n,\ n>n_0\ \forall\ \text{circuit}\ C\ \text{of size}\ \leq n^k\ \exists y,\ |y|=n,\ C(y)\neq f(y),$$ where $n_0$ is a constant and $C(y)\neq f(y)$ is a $\Sigma^b_2$ formula stating that a circuit $C$ on input $y$ outputs the opposite value of $f(y)$.

If we want to express $s(n)$-size lower bounds for $s(n)$ as big as $2^{O(n)}$, we add an extra assumption on $n$ stating that $\exists x,\ n=||x||$. That is, the resulting formula $\LBtt(f,s(n))$ has form `$\forall x,n; n=||x||\rightarrow\dots$'. Treating $x,n$ as free variables, $\LBtt(f,s(n))$ %srewritevpredc2vetach'That is, the' has form.. LBtt() ' %rewriteoxpredch' Treating x,n as free variables, '
 is $\Pi^b_1$ %rewriteoxpredch' Pib1 'naslmedzeraanasl'forallSigmab1'
 if $f$ is, for instance, \SAT because $n=||x||$ implies that the quantifiers bounded by $2^{O(n)}$ %on the (numbers whose binary representations code) $s(n)$-size circuits 
are sharply bounded. Moreover, allowing $f\in \mathsf{NE}$ lifts the complexity of $\LBtt(f,s(n))$ just to $\forall\Sigma^b_1$. The function $s(n)$ in $\LBtt(f,s(n))$ is assumed to be a \pv-function with input $x$ (satisfying $||x||=n$). %rewriteposlednavetaavdalsejvete'In terms.. , 'ktomu%preddalsouvetou
\medskip

%To indicate sizes of objects we employ the notation $x\in Log \leftrightarrow \exists y,\ x=|y|$ and $x\in LogLog\leftrightarrow \exists y,\ x=||y||$. For example,
In terms of the $Log$-notation, $\LB(f,n^k)$ implicitly assumes $n\in Log$ while $\LBtt(f,n^k)$ assumes $n\in LogLog$.
By chosing the scale of $n$ we are determining how big objects are going to be `feasible' for theories reasoning about the statement. %rewriteposlednavetaamedzera
 In the case $n\in LogLog$, the truth-table of $f$ (and everything polynomial in it) is feasible. Assuming just $n\in Log$ means that only the objects of polynomial-size in the size of the circuit are feasible. Likewise, the theory reasoning about the circuit lower bound is less %rewriteoxpredch' is less 'naslmedzeraanasl' than with '
powerful when working with $\LB(f,n^k)$ than with $\LBtt(f,n^k)$. (The scaling in $\LBtt(f,s)$ corresponds to the choice of parameters in natural proofs and in the formalizations by Razborov \cite{Rup}.)
\medskip%rewritedalsiparaajsmedskip

We can analogously define formulas $\LBtt(f,s(n),t(n))$ expressing an average-case lower bound for $f$, where $f$ is a free variable (with $f(y)$ being the $y$th bit of $f$ and $f(y)=0$ for $y>|f|$). %rewriteoxpredch' where f is..0.. y>|f|).  'snaslmedzerou 
 More precisely, $\LBtt(f,s(n),t(n))$ generalizes $\LBtt(f,s(n))$ by saying that each circuit of size $s(n)$ fails to compute $f$ on at least $t(n)$ inputs, for \pv-functions $s(n),t(n)$. Since $n\in LogLog$, $\LBtt(f,s(n),t(n))$ is $\Pi^b_1$.%rewriteoxpredch' LBtt.. is b_1. 'snaslmedzerou 
\medskip

\noindent {\bf Propositional version.} An $s(n)$-size circuit lower bound for a function $f:\{0,1\}^n\rightarrow \{0,1\}$ can be expressed by a $2^{O(n)}$-size propositional formula ${\sf tt}(f,s)$, $$\bigvee_{y\in\{0,1\}^n} f(y)\neq C(y)$$ where the formula $f(y)\neq C(y)$ says that an $s(n)$-size %rewriteoxpredch' an s(n)-size 'naslmedzeraanaslveta' The values.. bits. ' 
 circuit $C$ represented by $poly(s)$ variables does not output $f(y)$ on input $y$. The values $f(y)$ are fixed bits. That is, the whole truth-table of $f$ is hard-wired in ${\sf tt}(f,s)$. 

The details of the encoding of the formula $\ttable(f,s)$ are not important for us as long as the encoding is natural because systems like \EF considered in this paper can reason efficiently about them. We will assume that $\ttable(f,s)$ is the formula resulting from the translation of $\Pi^b_1$ formula $\LBtt(h,s)$, where $n_0=0$, $n,x$ are substituted after the translation by fixed constants so that $x=2^{2^n}$, %rewriteoxpredch', n,x.. 2^2^n, 'anaslmedzera
 and $h$ is a free variable (with $h(y)$ being the $y$th bit of $h$ and $h(y)=0$ for $y>|h|$%srewritepredch' and h(y)=0.. y>|h| 'smedzerou 
) which is substituted after the translation by constants defining $f$.%rewrite'where $n_0=0$ and'a'\ttable(f,s)' %rewritetentoparasmedzerami%rewriteoxpredch' by constants'

%A more succinct encoding follows from a result of Lipton and Young \cite{Lanti} who showed that whenever $f:\{0,1\}^n\rightarrow \{0,1\}$ is hard for circuits of size $poly(s(n))$, there is a set $S_n$ of $poly(s(n))$ $n$-bit strings such that each $s(n)$-size circuit fails to compute $f$ on some input from the ``anti-checking'' set $S_n$. The $s(n)$-size circuit lower bound for $f$ can be then expressed by a $poly(s(n))$-size formula ${\sf lb}_A(f,s)$, $$\bigvee_{y\in S_n} f(y)\neq C(y).$$%rewrite%nazaciatkutohtoparaanasledujucedefpricomuposlednejzatvorky}rewritepredchadzajucumedzeruvriadkuvyssiea}
\def\erasepartfor{
Even more feasible, uniform, encoding follows from translations of $\LB(f,n^k)$. This requires an efficient witnessing of existential quantifiers in $\LB(f,n^k)$ collapsing its complexity to $\forall\Sigma^b_0$. Such a p-time witnessing of $\LB(\SAT,n^k)$ follows, for example, from the existence of one-way permutations and a function in $\mathsf{E}$ hard for subexponential-size circuits, cf. \cite[Proposition 4.3]{clba} \footnote{Proposition 4.3 in \cite{clba} shows just the existence of an S-T protocol witnessing $\LB(\SAT,n^k)$ but the p-time witnessing easily follows. %but it is easy to modify the  the extra advice is actually not needed: if $S$ goes for an assignment inconsistent with $a_i$ we obtain an unsatisfiable formula. In fact, it is possible to generate a set of $poly(n)$ \SAT instances such that each circuit of size $n^k$ fails to compte \SAT on some of them.
}. Further, by the KPT theorem \cite{KPT}, whenever $\PV\vdash\LB(f,n^k)$ we get a sequence of finitely many p-time functions $\overline{w}=w_1,\dots,w_c$ witnessing the existential quantifiers in $\LB(f,n^k)$. %The same interactive witnessing follows from $\APC\vdash\LB(f,n^k)$ but with probabilistic p-time functions (or, alternatively, p-size circuits). 
$\LB(f,n^k)$ witnessed by $\overline{w}$ can be equivalently expressed by a sequence of $poly(n)$-size propositional formulas ${\sf lb}_{\overline{w}}(f,n^k)$. %Throughout this paper, we work only with formulas ${\sf lb}(f,s(n))$ resulting from the provability of $\LB(f,s(n))$, so the witnessing functions are coming out of the KPT theorem. This is also the case in the following observation concluding that a formalization of a circuit lower bound in \APC implies short propositional proofs of tautologies encoding the lower bound.
}

Analogously, we can express average-case lower bounds by propositional formulas $\ttable(f,s(n),t(n))$ obtained by translating $\LBtt(h,s(n),t(n)2^n)$, with $n_0=0$, fixed $x=2^{2^n}$ and $h$ substituted after the translation by $f$.%rewritetutovetusmedzerami%rewriteoxpredch', with n0=0.. f. '

\subsection{Learning algorithms}\label{ss:lform}
%rewriterestofsubsectionajejlabel
A circuit class $\mathcal{C}$ is defined by a \pv-formula if there is a \pv-formula defining the predicate $C\in\mathcal{C}$. Definition \ref{d:lear} can be formulated in the language of ${\sf HARD}^A$: A circuit class $\mathcal{C}$ (defined by a \pv-formula) is learnable over the uniform disribution by a circuit class $\mathcal{D}$ (defined by a \pv-formula) up to error $\epsilon$ with confidence $\delta$, %if there are oracle circuits $L^f$ from $\mathcal{D}$ such that for every Boolean function $f:\{0,1\}^n\mapsto\{0,1\}$ (represented by its truth-table) computable by a circuit from $\mathcal{C}$, when given oracle access to $f$ and input $1^n$, $L^f$ outputs the description of a circuit $(1-\epsilon)$-approximating $f$. 
if there are randomized oracle circuits $L^f$ from $\mathcal{D}$ such that for every Boolean function $f:\{0,1\}^n\mapsto\{0,1\}$ (represented by its truth-table) computable by a circuit from $\mathcal{C}$, for each $\gamma^{-1}\in Log$, when given oracle access to $f$, input $1^n$ and the internal randomness $w \in \{0,1\}^*$, $L^f$ outputs the description of a circuit satisfying 
  \begin{equation*}
    \Pr_w [ L^f(1^n, w) \text{ } (1-\epsilon) \text{-approximates } f ]_{\gamma} \geq \delta.%srewritepredch'.'medzerazaequationa', for each gamma in Log, 'vpredchpara
  \end{equation*}
The inner probability of approximability of $f$ by $L^f(1^n,w)$ is counted exactly. This is possible because $f$ is represented by its truth-table, which implies that $2^n\in Log$.\footnote{It could be interesting to develop systematically a standard theory of learning algorithms in \APC and \WF, but it is not our goal here. Note, for example, that when we are learning small circuits %rewriteoxpredch', for example..circuits'anaslmedzera
 it is not clear how to boost the confidence to 1 in \APC, because we don't have counting with exponential precision.}

\medskip
\noindent {\bf Propositional version.} %If circuit class $\mathcal{C}$ is defined by a \pv-function, the statement that a given oracle algorithm $L$ (given by a \pv-function with oracle queries) learns a circuit class $\mathcal{C}$ over the uniform distribution up to error $\epsilon$ with confidence $\delta$ is $\Pi^b_1$ except that it contains a $\PV(\alpha)$-function $Size$. 
In order, to translate the definition of learning algorithms to propositional formulas we need to look more closely at the definition of  ${\sf HARD}^A$.

%\begin{definition}[in \PV] Let $f:2^k\rightarrow 2$ be a truth-table of a Boolean function with $k$ inputs ($f$ is encoded as a string of $2^k$ bits, hence $k\in LogLog$). We say that $f$ is (worst-case) $\epsilon$-hard, written as $Hard_{\epsilon}(f)$ if no circuit $C$ of size $2^{\epsilon k}$ computes $f$. The function $f$ is average-case $\epsilon$-hard, written as $Hard^A_{\epsilon}(f)$, if for no circuit $C$ of size $\leq 2^{\epsilon k}$: $$|\{u<2^k|C(u)=f(u)\}|\geq (1/2+2^{-\epsilon k})2^k.$$
%\end{definition}

\PV can be relativized to $\PV(\alpha)$. The new function symbol $\alpha$ is then allowed in the inductive clauses for introduction of new function symbols. % in definition \ref{newp}. 
This means that the language of $\PV(\alpha)$, denoted also $\pv(\alpha)$, contains symbols for all p-time oracle algorithms.

\begin{proposition}[Je\v{r}\'{a}bek \cite{Jwphp}]\label{p:hardf} For every constant $\epsilon <1/3$ there exists a constant $n_0$ such that \APC proves: for every $n\in LogLog$ such that $n>n_0$, there exist a function $f:2^n\rightarrow 2$ such that no circuit of size $2^{\epsilon n}$ computes $f$ on $\ge (1/2+1/2^{\epsilon n})2^n$ inputs. %srewrite2xvpredchvete '\epsilon n'
\end{proposition}

\begin{definition}[Je\v{r}\'{a}bek \cite{Jwphp}]\label{d:hard} The theory ${\sf HARD}^A$ is an extension of the theory $\PV(\alpha)+dWPHP(\pv(\alpha))$ by the axioms
\smallskip

1. $\alpha(x)$ is a truth-table of a Boolean function in $||x||$ variables,
 
2. $\LBtt(\alpha(x),2^{||x||/4},2^{||x||}(1/2-1/2^{||x||/4}))$, for constant $n_0$ from Proposition \ref{p:hardf},

3. $||x||=||y||\rightarrow \alpha(x)=\alpha(y)$.
\end{definition}

By inspecting the proof of Lemma 2.14 in \cite{Japx}, we can observe that on each input $C,2^n,2^{\epsilon^{-1}}$ the $\PV(\alpha)$-function $Size$ calls $\alpha$ just once (to get the truth-table of a hard function %rewriteoxpredchanaslmedzera
 which is then used as the base function of the Nisan-Wgiderson generator). In fact, $Size$ calls $\alpha$ on input $x$ which %rewriteoxprech' Size calls.. which 'anadchmedzera
 depends only on $|C|$, the number of inputs of $C$ and w.l.o.g. also just on $|\epsilon^{-1}|$ (since decreasing $\epsilon$ leads only to a better approximation). In combination with the fact that the approximation $Size(C,2^n,2^{\epsilon^{-1}})\approx_{\epsilon} X$, for $X\subseteq 2^n$ defined by $C$, is not affected by a particular choice of the hard boolean function generated by $\alpha$, we get that \APC %srewritepredch' C,2^n2^e^-1 'predch2vetysmedzerami,vcelomparaajsdisplayedfla'\epsilon^-1'a'\approx_e X'vdisplayedfla
 proves $$\LBtt(y,2^{||y||/4},2^{||y||}(1/2-1/2^{||y||/4}))\wedge ||y||=S(C,2^n,2^{\epsilon^{-1}}) \rightarrow\ Sz(C,2^n,2^{\epsilon^{-1}},y)\approx_{\epsilon} X,$$ where $Sz$ is defined as $Size$ with the only difference that the call to $\alpha(x)$ on $C,2^n,2^{\epsilon^{-1}}$ is replaced by $y$ and $S(C,2^n,2^{\epsilon^{-1}})=||x||$ for a \pv-function $S$. 

This allows us to express $\Pr_{x<t}[x\in X]_{\epsilon}=a$, where $\epsilon^{-1}\in Log$ and $X\cap t\subseteq 2^{|t|}$ is defined by a circuit $C$, without a $\PV(\alpha)$ function, by formula%srewritevpredchvete'=a''\cap t'' by formula'celudisplayedflaokremLBtt()\wedge
 $$\forall y\ (\LBtt(y,2^{||y||/4},2^{||y||}(1/2-1/2^{||y||/4}))\wedge ||y||=S(C,2^{|t|},2^{\epsilon^{-1}})\rightarrow Sz(C,2^{|t|},2^{\epsilon^{-1}},y)/t=a).$$ %for each $||y||\le O(\log(|t|,\epsilon^{-1}))$. Since $\alpha(x)$ depends just on $||x||$, we are adding only $O(\log(|t|,\epsilon^{-1}))$ extra assumptions. %srewrite%zadisplayedflaavnasled2vetach'We denote.. in'avposlvete'/t'
We denote the resulting formula by $\Pr^y_{x<t}[x\in X]_{\epsilon}=a$. We will use the notation $\Pr^y_{x<t}[x\in X]_{\epsilon}$ in equations with the intended meaning that the equation holds for the value $Sz(\cdot,\cdot,\cdot,\cdot)/t$ under corresponding assumptions. For example, $t\cdot \Pr^y_{x<t}[x\in X]_{\epsilon}\preceq_{\delta} a$ stands for `$\forall y, \exists v,\exists$ circuit $\hat{C}$ (defining a surjection) which witnesses that $\LBtt(y,2^{||y||/4},2^{||y||}(1/2-1/2^{||y||/4}))\wedge ||y||=S(C,2^{|t|},2^{\epsilon^{-1}})$ implies $Sz(C,2^{|t|},2^{\epsilon^{-1}},y)\preceq_{\delta} a$'. %srewritepredchvetu%rewriteox3x',y'vSz()above1x',\cdot'vSzabove' hat{C} (defining a surjection) which witnesses 'anakoniec'\wedge 'vpredchvete

The definition of learning can be now expressed without a $\PV(\alpha)$ function: If circuit class $\mathcal{C}$ is defined by a \pv-function, the statement that a given oracle algorithm $L$ (given by a \pv-function with oracle queries) learns a circuit class $\mathcal{C}$ over the uniform distribution up to error $\epsilon$ with confidence $\delta$ can be expressed as before with the only difference that %srewrite' can be expressed.. that'
 we replace %the function $Size$ in
% \begin{equation*}
    $\Pr_w [ L^f(1^n, w) \text{ } (1-\epsilon) \text{-approximates } f ]_{\gamma} \geq \delta$ %srewritemedzerazadelta$
%  \end{equation*}
by 
 \begin{equation*}
    \Pr^y_w [ L^f(1^n, w) \text{ } (1-\epsilon) \text{-approximates } f ]_{\gamma} \geq \delta. %srewritetato'.'anasl3%
  \end{equation*}
%is an implication of the form $A\rightarrow B$ for $\Pi^b_1$-formulas $A,B$, where $A$ stands for the conjunction of
%the function $Sz$ and add 
%assumptions $\LBtt(y,2^{||y||/4},2^{||y||}(1/2-1/2^{||y||/4}))\wedge ||y||=S(\cdot,\cdot,\cdot)$, for each $||y||\le O(\log(\cdot,\gamma^{-1}))$. %Since $\alpha(x)$ depends just on $||x||$, we are adding only $2^{O(n)}$ extra assumptions.

Since the resulting formula $A$ defining learning %rewriteox' defining learning 'anaslmedzera
 is not $\Pi^b_1$ (because of the assumption $\LBtt$) %srewritevpredchvete' resulting '' A$ '' (..)'avnaslvete' obtained.. intact)'
 we cannot translate it to propositional logic. We will sidestep the issue by translating only the formula $B$ obtained from $A$ by deleting subformula $\LBtt$ (but leaving $||y||=S(\cdot,\cdot,\cdot)$ intact) and replacing the variables $y$ by %srewritepredch' by '
 fixed bits representing a hard boolean function. In more detail, $\Pi^b_1$ formula $B$ can be translated into a sequence of propositional formulas $\mathsf{lear}^y_{\gamma}(L,\mathcal{C},\epsilon,\delta)$ expressing that ``if $C\in\mathcal{C}$ is a circuit computing $f$, then $L$ querying $f$ generates a circuit $D$ such that $\Pr[D(x)=f(x)]\ge 1-\epsilon$ with probability $\ge \delta$, which is counted approximately with precision $\gamma$". Note that $C,f$ are represented by free variables and that there are also free variables for error $\gamma$ from approximate counting and for %$O(\log(2^{n},b))$%srewritetento%nasl' .%'smedzerou
 boolean functions $y$. %, where $b$ is the number of variables representing $\gamma$.
As in the case of \ttable-formulas, we fix $|f|=2^n$, so $n$ is not a free variable. %%rewriteoxpredchvetaaskipprednouzanouanasl' Importantly'
Importantly, $\mathsf{lear}^y_{\gamma}(L,\mathcal{C},\epsilon,\delta)$ does not postulate that $y$ is a truth-table of a hard boolean function. %srewrite' that $y$ is.. function.'smedzerou
 Nevertheless, for any fixed (possibly non-uniform) bits representing a sequence of boolean functions %srewrite'functions '
 $h=\{h_m\}_{m>n_0}$ %srewritetatomedzera
 such that $h_m$ is not $(1/2+1/2^{m/4})$-approximable by any circuit of size $2^{m/4}$, we can obtain formulas $\mathsf{lear}^h_{\gamma}(L,\mathcal{C},\epsilon,\delta)$ by substituting bits $h$ for $y$. 

Using a single function $h$ in $\mathsf{lear}^h_{\gamma}(L,\mathcal{C},\epsilon,\delta)$ does not ruin the fact that (the translation of function) $Sz$ approximates the respective probability with accuracy $\gamma$ because $Sz$ queries a boolean function $y$ which depends just on the number of atoms representing $\gamma^{-1}$ %rewriteoxpredch' $gamma^-1$ 'anaslmedzera
 and on the size of the circuit $D$ defining the predicate we count together with the number of inputs of $D$. The size of $D$ and the number of its inputs are w.l.o.g. determined by the number of inputs of $f$. %srewritepredchpparasmedzerami

If we are working with formulas $\mathsf{lear}^h_{\gamma}(L,\mathcal{C},\epsilon,\delta)$, where $h$ is a sequence of bits representing a hard boolean function, in a proof system which cannot prove efficiently that $h$ is hard, our proof system might not be able to show that the definition is well-behaved - it might not be able to derive some standard properties of the function $Sz$ %rewriteox' of the function Sz 'anaslmedzera
 used inside the formula. Nevertheless, in our theorems this will never be the case: our proof systems will always know that $h$ is hard.%Moreover, our systems will p-simulate WF so the standard proeprties of approx c will trans to the propositional case (for suitable error parameters depending only on the size of circ defining the predicates in question and on the # of their inputs) Add example P1.1ii)?
%srewritepredchpoznza%

In formulas $\mathsf{lear}^y_{\gamma}(L,\mathcal{C},\epsilon,\delta)$ we can allow $L$ to be a sequence of nonuniform circuits, with a different advice string for each input length. One way to see that is to use additional input to $L$ in $\Pi^b_1$ formula $B$, then translate the formula to propositional logic and substitute the right bits of advice for the additional input. Again, the precise encoding of the formula $\mathsf{lear}^y_{\gamma}(L,\mathcal{C},\epsilon,\delta)$ does not matter very much to us but in order to simplify proofs we will assume that $\mathsf{lear}^y_{\gamma}(L,\Circuit[n^k],\epsilon,\delta)$ has the from $\neg\ttable(f,n^k)\rightarrow R$, where $n, k$ are fixed, %rewriteoxpredch' n,k are fixed, 'anaslmedzera
$f$ is represented by free variables and $R$ is the remaining part of the formula expressing that $L$ generates a suitable circuit with high probability. %a \WF-proof of $\mathsf{lear}(L,\mathcal{C},\epsilon,1)$ and a \WF-proof of the fact that ``$\Pr[D(x)\ne f(x)]<\epsilon$ where $D$ is the circuit generated by $L$ on queries to $f$'', we obtain a \WF-proof of $\ttable(f,n^k)$ by a simple \WF-derivation.

\subsection{Automatability} 

Let $\Phi$ be a class of propositional formulas. We say that a proof system $P$ is automatable w.r.t. $\Phi$ up to proofs of size $s$ if there is a \pv-function $A$ such that for each $\phi\in\Phi$ and each $t$-size $P$-proof of $\phi$ with $t\le s$, $A(\phi,1^t)$ is a $P$-proof of $\phi$.
\smallskip %rewriteod\smallskipponext\medskip

In our main theorem we will need a slightly modified notion of automatability where the automating algorithm outputs a proof of a given tautology $\phi$ which is not much longer than a proof of an associated tautology $\psi$. (Formula $\psi$ will be closely related to $\phi$: while $\phi$ will express a worst-case lower bound, $\psi$ will express an average-case lower bound for the same function.)

Let $\Phi$ be a class of pairs of propositional formulas. We say that a proof system $P$ is automatable w.r.t. $\Phi$ up to proofs of size $s$ if there is a \pv-function $A$ such that for each pair $\langle\psi,\phi\rangle\in\Phi$ and each $t$-size $P$-proof of $\psi$ with $t\le s$, $A(\phi,1^t)$ is a $P$-proof of $\phi$.%rewriteoxvsetkyphiapsivpredchvete
\medskip

\noindent {\bf Propositional version.} If $\Phi$ is defined by a \pv-function and $s\in Log$ is a \pv-function%srewrite'is a pv-function'
, the statement that an algorithm $A$ (given by a \pv-function) automates system $P$ w.r.t. $\Phi$ up to proofs of size $s$ is $\Pi^b_1$. Therefore, it can be translated into a sequence of propositional formulas $\mathsf{aut}_{P}(A,\Phi,s)$. Again, in formulas $\mathsf{aut}_{P}(A,\Phi,s)$ we can allow $A$ to be a sequence of nonuniform circuits and $s$ to be arbitrary possibly nonuniform parameter%srewrite'and s .. parameter'
.
%rewrite'\pv-function'' (givenbya\pv-function)'alastsentencesmedzeramivtomtopara

\section{Lower bounds versus learning in proof complexity}

\def\skippedfacts{
For every reasonably strong proof system $P$, any super-polynomial lower bound for $P$ for arbitrary sequence of tautologies, impliess that $P$ cannot prove efficiently that $\NP\subseteq\Ppoly$, cf. \cite{Kpc}. Analogously, assuming circuit lower bouds are hard for $P$, we can derive $P$-consistency of lower bounds on learning. A formalization of this fact appears in the proof of Theorem \ref{t:main}.%rewritecelututosection

\begin{theorem}\label{t:lblear}
%Let $k\ge1$ be a constant and $P$ be a proof system which simulates \EF such that: I. for each $P$-proof $\pi_0$ of $\phi$ and a $P$-proof $\pi_1$ of $\phi\rightarrow \psi$, there is a $poly(|\pi_0|,|\pi_1|)$-size $P$-proof of $\psi$; II. for each $P$-proof $\pi$ of $\phi$ and  a possibly partial substitution $\rho$ of atoms of $\phi$ by arbitrary formulas, there is a $poly(|\pi|,|\phi|_{\rho}|)$-size $P$-proof of $\phi|_{\rho}$, where $\phi|_{\rho}$ is the formula $\phi$ after applying substituion $\rho$. 
Assume that $P$ proves efficiently $\ttable(h,2^{n/4},1/2-1/2^{n/4})$, for some boolean function $h=\{h_n\}_{n>n_0}$ and some $n_0$.  Further, assume that there is a boolean function $f=\{f_n\}_{n>n_1}$, for some $n_1$, such that $P$ does not prove efficietly $\ttable(f_n,n^k)$. Then there is no algorithm $A$ such that $P$ proves efficiently $\mathsf{lear}^h_{\delta/2}(A,\Circuit[n^k],\epsilon,\delta)$.
\end{theorem}

Prf.: follow the Prf of C5.1 in main T (with y'=h and f st no circuit approximates f').

Another property, which plays an essential role in the proof of Theorem \ref{t:main}} In this section we show how automatability together with efficient provability of a lower bound, or together with optimality, implies efficient learning. This shows some of the methods used in the proof of Theorem \ref{t:main}, but the proof of Theorem \ref{t:main} can be read independently. Then, we proceed with a formalization of the transformation of natural proofs into learning  algorithms, which is one of the cornerstones of Theorem \ref{t:main}.

\begin{theorem}\label{t:rkcikk} Let $k\ge1$ be a constant and $P$ be a proof system which simulates \EF such that: I. for each $P$-proof $\pi_0$ of $\phi$ and each $P$-proof $\pi_1$ of $\phi\rightarrow \psi$, there is a $poly(|\pi_0|,|\pi_1|)$-size $P$-proof of $\psi$; II. for each $P$-proof $\pi$ of $\phi$ and  a possibly partial substitution $\rho$ of atoms of $\phi$ by arbitrary formulas, there is a $poly(|\pi|,|\phi|_{\rho}|)$-size $P$-proof of $\phi|_{\rho}$, where $\phi|_{\rho}$ is the formula $\phi$ after applying substitution $\rho$. 

Assume that $P$ proves efficiently $\ttable(h,3n^{k})$, for some boolean function $h=\{h_n\}_{n>n_0}$ and some $n_0$. Then, automatability of $P$ implies that for each $\gamma\in (0,1)$, $\Circuit[n^{k\gamma/a}]$ is learnable by $\Circuit[2^{O(n^{\gamma})}]$ over the uniform distribution, with non-adaptive membership %rewriteoxpredch' membership 'anaslmedzera
 queries, confidence 1, up to error $1/n^{k\gamma/a}$, where $a$ is an absolute constant. \end{theorem}

\proof[Proof sketch] By Theorem \ref{t:cikkcore}, it suffices to construct a \Ppoly-natural property useful against $\Circuit[n^k]$. This is achieved by the proof of Theorem \ref{t:razbkra}, which we sketch - more details can be found in the proof of Theorem \ref{t:main} (direction 2. $\rightarrow$ 1.). Assume $P$ proves efficiently $\ttable(h_n,3n^k)$ for some boolean function $h=\{h_n\}_{n>n_0}$. If $P$ simulates \EF and satisfies properties I.-II., then $P$ proves efficiently also \begin{equation}\label{e:xorsketch}\ttable(g,n^k)\vee \ttable(h_n\oplus g,n^k),\end{equation} where $g$ is represented by free variables and $h_n\oplus g$ is a bitwise XOR of $h_n$ and $g$%srewrite' and hn\oplus g.. and g'
. This is because an $n^k$-size circuit $C_1$ computing $g$ and an $n^k$-size circuit $C_2$ computing $h_n\oplus g$ can be combined into a $3n^k$-size circuit $C_1\oplus C_2$ computing $h_n$. (Properties I.-II. can be used to simulate Frege rules, see Lemma~\ref{l:decent}.) Automatability of $P$ now implies the existence of a \Ppoly-natural property useful against $\Circuit[n^k]$: For each boolean functions $g$, we can either find efficiently a $P$-proof of $\ttable(g,n^k)$ or we recognize that $\ttable(h_n\oplus g,n^k)$ holds (if $\ttable(h_n\oplus g,n^k)$ did not hold, we could use properties I.-II. to substitute its falsifying assignment to the proof of (\ref{e:xorsketch}) and obtain a short $P$-proof of $\ttable(g,n^k)$ - formally, we use here also the fact that $P$-proves efficiently $\phi(b)$, whenever $b$ is a satisfying assignment of $\phi$, see Lemma \ref{l:decent}, Item 2.), and one of these options happens for at least $1/2$ of all functions $g$. \qed

\begin{theorem}[Optimality and automatability implies learning]\label{t:orkcikk} If there is an optimal proof system which is automatable, then for each $\gamma\in (0,1)$, each $k\ge 1/\gamma$, for infinitely many $n$, $\Circuit[n^{k\gamma}]$ is learnable by $\Circuit[2^{O(n^{\gamma})}]$ over the uniform distribution, with non-adaptive membership %srewrite 'membership'apredchadz'infinitely'
 queries, confidence $1/2^{4n^{\gamma}}$, up to error $1/2-1/2^{3n^{\gamma}}$.
\end{theorem}

\proof[Proof sketch] If $\SAT\in \Circuit[3n^k]$ for infinitely many $n$, then there is a \Ppoly-natural property useful against $\Circuit[n^{\log n}]$, for infinitely many $n$, and the conclusion of the theorem follows from the proof of Theorem \ref{t:cikkcore}, see Theorem \ref{t:cikk}. Assume $\SAT\not\in \Circuit[3n^k]$ holds for all sufficiently big $n$. Then there is a proof system $P$ which proves efficiently $\ttable(\SAT,3n^k)$ for all sufficiently big $n$: $P$ is by definition allowed to derive every substitutional instance of $\ttable(\SAT,3n^k)$ (which is a formula recognizable in $2^{O(n)}$-time) and, otherwise, it proceeds as \EF. Therefore, we can follow the proof of Theorem \ref{t:rkcikk}, noting that the automatability of $P$ can be replaced by the automatability of the optimal system, and obtain the desired conclusion. \qed
%rewriteoxrestofsubsectionsmedzeramiaposlvetavpredchproofsmedzerami

\begin{theorem}\label{t:unconditional} For each $\gamma\in (0,1)$, each $k\ge 1/\gamma$, there is a proof system $P$ such that $1.)$ $P$ is automatable if and only if $2.)$ for infinitely many $n$, $\Circuit[n^{k\gamma}]$ is learnable by $\Circuit[2^{O(n^{\gamma})}]$ over the uniform distribution, with non-adaptive membership %srewrite 'membership'apredchadz'infinitely'
 queries, confidence $1/2^{4n^{\gamma}}$, up to error $1/2-1/2^{3n^{\gamma}}$.
\end{theorem}

\proof Let $\gamma\in (0,1)$ and $k\ge 1$. If Item $2.)$ from the statement of the theorem holds, then let $P$ be a proof system with exponentially long proofs of all tautologies, so $P$ is trivially automatable and the equivalence holds.
Suppose Item $2.)$ does not hold. Then, similarly as in the proof of Theorem \ref{t:orkcikk}, we conclude that $\SAT\not\in \Circuit[3n^{ka}]$ for all sufficiently big $n$ and constant $a$ from Theorem \ref{t:rkcikk}. %(Otherwise, we could use p-size circuits for \SAT to find the smallest circuit consistent with sufficiently many random queries. Since, with high probability, random queries hit each $n^{k\gamma/a}$-size circuit, which errs on $1/n^{k\gamma/a}$ fraction of inputs, this would yield learning circuits contradicting the assumption.) To use previous the argument inside the bracket we would need to take into account n's where it works.. it should be ok.
It remains to observe that the proof system $P$ from the proof of Theorem \ref{t:orkcikk}, with $\ttable(\SAT,3n^{ka})$ instead of $\ttable(\SAT,3n^k)$, is not automatable. This follows by Theorem \ref{t:rkcikk}. Alternatively, in the case that Item $2.)$ fails, we can use the fact that $\Ptime\ne\NP$ implies the existence of a non-automatable proof system, cf. \cite[Lemma~5.3]{PSeps}. \qed
%thereisamoredirectproofofpreviousthm: in the second case we know that some pps is not aut by Pitass-Santhanam.
\medskip

The disadvantage of Theorem \ref{t:unconditional} in comparison to Corollary \ref{c:main} is that it does not provide an explicit definition of $P$. It is possible to get an explicit construction of $P$ based on a hardness assumption, but the hardness assumption would then itself falsify both sides of the desired equivalence. Another advantage of Corollary \ref{c:main} is that it applies to all well-behaved proof systems simulating a `ground' system $P$. 

\subsection{Learning algorithms from natural proofs in \APC}%rewriteoxpredchtitleinside{}%rewritecelusubsectionsmedzerami

An essential component of the transformation of natural proofs into learning algorithms is the Nisan-Wigderson generator and specific combinatorial designs on which it is based, cf. \cite{NW}. In order to formalize the transformation in \APC we would need to construct combinatorial designs in \APC. A construction of combinatorial designs has been formalized already in \cite{Jwphp}, but our transformation requires algebraic construction of designs obtained by evaluating polynomials on a finite field. A complication is that the algebraic construction uses Bertrand's postulate of the existence of a prime between $x$ and $2x$. We will bypass the problem of proving Bertrand's postulate in \APC simply by assuming the existence of such a prime.\footnote{It is possible that the desired formalization of Bertrand's postulate can be obtained from the work of Paris, Wilkie and Woods \cite{PWW}.} %rewriteoxpredchfootnotesmedzerami
 That is, we will not prove the existence of combinatorial designs unconditionally, but only under the assumption of the primality of a number in the interval $[x,2x]$. Fortunately, in our setting, we will have $2^x\in Log$, so once we translate the resulting statements to propositional logic, we will use Bertrand's postulate (even though we have not formalized it in \APC) to conclude that the assumption will have a trivial \WF-proof. The construction of Nisan and Wigderson otherwise does not require developing new methods, but we need to %inspect it and 
verify that each step is doable in \APC. In fact, \PV will suffice.
%rewriteoxnaslparasmedzerami
\smallskip

For $x\in\{0,1\}^n$ and $S\subseteq [n]$, denote by $x|S$ an $|S|$-bit string consisting of $x_i$'s such that $i\in S$ (in the natural order). 

\begin{lemma}[in \PV]\label{l:design}
Let $d\geq 2$. If $2^n\in Log$ and $n^d\le p\le 2n^d\in Log$ is a prime, there is a $2^n\times m$ 0-1 matrix $A$ with $n^{d}$ ones per row and $m=pn^d$ %$n^{2d}\le m\le 2n^{2d}$ 
which is also an $(n,n^{d})$-design meaning that for $J_i(A):=\{j\in  [m]; a_{i,j}=1\}$, for each $i\neq j$, $|J_i(A)\cap J_j(A)|\leq n$ and $|J_i(A)|=n^d$. Moreover, for sufficiently big $n$, there are $n^{9d}$-size circuits which given $i\in \{0,1\}^n$ and $w\in\{0,1\}^{m}$ output $w|J_i(A)$. %rewriteoxpredch' .', where $w|J_i(A)$ are $w_j$'s such that $j\in J_i(A)$. 
\end{lemma}

\proof  Let $n^d\le p\le 2n^{d}$ be a prime and $F$ a field of size $p$. $F$ can be constructed in \PV, cf. \cite[Section 4.3]{Jphd}. We construct the matrix $A$ so that the $i$-th row consists of positions $(u,v)$, for $u\in \{0,\dots,n^d-1\}, v\in F$, with 1's exactly on positions $(u,q(u))$, for $u\in \{0,\dots,n^d-1\}$, where $q$ is a polynomial of degree $n$ with binary coefficients corresponding to the binary representation of $i$. Formally, an $n$-degree polynomial is represented by a sequence of its coefficients. The evaluation of an $n$-degree polynomial on an element from $F$ can be done in $poly(n)$-time and is well-defined in \PV, cf. \cite[Section 4.3]{Jphd}. By definition, $A$  is a $2^n\times m$ 0-1 matrix with $n^{d}$ ones per row. $|J_i(A)\cap J_j(A)|\le n$, for $i\ne j$, follows from the fact that a non-zero $n$-degree polynomial over $F$ has $\le n$ roots, which is provable in \PV, cf. \cite[Lemma 4.3.6]{Jphd}. \def\alreadydoneinJ{results from the following claim.

\begin{claim}[in \PV] Each non-zero $n$-degree polynomial over F has $\le n$ roots. (Lemma 4.3.6 in Jphd?)
\end{claim}

The claim is proved by $\Sigma^b_0(\pv)$-induction on $n$. We express the statement that an $n$-degree polynomial has $\le n$ roots in $F$ by a \pv-formula which counts the number of roots by evaluating the polynomial on each element from $F$. The base case $n=1$ holds trivially. For the inductive step, it suffices to show that if $a$ is a root of an $n$-degree polynomial $r$, then $r=(x-a)r'$, where $r'$ is a polynomial of degree $n-1$. %% and the equality holds even between formal expressions obtained by expanding both sides into sums of monomials. 
Let $r=\sum_{i=0}^n b_ix^i$. We want to find $c_0,\dots, c_{n-1}$ such that $r=(x-a)\sum_{i=0}^{n-1} c_ix^i$. That is, $b_n=c_{n-1}$, $b_0=-ac_0$ and $b_{i}=c_{i-1}-ac_{i}$, for $1\le i\le n-1$. This system of linear equations can be solved by setting $c_{n-1}:=b_n$ and $c_{i-1}:=b_i+ac_i$ for $1\le i\le n-1$, which implies $c_0=\sum_{i=1}^{n}b_ia^{i-1}$ and, since $r(a)=0$, satisfies $b_0+ac_0=0$. }
\smallskip

It remains to prove the `moreover' part. The $n^{9d}$-size circuit first evaluates the $i$-th polynomial on inputs $0,\dots,n^d-1$. This way it obtains $J_i(A)$ and $w|J_i(A)$. Evaluating an $n$-degree polynomial on an input from $F$ can be done \PV-provably by a circuit of size $n\cdot poly(d\log n)$. (In more detail, we compute $x,x^2,\dots,x^n$ over $F$ in a standard way by an $n\cdot poly(d\log n)$-size circuit, then multiply $x^i$'s with the corresponding coefficients and sum the results over $F$, which takes another $n\cdot poly(d\log n)$-size circuit.) Thus, given $i$, the $n^d$ indices from $J_i(A)$ can be generated by a circuit of size $n^{d+1}poly(d\log n)$. Given $w$ and indices from $J_i(A)$, the bits $w|J_i(A)$ can be generated by a circuit of size $O(n^dpd\log n)$ so the total size of the circuit generating $w|J_i(A)$ on $i$ and $w$ is $\le n^{d+2}+O(dn^{2d+1})\le n^{9d}$.
\qed

\bigskip

The formalization of the transformation of natural proofs into learning algorithms follows from a straightforward inspection of the original proof as well.

\begin{theorem}\label{t:cikk}
There is a \pv-function $L$ such that \APC proves: For $k\ge 1$, $d\ge 2$, $2^{n^d}, n^{dk},\delta^{-1}\in Log$, $\delta<1/N^3$ and a prime $n^d\le p\le 2n^d$, let $R_N$ be a circuit with $N=2^n$ inputs such that for sufficiently big $N$,
\begin{itemize}
\item[1.] $R_N(x)=1$ implies that $x$ is a truth-table of a boolean function with $n$ inputs hard for $\Circuit[n^{10dk}]$,
\item[2.] $\{x\mid R_N(x)=1\}\succeq_{\delta} 2^N/N$.
\end{itemize}
Then, circuits with $n^d$ inputs and size $n^{dk}$ are learnable by circuit $L(R_N,p)$ over the uniform distribution with %non-adaptive 
membership queries, confidence $1/N^4$, up to error %$1/n^{dk}$
$1/2-1/N^3$. Here, the confidence is counted approximately with error $\delta$ using \pv-function $Sz$ and the corresponding assumptions \LBtt expressing hardness of a boolean function $y$, i.e. using formulas $\Pr^y[\cdot]_{\delta}$. %, where %$k = \frac{d\gamma}{a}$ and 
%$a$ is an absolute constant.
\end{theorem}

\proof
We reason in \APC. Consider a Nisan-Wigderson generator based on a circuit $C$ which we aim to learn. Specifically, for $d\ge 2$ and $n^{2d}\le m=pn^d\le 2n^{2d}$, let $A=\{a_{i,j}\}^{i\in [N]}_{j\in [m]}$ be an $N\times m$ 0-1 matrix with $n^{d}$ ones per row. %$J_i(A):=\{j\in  [m]; a_{i,j}=1\}$. 
Then define an NW-generator $NW_{C}:\{0,1\}^{m}\mapsto\{0,1\}^{N}$ as $$(NW_{C}(w))_i=C(w|J_i(A)).$$ %where $w|J_i(A)$ are $w_j$'s such that $j\in J_i(A)$. 
We can assume that $A$ is, in addition, a combinatorial design from Lemma \ref{l:design}. Therefore, if $C$ has $n^{d}$ inputs and size $n^{dk}$, then for each $w\in \{0,1\}^{m}$, $(NW_{C}(w))_x$ is a function on $n$ inputs $x$ computable by circuits of size $n^{10dk}$, for sufficiently big $n$. We want to learn $C$ by a circuit $L(R_N,p)$ of size $2^{O(n)}$. 
\smallskip

We will use circuits $R_N$ which function as distinguishers for $NW_C$: By the assumption of the theorem, a trivial surjection witnesses that $\{w\mid R_N(NW_C(w))=1\}\preceq_{0} 0$. Hence, by Proposition \ref{lem} $ii)$, $2^m\cdot \Pr_w^y[ R_N(NW_C(w))=1]_{\delta}\preceq_{\delta} 0$, and by Proposition \ref{really} $1. ii)$, $\Pr_w^y[R_N(NW_C(w))=1]_{\delta}<2\delta$, for universally quantified $y$. %srewrite ', for univers quantified y. 'smedzerou
Similarly, by the assumption of the theorem $\{u\mid R_N(u)=1\}\succeq_{\delta} 2^N/N$, and $\Pr_u^{y'}[R_N(u)=1]_{\delta}>1/N-3\delta$. Therefore, $$\Pr_u^{y'}[R_N(u)=1]_{\delta}-\Pr_w^{y}[R_N(NW_C(w))=1]_{\delta}>1/N-5\delta.$$%rewriteoxvpredchdisplayedfla'{y'}''{y}'
%\smallskip

$L(R_N,p)$ chooses a random $i\in [N]$, random bits $r_{1},\dots,r_N$, random $w'\in \{0,1\}^{m-n^d}$ and queries the bits $C(w|J_1(A)),\dots, C(w|J_{i-1}(A))$, for all $w\in \{0,1\}^m$ such that $w|J_i(A)=w'$. Since $A$ is an $(n,n^d)$-design, there are just $2^{O(n)}$ such queries. For $w\in\{0,1\}^m$, let $p_i:=R_N(C(w|J_1(A)),\dots,C(w|J_{i-1}(A)),r_i,\dots,r_N)$. Then $L(R_N,p)$ outputs a circuit $L'$ which on $x\in \{0,1\}^{n^d}$ constructs $w\in \{0,1\}^m$ such that $w|J_i(A)=x$ while $w|[m]\backslash J_i(A)=w'$ and predicts the value $C(x)$ by outputting $\neg r_i$ iff $p_i=1$. % and $r_i$ otherwise. 
\smallskip

We want to show that $L'$ approximates $C$ with high probability.

First, note that by Proposition \ref{lem} $iii)$, inequality $\{u\mid R_N(u)=1\}\succeq_{\delta} 2^N/N$ implies $\{w,r_1,\dots,r_N\mid p_1=1\}\succeq_{\delta} 2^{m+N}/N$, so we have $\Pr_{w,r_1,\dots,r_N}^{y'}[p_1=1]_{\delta}>1/N-3\delta$ and $\Pr_{w,r_1,\dots,r_N}^y[p_{N+1}=1]_{\delta}<2\delta$. Consequently, there exists $y_1,\dots,y_{N+1}$ and $i\in [N]$ such that \begin{equation}\label{e:cikk2}\Pr_{w,r_1,\dots,r_N}^{y_i}[p_{i}=1]_{\delta}-\Pr_{w,r_1,\dots,r_N}^{y_{i+1}}[p_{i+1}=1]_{\delta}>1/N^2-5\delta/N.\end{equation} Otherwise, for all $y_1,\dots,y_{N+1}$, for all $i$, $\Pr^{y_i}[p_{i}=1]_{\delta}-\Pr^{y_{i+1}}[p_{i+1}=1]_{\delta}\le 1/N^2-5\delta/N$, and by $\Sigma^b_0(\pv)$-induction $\Pr^{y_1}[p_{i}=1]_{\delta}-\Pr^{y_{N+1}}[p_{i+1}=1]_{\delta}\le 1/N-5\delta$. As \APC proves that some $y_1,\dots,y_{N+1}$ satisfy the assumptions of $\Pr^{y_j}[\cdot]_{\delta}$, for all $j=1,\dots, N+1$, this would be a contradiction. (The existence of $y_1,\dots,y_{N+1}$ is proved analogously as Proposition \ref{p:hardf} and the proof does not require %rewriteox' and the proof.. require 'anaslmedzera
 sharply bounded collection scheme - we can construct $y_1,\dots,y_{N+1}$ from the string $w$ in \cite[Lemma 4.1.8]{Jphd}.) That is, (\ref{e:cikk2}) means that $y_i$'s satisfy  formulas $\LBtt$ (with suitable parameters), have the right length $||y_i||$ and the corresponding functions $Sz$ witness that the difference of the respective probabilities is big.
%Here, we can use exact counting to conclude the existence of $i$ because $N\in Log$. %with probability $1/N$. Here, the outermost probability is counted exactly (because $N\in Log$). %srewriteall'{y'}''{yj}'vcelomdokazedalej'thereexy1.. such that'vpredchpara,2xpredch%a'Otherwise.. big.)'

Since trivial surjections witness that, for $i\in [N]$, $z=w',x,r_1,\dots,r_N<2^{m+N}$,
$$\{z\mid L'(x)=C(x)\}\succeq_{0}\{z\mid p_i=1\wedge r_i\ne C(x)\}\cup\{z\mid p_i\ne 1\wedge r_i= C(x)\},$$
and $\{z\mid p_i=1\wedge r_i\ne C(x)\}\cap\{z\mid p_i\ne 1\wedge r_i= C(x)\}\approx_0 0$, by Proposition \ref{really} $1.iv)$ and Proposition \ref{lem} $ii)$,
$$2^{m+N}\Pr_{z}^y[L'(x)=C(x)]_{\delta}\succeq_{4\delta}2^{m+N}\Pr_z^{y'}[p_i=1\wedge r_i\ne C(x)]_{\delta}+ 2^{m+N}\Pr_z^{y''}[p_i\ne 1\wedge r_i= C(x)]_{\delta}.$$ Notably, when we applied Proposition \ref{really} $1. iv)$, we switched the domain of surjections from $2^{m+N}$ to $2^{m+N+1}$. This does not affect our inequalities because surjections witnessing $X \preceq_{\delta} Y$, for $X,Y\subseteq 2^{m+N}$ can be used to witness $X\preceq_{\delta} Y$, where we see $X,Y$ as subsets of $2^{m+N+1}$ (and consider the error $\delta$ w.r.t. $2^{m+N+1}$, not w.r.t. $2^{m+N}$), but we need to take it into account when we now apply Proposition \ref{really} $1.ii)$ to conclude that\footnote{If we worked on the domain $2^{m+N}$ the resulting error would not be $10\delta$ but $5\delta$.}
$$\Pr_{z}^y[L'(x)=C(x)]_{\delta}>\Pr_z^{y'}[p_i=1\wedge r_i\ne C(x)]_{\delta}+\Pr_z^{y''}[p_i\ne 1\wedge r_i= C(x)]_{\delta}-10\delta.$$ Further, we have $\{z\mid p_i=1\wedge r_i=C(x)\}\cup\{z\mid p_i\ne 1\wedge r_i= C(x)\}\approx_0 2^{m+N}/2$, so $\Pr_z^{y''}[p_i\ne1\wedge r_i=C(x)]_\delta>1/2-\Pr_z^{y'''}[p_i=1\wedge r_i=C(x)]_{\delta}-8\delta$ and 
\begin{equation}\label{e:cikk3}\Pr_{z}^y[L'(x)=C(x)]_{\delta}>\Pr_z^{y'}[p_i=1\wedge r_i\ne C(x)]_{\delta}+\frac{1}{2}-\Pr_z^{y'''}[p_i=1\wedge r_i= C(x)]_{\delta}-18\delta.\end{equation} Next, we similarly derive $\{z\mid p_i=1\}\approx_0 \{z\mid p_i=1 \wedge r_i=C(x)\}\cup\{z\mid p_i= 1\wedge r_i\ne C(x)\}$ and \begin{equation}\label{e:cikk4}\Pr_z^{y_i}[p_i=1]_{\delta}-10\delta<\Pr_z^{y'''}[p_i=1\wedge r_i=C(x)]_{\delta}+\Pr_z^{y'}[p_i=1\wedge r_i\ne C(x)]_{\delta}.\end{equation} Now, observe that $\{z\mid p_i=1\wedge r_i=C(x)\}\approx_0 \{w,r_1,\dots,r_N\mid p_{i+1}=1\wedge r_i=1\}$. % where %$w$ used by $p_i$ is obtained from $w', x$ as in $L'$ and 
%the sets are compared as subsets of $2^{m+N}$. 
This yields 
%$2^{m+N}\Pr_{z}^h[p_i=1\wedge r_i=C(x)]_{\delta}\preceq_{\delta}\{w,r_1,\dots,r_N\mid p_{i+1}=1\wedge r_1=1\}$ and 
$2^{m+N}\Pr_{z}^{y'''}[p_i=1\wedge r_i=C(x)]_{\delta}\preceq_{2\delta} 2^{m+N}\Pr_{w,r_1,\dots,r_N}^{y''''}[p_{i+1}=1\wedge r_i=1]_{\delta}$. %srewritepredchanasl'{i+1}'
 Analogously, $2^{m+N}\Pr_{z}^{y'''}[p_i=1\wedge r_i=C(x)]_{\delta}\preceq_{2\delta} 2^{m+N}\Pr_{w,r_1,\dots,r_N}^{y'''''}[p_{i+1}=1\wedge r_i=0]_{\delta}$. As $\{w,r_1,\dots,r_N\mid p_{i+1}=1\wedge r_i=1\}\cup \{w,r_1,\dots,r_N\mid p_{i+1}=1\wedge r_i=0\}\approx_0 \{w,r_1,\dots,r_N\mid p_{i+1}=1\}$, we have also $2^{m+N}\Pr_{w,r_1,\dots,r_N}^{y''''}[p_{i+1}=1\wedge r_i=1]_{\delta}+2^{m+N}\Pr_{w,r_1,\dots,r_N}^{y'''''}[p_{i+1}=1\wedge r_i=0]_{\delta}\approx_{4\delta}2^{m+N}\Pr_{w,r_1,\dots,r_N}^{y_{i+1}}[p_{i+1}=1]_{\delta}$. It follows that
%$s\approx_{\delta} \{w,1,r_{2},\dots,r_N\mid p_{i+1}=1\}$ implies $2s\approx_{\delta} \{w,1,r_{1},\dots,r_N\mid p_{i+1}=1\}$. As a consequence we obtain $2^{m+N+1}\Pr_{w,1,r_2,\dots,r_N}^h[p_{i+1}=1]_{\delta}\approx_{2\delta}\Pr_{w,1,r_1,\dots,r_N}^h[p_{i+1}=1]_{\delta}2^{m+N}$ and 
%$2\Pr_z^h[p_i=1\wedge r_i=C(x)]_{\delta}-9\delta<\Pr_{w,r_{1},\dots,r_N}^h[p_{i+1}=1]_{\delta}$ and
\begin{equation}\label{e:cikk6} 
2\Pr_z^{y'''}[p_i=1\wedge r_i=C(x)]_{\delta}-18\delta<\Pr_{w,r_{1},\dots,r_N}^{y_{i+1}}[p_{i+1}=1]_{\delta}.\end{equation}

Combining (\ref{e:cikk2}) - (\ref{e:cikk6}) shows that for some $i\in [N]$, \begin{equation}\label{e:cikk1}\Pr_{w',x,r_1,\dots,r_N}^y[L'(x)=C(x)]_{\delta}>1/2+1/N^2-5\delta/N-46\delta,\end{equation} %This gives us a surjection witnessing that $$\{w',x,r_i,\dots,r_N\mid L'(x)=C(x)\}\succeq_{\delta} (1/2+1/N^2-27\delta)2^{m+N-i+1},$$ which can be used to witness $\{w',x,r_1,\dots,r_N\mid L'(x)=C(x)\}\succeq_{\delta} (1/2+1/N^2-27\delta)2^{m+N}$ and \begin{equation}\label{e:cikk1}\Pr_{w',x,r_1,\dots,r_N}^h[L'(x)=C(x)]_{\delta}>1/2+1/N^2-30\delta,\end{equation} 
where $L'$ is generated by $L(R_N,p)$ on $w',x,r_1,\dots,r_N$ and $i$. As in the case of (\ref{e:cikk2}), $y$ is quantified existentially in (\ref{e:cikk6}) and satisfies the assumptions of $\Pr^y[\cdot]_{\delta}$. %srewritepredchvetaa' (for univers quantified y) 'vnaslvete

It remains to observe that (for universally quantified $y$) $$\Pr_{w',i,r_1,\dots,r_N}^y[L(R_N,p)\ (1/2+1/N^3)\text{-approximates }C]_{\delta}>1/N^4.$$ For the sake of contradiction, assume this is not the case. Then $$\{w',i,r_1,\dots,r_N\mid L'\ (1/2+1/N^3)\text{-approximates }C\}\preceq_{\delta}2^{m-n^d+n+N}/N^4$$ and by averaging (Proposition \ref{really}, Item 3), $$\{w',x,i,r_1,\dots,r_N\mid L'\ (1/2+1/N^3)\text{-approximates }C\}\preceq_{2\delta}2^{m+n+N}/N^4.$$ Therefore, for each $i\in [N]$, \begin{equation}\label{e:cikk5}\{w',x,r_1,\dots,r_N\mid L'\ (1/2+1/N^3)\text{-approximates }C\}\preceq_{2\delta}2^{m+N}/4N^2.\end{equation} %where $L'$ is generated by $L(R_N,p)$ on $w',x,r_1,\dots,r_N$ and $i$. 
(Otherwise, by Proposition \ref{really} $1.i)$, there is a surjection witnessing the opposite inequality, which can be used to witness also $\{w',x,i,r_1,\dots,r_N\mid L'\ (1/2+1/N^3)\text{-approximates }C\}\succeq_{2\delta}2^{m+N}/4N^2$ and $2^{m+N}/4N^2\preceq_{4\delta} 2^{m+n+N}/N^4=2^{m+N}/N^3$, contradicting Proposition \ref{really} $1.ii)$ for $\delta<1/N^3$.)

On the other hand, for each $w',i,r_1,\dots,r_N$, $$\{x\mid L'(x)=C(x)\wedge L'\ <(1/2+1/N^3)\text{-approximates }C\}\preceq_{0}(1/2+1/N^3)2^{n^d},$$ which can be counted exactly because $2^{n^d}\in Log$. Hence, by averaging, for each $i\in [N]$, $$\{w',x,r_1,\dots,r_N\mid L'(x)=C(x)\wedge L'\ <(\frac{1}{2}+\frac{1}{N^3})\text{-approximates }C\}\preceq_{\delta}(1/2+1/N^3)2^{m+N}.$$ The last approximation together with (\ref{e:cikk5}) imply that for each $i\in [N]$, $$\{w',x,r_1,\dots,r_N\mid L'(x)=C(x)\}\preceq_{3\delta}(1/2+1/2N^2)2^{m+N},$$ which in turn implies that (for universally quantified $y$) %srewrite' (for univers quantified y) 'smedzerou
 $\Pr_{w',x,r_1,\dots,r_N}^y[L'(x)=C(x)]_{\delta}<1/2+1/2N^2+5\delta$, contradicting (\ref{e:cikk1}) if $\delta<1/N^3$ and $N$ is sufficiently big.
%\smallskip
%The last prob: $Pr[]\succeq_0 1/2+1/2ms$. However, since in Log, $Pr[L(D)(x)=f(x)]$ can be counted exactly. Otherwise, $\{L(D)(x)=f(x)\}\preceq (1/2+1/4ms)2^n$, which would yield a contradiction.\textcolor{red}{(Doublecheck this step 'moving from approx counting to exact counting'.)}
\qed

\section{Main theorem}\label{s:main}
%rewritefollowingtextuntilmainthm
Our main theorem holds for any `decent' proof system p-simulating \WF, which is well-behaved in the sense that it \APC-provably satisfies some basic properties.

\begin{definition}[\APC-decent proof system]\label{d:decent} A propositional proof system $P$ is \APC-decent if the language $L$ of $P$ is finite and complete, i.e. $L$ consists of connectives of constant arity such that each boolean function of every arity can be expressed by an $L$-formula, $P$ proves efficiently its own reflection principle, i.e. formulas %of the form $Prf_P(\pi,\phi)\rightarrow SAT(\phi,x)$ 
stating that if $\pi$ is a $P$-proof of $\phi$ then $\phi$ holds, %is satisfied by $
cf.~\cite{Kpc}, and there is a \pv-function $F$ such that \APC proves:
\begin{itemize}
\item[1.] $P$ p-simulates \WF, i.e. $F$ maps each \WF-proof of $\phi$ to a $P$-proof of $\phi$.
\item[2.] $P$ admits substitution property: $F$ maps each triple $\langle\phi,\rho,\pi\rangle$ to a $P$-proof of $\phi|_{\rho}$, where $\pi$ is a $P$-proof of $\phi$ and $\phi|_{\rho}$ is formula $\phi$ after applying substitution $\rho$ which replaces atoms of $\phi$ by formulas.
\item[3.] $F$ maps each pair $\langle\pi,\pi'\rangle$, where $\pi$ is a $P$-proof of $\phi$ and $\pi'$ is a $P$-proof of $\phi\rightarrow \psi$, to a $P$-proof of $\psi$.
\end{itemize}
\end{definition}

In Definition \ref{d:decent}, \WF refers to some %rewriteoxpredch' some'anaslmedzera
 fixed system from the set of all \WF systems. It follows from the proof of Lemma \ref{l:decent} that if \APC proves that $P$ p-simulates a \WF-system $Q$, then for every \WF-system $R$, \APC proves that $P$ p-simulates $R$, so the particular choice of the \WF-system does not matter. %rewriteox' It follows..  R.'anaslmedzera
 When we use connectives $\wedge,\vee,\neg,\rightarrow$ in an \APC-decent system $P$, we assume that these are expressed in the language %rewriteoxpredch' language 'anaslmedzera
 of $P$.

%Note that when we speak about a pps P in APC, all we need to define is the p-time predicate expressing that a given bit-string is a $P$-proof. In particular, we do not express other properties of the system such as that the language of P is complete or that P is implicationaly complete. 

\begin{lemma}\label{l:decent}  Each \WF system %and $\mathsf{ZFC}$ (interpreted as a propositional proof system) 
is \APC-decent. Moreover, for each \APC-decent proof system $P$ %rewriteoxpredch' P 'snaslmedzerou
 the following holds.
\begin{itemize}
%\item[1.] \WF and $\mathsf{ZFC}$ (interpreted as a propositional proof system) are \APC-decent proof systems.
\item[1.] %Let $P$ be an \APC-decent proof system. Them 
For every Frege rule %rewriteoxpredchanaslmedzera
 which derives $\phi$ from $\phi_1,\dots,\phi_k$, there is a \pv-function $F$ such that \APC proves that $F$ maps each $(k+1)$-tuple $\langle \pi_1,\dots,\pi_k,\rho\rangle$ to a $P$-proof of $\phi|_{\rho}$, where $\pi_i$ is a $P$-proof of $\phi_i|_{\rho}$ for a substitution $\rho$ replacing each atom of $\phi,\phi_1,\dots,\phi_k$ by a formula.
\item[2.] There is a \pv-function $F$ such that \APC proves that $F$ maps each pair $\langle \phi,b\rangle$, for assignment $b$ satisfying formula $\phi$, to a $P$-proof of $\phi(b)$.
\item[3.] Let $\pi$ be a $P$-proof of $E\rightarrow \phi$, where $E$ defines a computation of a circuit which is allowed to use atoms from $\phi$ as inputs but other atoms of $E$ do not appear in $\phi$, i.e. $E$ is the conjunction of extension axioms of \EF built on atoms from $\phi$. Then, there is a $poly(|\pi|)$-size $P$-proof of $\phi$. %rewriteoxpredchitemsmedzeramiaskipom
\end{itemize}
\end{lemma}

%(- from a P-proof of nA->nB and a P-proof of nA->B we get a P-proof of A, (proof: P>WF>(nA->nB ->(nA->B)>A))
%(- from a P-proof of neg B(b) or A we get a P-proof of A. (proof: P>WF>(nB(b) or A >A))
%(- APC prves that for each circuit A and assignment st A(a)=1, WF/P proves efficiently A(a) (Note that WF operates with circuits) (proof: induction if right gation/negation of subcircuits derivable, then conjunction or disjunction derivable)

\proof \WF is known to prove efficiently its own reflection principle, cf. \cite{Jwphp}. In order to show that it is \APC-decent, it thus suffices to prove that it satisfies Items 1-3 from Definition \ref{d:decent}. 

Item 2 is established already in \PV by $\Sigma^b_1$-induction on the length of the proof $\pi$ (which can be used because of $\forall\Sigma^b_1$-conservativity of \SB over \PV): $F$ replaces each circuit $C$ from $\pi$ by $C|_{\rho}$ and preserves all \WF-derivation rules. 

Item 1 holds trivially if the given \WF-system $P$ is the \WF-system $P'$ from Definition~\ref{d:decent}. Otherwise, we use implicational completeness of $P$ and the completeness of the language of $P$ to simulate all $O(1)$ Frege rules of $P'$ by $O(1)$ steps in $P$. (This does not require that the implicational completeness of $P$ is provable in $\APC$ because we need to simulate only $O(1)$ Frege rules of finite size). Similarly, by $\Sigma^b_1$-induction and the completeness of the language %rewriteoxpredch' language 'anaslmedzera 
 of $P$, we simulate each circuit in the language of $P'$ by a circuit in the language of $P$ and show that this simulation preserves the similarity rule. Then, given an $s$-size $P'$-proof of $\phi$, we obtain a $poly(s)$-size $P$-proof of $\phi$ using the simulation of Frege rules of $P'$, the similarity rule and dWPHP axiom, together with substituting the right circuits in Frege rules. This is done again in \PV by $\Sigma^b_1$-induction on the length of the $P'$-proof.

Item 3 follows by simulating modus ponens as in the proof of Item 1.
\medskip

For the `moreover' part, we consider three %rewriteoxpredch' three 'snaslmedzerou
 cases:

\noindent Item 1: As in the case of \WF, observe that by completeness, $P$ proves $\phi_1\rightarrow \dots \phi_k\rightarrow \phi$ and that this fact is provable in \PV. By Definition \ref{d:decent}, Item 2, \APC can construct a $P$-proof of $\phi_1|_{\rho}\rightarrow \dots \phi_k|_{\rho}\rightarrow \phi|_{\rho}$. The claim then follows from $k$ applications of Definition~\ref{d:decent}, Item~3.
\smallskip

\noindent Item 2: By Definition \ref{d:decent}, Item 1, it suffices to prove the claim for \WF. This follows from a $\Sigma^b_1$-induction on the complexity of $\phi$, where we strengthen the claim to: ``For each multi-output circuit $C$ and complete assignment $b$, $F$ outputs a $k|C|^2$-size \WF-proof %rewriteoxpredchanaslmedzera
 which contains every single-output circuit $C'(b)$ such that $C'$ is a subcircuit of $C$ satisfied by $b$ or $C'$ is $\neg C''$ for a subcircuit $C''$ of $C$ falsified by $b$. Here, $k$ is an absolute constant". The strengthened claim holds for literals by simulating $O(1)$ Frege rules, which we have by Item 1. Further, $O(1)$ Frege rules (and their substitutional instances) suffice to prove that if the claim holds for a multi-output circuit $B$ and we extend $B$ by one gate to a multi-output circuit $B'$, then the claim holds for $B'$ as well. (The length of the proof corresponding to $B'$ is $\le k(|B|)^2+O(|B'|)<k|B'|^2$, for sufficiently big $k$, where we use the choice of \WF %rewriteoxpredchanaslmedzera
 which guarantees linear increase of proof-size when applying substitutions and modus ponens.) 
\smallskip

%rewriteoxpredchmedzerasmallskipprednoumedzerapredsmallskipananaslpara
\noindent Item3: This is easy to see for $P=\EF$, since \EF can introduce extension axioms. For \APC-decent system $P$, observe that a $P$-proof $\pi$ of $E\rightarrow\phi$ implies the existence of a $poly(|\pi|)$-size \EF-proof of $Ref_P\wedge E\rightarrow\phi$, where $Ref_P$ postulates the reflection principle for $P$ instantiated by $\pi$, which further implies the existence of a $poly(|\pi|)$-size \EF-proof of $Ref_P\rightarrow\phi$, and finally a $poly(|\pi|)$-size $P$-proof of $\phi$. \qed
\bigskip

\APC-decent proof systems can be much stronger than \WF. For example, consider $\mathsf{ZFC}$ as a propositional proof system: a $\mathsf{ZFC}$-proof of propositional formula $\phi$ is a $\mathsf{ZFC}$-proof of the statement encoding that $\phi$ is a tautology. We can add the reflection of $\mathsf{ZFC}$ to \WF, i.e. we will allow \WF to derive (substitutional instances of) formulas stating that ``If $\pi$ is a $\mathsf{ZFC}$-proof of $\phi$, then $\phi$ holds." The new system is as strong as $\mathsf{ZFC}$ w.r.t. tautologies and it is easy to see that it is \APC-decent. (The reflection of the system can be proved in \APC extended with an axiom postulating the reflection for $\mathsf{ZFC}$.)

\begin{theorem}[Learning versus automatability]\label{t:main}
Let $P$ be an \APC-decent proof system and assume there is a sequence of boolean functions $h=\{h_n\}_{n>n_1}$, for a constant $n_1$, %srewrite'a constant n1, 'tiez'n1'vdefiniciihanaslmedzera
 such that $P$ proves efficiently $\ttable(h_n,2^{n/4},1/2-1/2^{n/4})$. Then, for each constant $K$ and constant $\gamma<1$, the following statements are equivalent.
\begin{itemize}
\item[1.] {\bf Provable learning.} For each $k\ge 1$ and $\ell\ge K+1$, %rewriteoxpredch'\ge K+1 'anaslmedzera
 there are $2^{Kn^{\gamma}}$-size circuits $A$ such that for each sufficiently big $n$, $P$ proves efficiently $$\mathsf{lear}^h_{1/2^{\ell n^{\gamma}}}(A,\Circuit[n^k],1/2-1/2^{Kn^{\gamma}},1/2^{Kn^{\gamma}}).$$%srewrite2x$$vdisplayedflaapredch'.'
\item[2.] {\bf Provable automatability.} For each $k\ge 1$, for each function $s(n)\ge 2^n$, there is a constant $K'$ and $s^{K'}$-size circuits $B$ such that $P$ proves efficiently $$\mathsf{aut}_{P}(B, \Phi,s),$$ where $\Phi$ is the set of pairs $\langle \ttable(f,2^{Kn^{\gamma}},1/2-1/2^{Kn^{\gamma}}), \ttable(f,n^k)\rangle$ for all boolean functions $f$ with $n$ inputs.%srewritevite2'k\ge..2^n,'s^K;'$$aut(,s),$$'%rewritethisthm
%\item[1'.] There is a subexponential-size circuit $A$ such that \WF proves efficiently that $A$ learns circuits of size $n^{\Omega(l)}$.
%\item[2'.] There are p-size circuits $B$ such that \WF proves efficiently $\mathsf{aut}_{\WF}(B,\Phi,2^{O(n)})$ where $\Phi$ is the set of formulas $\ttable(f,n^l)$ for all boolean functions $f$ with $n$ inputs.
\end{itemize}
\end{theorem}

\proof 

\smallskip
\noindent {\bf ($1.\rightarrow 2.$)} We first prove the following statement in \APC.

\begin{claim}[in \APC]\label{c:mpv} Assume that $\pi$ is a $P$-proof of $\mathsf{lear}^y_{1/2^{\ell n^{\gamma}}}(A,\Circuit[n^k],1/2-1/2^{Kn^{\gamma}},\delta)$ for a circuit $A$ and a boolean function $y$ represented by fixed bits in formula $\mathsf{lear}^y_{1/2^{\ell n^{\gamma}}}(\cdot,\cdot,\cdot,\cdot)$. Further, assume that the probability that $A$ on queries to $f$ outputs a circuit $D$ such that $\Pr[D(x)=f(x)]\ge 1/2+1/2^{Kn^{\gamma}}$ is $<\delta$, where the outermost probability is counted approximately with error $1/2^{\ell n^{\gamma}}$ using \pv-function $Sz$ and the corresponding assumptions \LBtt expressing hardness of $y$, i.e. using formulas $\Pr^y[\cdot]_{1/2^{\ell n^{\gamma}}}$ for the same $y$ as above - we treat $y$ as a free variable here. %srewrite' - we treat .. here.'smedzerou
 Then there is a $poly(|\pi|)$-size $P$-proof of $\ttable(f,n^k)$ or $y$ does not satisfy the assumptions of $\Pr^y[\cdot]_{1/2^{\ell n^{\gamma}}}$. %srewrite' or y does.. gamma}}}.'smedzerou
% or there is a $2^{||y||/4}$-size circuit $(1/2+1/2^{||y||/4})$-approximating $y$, for some $||y||>n_0$.
\end{claim}

To see that the claim holds, we reason in \APC as follows. Assume $\pi$ is a $P$-proof of $\mathsf{lear}^y_{1/2^{\ell n^{\gamma}}}(A,\Circuit[n^k],1/2-1/2^{Kn^{\gamma}},\delta)$ but $A$ on queries to $f$ outputs a circuit $(1/2+1/2^{Kn^{\gamma}})$-approximating $f$ with probability $<\delta$. Then, either $y$ does not satisfy the assumptions of $\Pr^y[\cdot]_{1/2^{\ell n^{\gamma}}}$ or %srewrite' or y does.. gamma}}} or 'smedzerou
 %either for some $||y||>n_0$, there is a $2^{||y||/4}$-size circuit $(1/2+1/2^{||y||/4})$-approximating $y$ or 
there is a trivial $2^{O(n)}$-size $P$-proof of $\neg\ttable(f,n^k)\rightarrow\neg R(b)$, for predicate $R$ from the definition of $\mathsf{lear}^y_{1/2^{\ell n^{\gamma}}}(A,\Circuit[n^k],1/2-1/2^{Kn^{\gamma}},\delta)$ and a complete assignment $b$. The $P$-proof is obtained by evaluating function $Sz$ which counts the confidence of $A$ - note that functions $f,y$ and algorithm $A$ are represented inside $P$ by fixed bits so the $P$-proof just evaluates a $2^{O(n)}$-size circuit on some input, which is possible by Lemma \ref{l:decent}, Item 2. (We use here also the fact that \APC knows that the probability statement expressed by function $Sz$ translates to $\neg R$ in the negation normal form.) %rewriteoxpredchvetavzatvsmedzerami
 The formula $\neg\ttable(f,n^k)\rightarrow\neg R(b)$ is obtained from $\neg R(b)$ by an instantiation of a single Frege rule, which is available by Lemma \ref{l:decent}, Item 1. %exhaustively evaluating $\Pr[f(x)=D(x)]$ in \WF. 
Applying again Lemma \ref{l:decent}, Item 1, from a $P$-proof of $\mathsf{lear}^y_{1/2^{\ell n^{\gamma}}}(A,\Circuit[n^k],1/2-1/2^{Kn^{\gamma}},\delta)$ and a $P$-proof of $\neg\ttable(f,n^k)\rightarrow\neg R(b)$, we construct a $poly(|\pi|)$-size $P$-proof of $\ttable(f,n^k)$. This proves the claim.%\footnote{Note that our encoding of $\mathsf{lear}^h_{1/2^{\ell n^{\gamma}}}(L,\Circuit[n^k],\epsilon,\delta)$ by $\neg\ttable(f,n^k)\rightarrow R$ made it easy to construct a \WF-proof of $\ttable(f,n^k)$ in \APC without constructing \APC-proofs in \APC and then translating them to \WF inside \PV.}
\medskip

Next, observe that \APC proves that ``If for a sufficiently big $n$ and $\ell\ge K+1$ %rewriteoxpredch'\ge K+1 'anaslmedzera
 the probability that a circuit $A$ on queries to $f$ outputs a circuit $(1/2+1/2^{Kn^{\gamma}})$-approximating $f$ is $\ge 1/2^{Kn^{\gamma}}$, where the probability is counted approximately with error $1/2^{\ell n^{\gamma}}$ using \pv-function $Sz$ and the corresponding assumptions \LBtt, then there is a circuit of size $|A|$ $(1/2+1/2^{Kn^{\gamma}})$-approximating $f$ or $y$ does not satisfy the assumptions of  $\Pr^y[\cdot]_{1/2^{\ell n^{\gamma}}}$.'' %srewrite' or y does.. gamma}}}.'''smedzerou 
%or there is a $2^{||y||/4}$-size circuit $(1/2+1/2^{||y||/4})$-approximating $y$, for some $||y||\le n_0$." 
This is because, if such a circuit did not exist, a trivial surjection would witness that $2^m$ times the probability that $A$ outputs a circuit $(1/2+1/2^{Kn^{\gamma}})$-approximating $f$, counted approximately with error $1/2^{\ell n^{\gamma}}$ using function $Sz$, is $\preceq_{1/2^{\ell n^{\gamma}}} 0$. Here, $2^m$ is the domain of the surjection. By Proposition \ref{really} $1. ii)$, this would imply $2^m/2^{Kn^{\gamma}}<2^{m+1}/2^{\ell n^{\gamma}}$, which is a contradiction for $\ell\ge K+1$ %rewriteoxpredch'\ge K+1 'anaslmedzera
 and sufficiently big $n$. 

Therefore, Claim \ref{c:mpv} implies that \APC proves that ``For sufficiently big $n$ and $\ell\ge K+1$, %rewriteoxpredch'\ge K+1 'anaslmedzera
 if $\pi$ is a $P$-proof of $\mathsf{lear}^y_{1/2^{\ell n^{\gamma}}}(A,\Circuit[n^k],1/2-1/2^{Kn^{\gamma}},1/2^{Kn^{\gamma}})$ for circuits $A$ of size $2^{Kn^{\gamma}}$, then there is a $P$-proof of $\ttable(f,n^k)$ or there is a $2^{Kn^{\gamma}}$-size circuit $(1/2+1/2^{Kn^{\gamma}})$-approximating $f$ or there is a $2^{||y||/4}$-size circuit $(1/2+1/2^{||y||/4})$-approximating $y$ or $||y||\le n_0$ or $||y||\ne S(\cdot,2^m,2^{2^{\ell n^{\gamma}}})$;" %rewriteoxpredch'2^{2^ln^gamma}"; 'snaslmedzerou
 for $n_0$ from Definition \ref{d:hard}. %srewritevpredchvete' or ||y||<n0.. hard}. 'smedzerou
 Since this is a $\Sigma^b_1$-statement, by Lemma \ref{singlehard}, \PV proves the same statement with the existential quantifiers witnessed by \pv-functions assuming they are given a boolean function $h'$ which is hard for circuits of size $2^{||h'||/4}$, for sufficiently big $|h'|$.

The last statement provable in \PV is $\Pi^b_1$ so we can translate it to \EF. This gives us $poly(|\pi|,2^{n})$-size circuits $B_0$ such that for sufficiently big $n$, \EF proves efficiently 
\medskip

\leftline{``If $\ell\ge K+1$,}%rewriteoxpredch' If l\ge K+1,'
\leftline{\ $h'$ is not computable by a particular circuit of size $2^{||h'||/4}$, $|h'|$ is sufficiently big,} 
\leftline{\ $y$ is not $(1/2+1/2^{||y||/4})$-approximable by a particular circuit of size $2^{||y||/4}$, $||y||>n_0$,}
\leftline{\ $||y||=S(\cdot,2^m,2^{2^{\ell n^{\gamma}}})$}%srewritetentolinesmedzerami%rewriteoxpredch'2^{2^ln^gamma}}}'
\leftline{ and $\pi$ is a $P$-proof of $\mathsf{lear}^y_{1/2^{\ell n^{\gamma}}}(A,\Circuit[n^k],1/2-1/2^{Kn^{\gamma}},1/2^{Kn^{\gamma}})$ for $2^{Kn^{\gamma}}$-size $A$,}
\rightline{then $B_0$ (given $\pi,h'$ and formula $\ttable(f,n^k)$) outputs a $P$-proof of $\ttable(f,n^k)$}
\rightline{ or $B_0$ outputs a $2^{Kn^{\gamma}}$-size circuit $(1/2+1/2^{Kn^{\gamma}})$-approximating $f$.".\footnote{Formally, the statement `If a particular assignment $a$ satisfies formula $\phi$, then formula %rewriteoxpredch' fomula'arestoffootnote
 $\psi$ holds' means that `If $a$ is the output of a computation of a specific circuit $W$ (where $W$ is allowed to use as inputs atoms from $\psi$, but other atoms of $W$ do not appear in $\psi$), and $a$ satisfies $\phi$, then $\psi$'. By Lemma \ref{l:decent}, Item~3, if we assume that the statement is efficiently provable in $P$ and that $P$ proves efficiently $\phi$, then $P$ proves efficiently $\psi$. Note also that for $A,B\in\Sigma^b_0$, the translation $||A\rightarrow B||$ is $\neg ||\neg A||\rightarrow ||B||$, which might not be the same formula as $||A||\rightarrow ||B||$. Nevertheless, \EF proves efficiently that $E\rightarrow (||A||\leftrightarrow \neg ||\neg A||)$, where $E$ postulates that auxiliary variables of $||A||$ encode the computation of a suitable circuit. Therefore, in systems like \EF or $P$, if we have a proof of $||A||$ and $||A\rightarrow B||$, we can remove the assumption $E$ after proving $E\rightarrow ||B||$, assuming `non-input' variables of $E$ do not occur in $||B||$, and ignore the difference between $||A||$ and $\neg ||\neg A||$.}}%srewritepredchfootnotesdvoma}}nakonci
\medskip

\noindent If we now assume that $P$ proves efficiently $\ttable(h_n,2^{n/4},1/2-1/2^{n/4})$ and that Item~1 holds, then by Definition \ref{d:decent}, Items 1-3, for each $k$, there are p-size circuits $B_1$ such that for each sufficiently big $n$, $P$ proves efficiently ``$B_1$ (given just formula $\ttable(f,n^k)$) outputs a $P$-proof of $\ttable(f,n^k)$ or $B_1$ outputs a $2^{Kn^{\gamma}}$-size circuit $(1/2+1/2^{Kn^{\gamma}})$-approximating $f$." (We use here also the fact that \PV knows that $S(\cdot,2^m,2^{2^{\ell n^{\gamma}}})$ depends just on $n$.) %rewriteoxpredchvetavzatvsmedzerami
%Remark: WF>'E is a computation of the particular circuit and tt[E]>..' > P>'same' > P>'E is not a computation of the particular circuit or ..' >(substitution)> P>'..'
Consequently, since $P$ proves efficiently its own reflection, for each sufficiently big $n$, $P$ proves efficiently that ``if $\pi$ is a $P$-proof of $\ttable(f,2^{Kn^{\gamma}},1/2-1/2^{Kn^{\gamma}})$ then $B_1$ outputs a $P$-%rewriteoxpreedch'-'anaslmedzera
proof of $\ttable(f,n^k)$".\footnote{It is assumed that the encoding of the statement coincides with the encoding of $\mathsf{aut}_P$.} %\textcolor{red}{(Here, we need to employ formalized error reduction.)} 
Finally, we make the $P$-proofs work for all $n$ by increasing the size of $B_1$ by a constant. This finishes the proof of case $(1.\rightarrow 2.)$.%rewriteproofof(1->2)a(2->1)smedzerami
\bigskip

\noindent {\bf ($2.\rightarrow 1.$)} The opposite implication can be obtained from Lemma \ref{l:apcautlear} and \ref{l:razxor} %rewriteoxpredchanaslmedzera
 which formalize Theorem \ref{t:intro2}.

\begin{lemma}\label{l:apcautlear} For each $d\ge 2$, each $k\ge 10d$ and each sufficiently big $c$, there is a \pv-function $L$ such that for each \pv-function $B$ the theory \APC proves: Assume the reflection principle for $P$ holds, $\pi$ is a $P$-proof of \begin{equation}\label{e:xor}\ttable(h_n\oplus g,2^{Kn^{\gamma}},1/2-1/2^{Kn^{\gamma}})\vee \ttable(g,2^{Kn^{\gamma}}),\end{equation} where $g$ is represented by free variables, and that $B$ automates $P$ on $\Phi$ up to size $|\pi|^c$. Then, for prime $n^d\le p\le 2n^d$, where $2^{n^d}\in Log$, for $\delta^{-1}\in Log$ such that $\delta<1/N^3=2^{3n}$, $L(B,\pi,p)$ is a $poly(2^n,|\pi|)$-size circuit learning circuits with $m=n^d$ inputs and size $m^{k/10d}$, with confidence $1/N^4$, up to error $1/2-1/N^3$, where the confidence is counted approximately with error $\delta$ using \pv-function $Sz$ and the corresponding assumptions \LBtt expressing hardness of a boolean function $y$, i.e. using formulas $\Pr^y[\cdot]_{\delta}$.
\end{lemma}

\begin{lemma}[%rewriteoxnaslmedzera,2xvnaslleme'2^n(..)'a' for sufficiently big n '
`XOR trick']\label{l:razxor} \PV proves that for all boolean functions $g,h''$ with $n$ inputs, for sufficiently big $n$,  $\LBtt'(h'',3\cdot 2^{Kn^{\gamma}},2^n(1/2-1/2^{Kn^{\gamma}}))$ implies $\LBtt'(h''\oplus g,2^{Kn^{\gamma}},2^n(1/2-1/2^{Kn^{\gamma}}))\vee \LBtt'(g,2^{Kn^{\gamma}})$, where $\LBtt'$ is obtained from $\LBtt$ by setting $n_0=0$ and skipping the universal quantifier on $n$, i.e. all formulas $\LBtt'$ refer to the same $n$.%srewritevtejtolemme'vLBttflacha'where LBtt' is.. same n.'
\end{lemma}

The proof of Lemma \ref{l:razxor} is almost immediate: By $\Sigma^b_1$-induction, a $2^{Kn^{\gamma}}$-size circuit $C_1$ computing $g$ and a $2^{Kn^{\gamma}}$-size circuit $C_2$ $(1/2+1/2^{Kn^{\gamma}})$-approximating $h''\oplus g$ can be combined into a circuit $C_1\oplus C_2$ of size $3\cdot 2^{Kn^{\gamma}}$ which $(1/2+1/2^{Kn^{\gamma}})$-approximates $h''$.
\bigskip

The implication ($2.\rightarrow 1.$) can be derived from Lemma \ref{l:apcautlear} and \ref{l:razxor} as follows. Since the \APC-provable statement from Lemma \ref{l:apcautlear} is $\Sigma^b_1$, similarly as above, we can witness it and translate to \EF at the expense of introducing an additional assumption about the hardness of a boolean function $h'$. That is, for each p-size circuit $B$ there are $poly(|\pi|,2^{n^d})$-size circuits $A$ and %srewritevtejtovete', for'' there are.. and 'smedzerami
 $poly(|\pi|,2^{n^d})$-size \EF-proofs of
\medskip

\leftline{``If the reflection principle for $P$ is satisfied by a particular assignment,} 
\leftline{\ $\pi$ is a $P$-proof of ($\ref{e:xor}$),} 
\leftline{\ $h'$ is not computable by a particular circuit of size $2^{||h'||/4}$, $|h'|$ is sufficiently big,} 
\leftline{\ $y$ is not $(1/2+1/2^{||y||/4})$-approximable by a particular circuit of size $2^{||y||/4}$, $||y||>n_0$,}
\leftline{\ $||y||=S(\cdot,\cdot,2^{|\delta^{-1}|})$}%srewriteoxtentolinesmedzerami
\leftline{ and $n^d\le p\le 2n^d$ is a prime,}
\rightline{then, for $\delta<1/N^3$, $\mathsf{lear}^y_{\delta}(L(B,\pi,p),\Circuit(m^{k/10d}),1/2-1/N^3,1/N^4)$}
\rightline{or $A(B,\pi,h')$ outputs a falsifying assignment of $\mathsf{aut}_{P}(B,\Phi,|\pi|^{c})$.".}
\medskip

Analogously, \PV-proof from Lemma \ref{l:razxor} yields p-size \EF-proofs of the implication ``$\ttable(h_n,3\cdot 2^{Kn^{\gamma}},1/2-1/2^{Kn^{\gamma}})$ is falsified by a particular assignment or ($\ref{e:xor}$) holds". By the assumption of the theorem, there are p-size $P$-proofs of $\ttable(h_n,3\cdot 2^{Kn^{\gamma}},1/2-1/2^{Kn^{\gamma}})$ for sufficiently big $n$. %srewritevpredchanaslvete' for sufficiently big n. 'smedzerami
 Hence, by Definition~\ref{d:decent}, Items 1-3, there are p-size $P$-proofs of ($\ref{e:xor}$) for sufficiently big $n$. As $P$ proves efficiently also its own reflection, this yields $poly(2^{n^d})$-size $P$-proofs of
\medskip

\leftline{``If $h'$ is not computable by a particular circuit of size $2^{||h'||/4}$, $|h'|$ is sufficiently big,} 
\leftline{\ $y$ is not $(1/2+1/2^{||y||/4})$-approximable by a particular circuit of size $2^{||y||/4}$, $||y||>n_0$,}
\leftline{\ $||y||=S(\cdot,\cdot,2^{|\delta^{-1}|})$}%srewriteoxtentolinesmedzerami
\leftline{ and $n^d\le p\le 2n^d$ is a prime,}
\rightline{then, for $\delta<1/N^3$, $\mathsf{lear}^y_{\delta}(L(B,\pi,p),\Circuit(m^{k/10d}),1/2-1/N^3,1/N^4)$}
\rightline{or $A(B,\pi,h')$ outputs a falsifying assignment of $\mathsf{aut}_{P}(B,\Phi,|\pi|^{c})$.".}
\medskip

\noindent By Bertrand's postulate there is a prime $n^d\le p\le 2n^d$, so \EF proves that $p$ is a prime by a  trivial $2^{O(n^d)}$-size proof which verifies all possible divisors. Therefore, choosing $d>1/\gamma$, Item 2 and p-size $P$-proofs of $\ttable(h_n,2^{n/4},1/2-1/2^{n/4})$ imply Item 1. %\textcolor{red}{(KPT needed or provability of lb in \PV which yields \PV=\APC.)}
\bigskip

It remains to prove Lemma \ref{l:apcautlear}.

Suppose $\pi$ is a $P$-proof of (\ref{e:xor}). Assuming that $B$ automates $P$ on $\Phi$, we want to obtain a \Ppoly-natural property useful against $\Circuit[n^{k}]$. To do so, observe (first, without formalizing it in \APC) that for each $g$, $B$ can be used to find a proof of $\ttable(h_n\oplus g,n^k)$ or to recognize that $\ttable(g,2^{Kn^{\gamma}})$ holds - if $\ttable(g,2^{Kn^{\gamma}})$ was falsifiable, there would exist a $poly(|\pi|)$-size $P$-proof of $\ttable(h_n\oplus g,2^{Kn^{\gamma}},1/2-1/2^{Kn^{\gamma}})$ obtained %rewriteoxpredch' obtained 'naslmedzeraanasl'assignment'
 by substituting the falsifying assignment to the proof of (\ref{e:xor}) and thus $B$ would find a short proof of $\ttable(h_n\oplus g,n^k)$, for sufficiently big $c$. Since for random $g$, both $h_n\oplus g$ and $g$ are random functions, we know that with probability $\ge 1/2$ $B$ finds a proof of $\ttable(h_n\oplus g,n^k)$ or with probability $\ge 1/2$ it recognizes that $\ttable(g,2^{Kn^{\gamma}})$ holds. In both cases, $B$ yields a \Ppoly-natural property useful against $\Circuit[n^k]$. 

Let us formalize reasoning from the previous paragraph in \APC. Let $N=2^n$ and $B'$ be the algorithm which uses $B$ to search for $P$-proofs of $\ttable(h_n\oplus g,n^k)$ %srewritepredch' n^k)% 'smedzerou
 or to recognize that $\ttable(g,2^{Kn^{\gamma}})$ holds. $B'$ uses $\pi$ to know how long it needs to run $B$. Assume for the sake of contradiction that $$G_0:=\{g\oplus h_n\mid B'(g)\text{ outputs a }P\text{-proof of }\ttable(h_n\oplus g,n^k)\}\preceq_{0} 2^N/3$$ $$G_1:=\{g\mid B'(g)\text{ recognizes that }\ttable(g,2^{Kn^{\gamma}})\text{ holds}\}\preceq_{0} 2^N/3.$$ It is easy to construct a surjection $S$ witnessing that $2^N\preceq_0 G_0\cup G_1$: $S$ maps $g\in G_1$ to $g$ and $g\in G_0$ to $g\oplus h_n$. Following the argument above we conclude that $S$ is a surjection: for each $g$, either $g\in G_1$ (and $S(g)=g$) or $g\oplus h_n\in G_0$ (and $S(g\oplus h_n)=g$). Here, we use the assumption that \APC knows that $P$ admits the substitution property and simulates Frege rules. Thus, by Proposition \ref{lem} $iv)$, $2^N\preceq_0 2\cdot 2^N/3$, which yields a contradiction by Proposition \ref{really} $1. ii)$. Consequently, by Proposition \ref{really} $1. i)$, $G_0\succeq_{\delta} 2^N/3$ or $G_1\succeq_{\delta} 2^N/3$ for $\delta^{-1}\in Log$. %Hence, $G_0\succeq_0 N/4$ or $G_1\succeq_0 N/4$
Since $g\in G_0$ and $g\in G_1$ are decidable by p-size circuits and we assume the reflection principle for $P$ (which implies that $G_0$ is useful), this means that either $G_0$ or $G_1$ defines a \Ppoly-natural property useful against $\Circuit[n^k]$.

Finally, by the \APC-formalization of \cite{CIKK}, Theorem \ref{t:cikk}, we obtain $poly(2^n,|\pi|)$-size circuit $L(B,\pi,p)$ learning circuits with $m=n^d$ inputs and size $n^{k/10}$, over the uniform distribution, with membership queries, confidence $1/N^4$, up to error $1/2-1/N^3$.
 \qed

%rewriterestsekcie
\begin{corollary}\label{c:main}
Assume there is a $\mathsf{NE}\cap \mathsf{coNE}$-function $h_n:\{0,1\}^n\mapsto \{0,1\}$ such that for each sufficiently big $n$, $h_n$ is not $(1/2+1/2^{n/4})$-approximable by $2^{n/4}$-size circuits. Then there is a proof system $P$ (which can be described explicitly\footnote{More formally, there is a p-time algorithm $R$ such that given predicates $H_0,H_1$ defining $h_n$ (see the proof of Corollary \ref{c:main}), $R$ outputs a p-time algorithm defining system $P$.} given the definition of $h_n$) %rewriteoxpredch' (..) 'anaslmedzera
such that for each constant $K$ and $\gamma<1$, Items 1 and 2 from Theorem \ref{t:main} are equivalent. Moreover, the equivalence holds for each \APC-decent system simulating $P$.%rewriteoxpredchvetasmedzerou
\end{corollary}

\proof We want to construct an \APC-decent proof system $P$ which proves efficiently $\ttable(h_n,2^{n/4},1/2-1/2^{n/4})$, for some $h_n$. By the assumption, there is $h_n\in \mathsf{NE}\cap \mathsf{coNE}$, which is hard to approximate. Let $H_{\epsilon}$, for $\epsilon\in\{0,1\}$, be the \Ptime-time predicates defining $h_n$, i.e. $h_n(x)=\epsilon\leftrightarrow \exists y, |y|\le 2^{O(|x|)}, \ H_{\epsilon}(x,y)$. %srewritepredchvetasmedzerami
 Define $P$ as \WF extended by a rule, which allows to derive every substitutional instance of $||L||$, where $||\cdot ||$ is the propositional translation from Section \ref{sba} and $L$ is the formula %rewriteoxpredch' ||L||,.. fla 'anaslflaazpo'LBtt. '
\begin{equation*} \left[\forall x<2^{||z||},\ \left((H_{0}(x,y^{0}_{x})\vee H_{1}(x,y^{1}_{x}))\wedge \bigwedge_{\epsilon=0,1}(H_{\epsilon}(x,y^{\epsilon}_{x})\rightarrow
 z_x=\epsilon)\right)\right]\end{equation*}
\begin{equation*}\rightarrow\LBtt(z,2^{||z||/4},2^{||z||}(1/2-1/2^{||z||/4})),\end{equation*} for sufficiently big $n_0$ in $\LBtt$. %srewritenas%apredchdisplayedflaazanou' where.. big. '
% whenver it is given nondeterministic bits witnessing the values of $h_n$ on all $2^n$ inputs. (The non-deterministic bits guarantee that we can efficiently recognize correct $P$-proofs and formulate the definition of a $P$-proof in \APC.) %srewritemedzeraprednaslvetou
By definition, $P$ proves efficiently $\ttable(h_n,2^{n/4},1/2-1/2^{n/4})$ for each sufficiently big $n$ (we can hardwire the hardness of a boolean function for remaining $n$, if needed%rewriteoxpredch'n if needed'anaslmedzera
) and satisfies Items 1-3 from the definition of \APC-decent systems. To see that $P$ proves its own reflection principle we reason in \APC: given a $P$-proof $\pi$, each circuit in $\pi$ is derived either by a \WF-rule or it is a substitutional instance of $||L||$%srewritevpredchvete'substitutional''||L||'smedzer%rewriteoxpredch' ||L||'
, so by $\Sigma^b_1$-induction on the length of $\pi$ and the \APC-provability of the reflection of \WF, cf. \cite{Jwphp}, $L$ %srewrite' L '%rewriteox' L'
 implies that each circuit from $\pi$ holds. %srewritepredchmedzera%rewriteoxpredchmedzera
Similarly as in the proof of Theorem~\ref{t:main}, we can now translate the resulting $\Sigma^b_1$ theorem of \APC to \EF and remove the assumptions %srewrite' the assumptions 'smedzerou
 after moving to $P$. This shows that $P$ proves efficiently its own reflection and is \APC-decent. \qed
%\bigskip

%The proof of Theorem \ref{t:main} reveals also a conditional proof complexity collapse.

\begin{corollary}\label{c:collapse}
Let $P, P_0$ be \APC-decent proof systems and assume there is a sequence of boolean functions $h=\{h_n\}_{n>n_1}$, for a constant $n_1$, such that systems $P,P_0$ prove %srewrite'systems P,P0 prove ''for a constant n1'a'n1'vdefiniciih
 efficiently $\ttable(h_n,2^{n/4},1/2-1/2^{n/4})$. Then, for each constant $K$ and constant $\gamma<1$, Item~1 implies Item 2:
\begin{itemize}
\item[1.] {\bf $P$-provable automatability.} For each $k\ge 1$, for each function $s(n)\ge 2^n$, there is a constant $K'$ and $s^{K'}$-size circuits $B$ such that $P$ proves efficiently $\mathsf{aut}_{P}(B, \Phi,s)$, where $\Phi$ is the set of pairs $\langle \ttable(f,2^{Kn^{\gamma}},1/2-1/2^{Kn^{\gamma}}), \ttable(f,n^k)\rangle$ for all boolean functions $f$ with $n$ inputs.%srewritevitem1'k\ge..2^n,''s^K'''s)'
\item[2.] {\bf $P_0$-provable proof search.} For each $k\ge 1$, there is a constant $K'$ and $2^{K'n}$-size circuits $B$ such that $P_0$ proves efficiently ``$B$ (given just $\ttable(f,n^k)$) outputs a $P$-proof of $\ttable(f,n^k)$ or $B$ outputs a $2^{Kn^{\gamma}}$-size circuit $(1/2+1/2^{Kn^{\gamma}})$-approximating $f$.". %rewritethisthm
%\item[1'.] There is a subexponential-size circuit $A$ such that \WF proves efficiently that $A$ learns circuits of size $n^{\Omega(l)}$.
%\item[2'.] There are p-size circuits $B$ such that \WF proves efficiently $\mathsf{aut}_{\WF}(B,\Phi,2^{O(n)})$ where $\Phi$ is the set of formulas $\ttable(f,n^l)$ for all boolean functions $f$ with $n$ inputs.
\end{itemize}\end{corollary}

\proof Suppose Item 1 holds. By Theorem \ref{t:main}, Item 1 of Theorem \ref{t:main} holds. Then, following the proof of ($1.\rightarrow 2.$) of Theorem \ref{t:main} with $P_0$ instead of $P$ in the last paragraph, we obtain Item 2. \qed

\bigskip%srewritebigskipanasl2para

\noindent {\bf Remark on the collapse.} Denote by $P\vdash\phi_n$ the existence of a p-time algorithm which given $\phi_n$ generates a $P$-proof of $\phi_n$. Corollary \ref{c:collapse} exploits the fact (captured by Lemma \ref{l:decent}, Item 2) that for \PV-decent proof systems $P$ (defined analogously as \APC-decent systems, with \APC replaced by \PV and \WF replaced by \EF%rewriteoxpredch', with.. EF'snaslmedzerou
) there is a p-time algorithm $B$ such that \begin{equation}\label{e:corecol}\EF\vdash SAT(x,y)\rightarrow Prf_{P}(B(x,y),\lceil SAT(x,y)\rceil),\end{equation} where formula $SAT(x,y)$ says that propositional formula (encoded by) $x$ is satisfied by assignment $y$, $Prf_P(z,x)$ says that $z$ is a $P$-proof of $x$, and $\lceil \phi\rceil$ is a code of formula $\phi$. %, and $y'$ are fixed bits determined by free atoms $y$. 
Importantly, while $y$ stands for free atoms in the assumption $SAT(x,y)$, it represents fixed bits (determined by $y$)
w.r.t. $P$ in $\lceil SAT(x,y)\rceil$. That is, there is a p-time algorithm %rewriteoxpredchanaslmedzera
 which given $1^{|x|}$ %rewritepredch' 1^|x| 'smedzerami
 generates an \EF-proof of $SAT(x,y)\rightarrow Prf_{P}(B(x,y),\lceil SAT(x,y)\rceil)$.

Using (\ref{e:corecol}), it is possible to obtain a collapse similar to Corollary \ref{c:collapse} %rewriteoxpredchanaslmedzera
 which is essentially %rewriteoxpredch' essentially '' (captured by Lemma 3,Item2) 'anaslmedzera
 {\em unconditional}: Assume that \PV proves that a p-time algorithm efficiently generates $P$-proofs of the reflection principle for $P$. %(see below for a more precise formulation).%rewriteoxpredch'P. ..formulation).'anaslmedzera
 If there is a p-time algorithm $A$ such that $P\vdash \neg SAT(x,A(x))\vee Prf_P(A(x),x)$ %where $\lceil x(y)\rceil$ encodes formula $x$ with free atoms $y$ 
(in other words, we can efficiently generate $P$-proofs of $P$ being p-bounded and automatable), then there is a p-time algorithm $A'$ such that %rewriteoxpredch' there is a p-time algorithm A' such that 'naslmedzeraa'A''vnaslfle
$\EF\vdash  \neg SAT(x,A'(x))\vee Prf_P(A'(x),x)$. %rewriteoxpredch' x). 'anaslmedzera
 For example, if we can efficiently generate {\sf ZFC}-proofs of {\sf ZFC} being p-bounded and automatable, then we can efficiently generate \EF-proofs of {\sf ZFC} being p-bounded and automatable. %rewriteoxpredch'e. ' 

Intuitively, the proof proceeds as follows. Assume that, for some p-time algorithm $A$, $$\mathsf{ZFC}\vdash \neg  SAT(x,A(x))\vee Prf_{\mathsf{ZFC}}(A(x),x).$$
Then, \begin{equation}\label{e:collapse1}\EF\vdash Prf_{\mathsf{ZFC}}(\pi , \lceil\neg SAT(x,A(x))\vee Prf_{\mathsf{ZFC}}(A(x),x)\rceil),\end{equation} for an assignment $\pi$ %rewriteoxpredchanaslmedzera
 which is efficiently generable from $1^{|x|}$%rewriteoxpredch' 1^|x| 'anaslmedzera
. Similarly, since \PV proves that a p-time algorithm $C$ %rewriteoxpredch' C 'anaslmedzera
 efficiently generates $\mathsf{ZFC}$-proofs of the reflection principle for $\mathsf{ZFC}$, i.e. \PV proves that $C$ %rewriteoxpredch' C 'anaslmedzera
 given $x$ %rewriteoxpredch' x 'naslmedzeraanaslflaodrightarrow
 outputs a $\mathsf{ZFC}$-proof of $Prf_{\mathsf{ZFC}}(z,x)\rightarrow \phi$, where $\lceil\phi\rceil=x$, %rewriteoxpredch' where phi=x, 'anaslmedzera
we have \begin{equation}\label{e:collapse2}\EF\vdash Prf_{\mathsf{ZFC}}(C(x),\lceil Prf_{\mathsf{ZFC}}(A(x),x)\rightarrow \phi\rceil),\end{equation} %rewriteoxpredchanaslmedzerapredchequationanasl%
% In more detail, there are free atoms in outer formula $Prf_{\mathsf{ZFC}}$ determining a formula $\phi$ which is fixed w.r.t. $\pi'$, %rewriteoxpredch' formula.. $, 'naslmedzeraanasl%
% i.e. \EF proves that `for each $\phi$, $Prf_{\mathsf{ZFC}}(\pi',\lceil Prf_{\mathsf{ZFC}}(A(x),x)\wedge \lceil\phi\rceil=x\rightarrow \phi\rceil)$'. %rewriteoxpredchanaslmedzera
By (\ref{e:corecol}), 
\begin{equation}\label{e:collapse3}\EF\vdash \neg SAT(x,A(x))\vee Prf_{\mathsf{ZFC}}(B(x,A(x)),\lceil SAT(x,A(x))\rceil).\end{equation}%rewriteoxcode'beginequationlabelendequation'inprevious3displayedflasanasl' by ()-(). 'smedzerou
Therefore, by (\ref{e:collapse1})-(\ref{e:collapse3}), there is a p-time algorithm $B'$ such that $$\EF\vdash \neg SAT(x,A(x))\vee Prf_{\mathsf{ZFC}}(B'(x),x).$$%rewriteoxpredch'x'
\def\notes{%srewrite'\def\notes 'smedzerou
%rewritenaslsekcia
\section{Notes (To be deleted)}

Theorem \ref{t:main} holds unconditionally for some proof system? Proof: If SAT hard for $\Circuit[2^{n^{o(1)}}]$ then there is a pps P which proves it. For this pps, the equivalence still holds. If SAT easy, then both efficient automatability and lear exists, so they are equivalent (A complication: does the equivalence hold even w.r.t. $P$-provability?). If we consider lear with confidence 1, then $2^{n^{o(1)}}$ lower bounds might indeed suffice to get the equivalence. The confidence 1 might be ok because the proof system can be wlog strong enough to count with exponential precision. A complication with getting confidence 1: unclear how to witness that natural proofs yield lear, ie given natproperty, what is going to be the lear algorithm.

Include T t:lblear? Did Emil want to specify C1,.., D1,.. in the dWPHP axiom?
%srewritezvysoksekcie
Can we prove that if P is provably automatable on average, then P is automatable in worst-case (Razborov's XOR trick does not seem to work for average-case automatability)? Mention automatability of WF on weak clbs [MP'17]? Can we formulate the equivalence for APC instead of WF? Use Emil's/Kortens HM to get worst-case hardness assumption on h in main thm? Remark in Sec 4(?):There exists a nonuniform pps P such that P aut iff lear in P/poly: Case I: if lear in P/poly, we can define P which proves that and is aut on tt. Case II: if lear notin P/poly, define P=WF+tt(f), i.e. if P not aut (since it would yield natural property).}

\def\previouspaper{%rewritedefa}farbelow
\section{Learning from breaking pseudorandom generators}\label{s:leargen}

%rewritefromhereuntildefinitionofgcsp
Circuit lower bounds can be used to construct PAC learning algorithms also if we assume that they break pseudorandom generators. The construction goes back to a %well-known 
relation between predictability and pseudorandomness which can be interpreted in terms of learning algorithms, as shown by Blum, Furst, Kearn and Lipton \cite{BFKL} and later extended by several other works. In this section we survey some of these connections, derive a construction of learning algorithms from the non-existence of succinct nonuniform pseudorandom function families and show how these connections relate to a question of Rudich about turning demibits to superbits.
\bigskip

We start by recalling the construction from \cite{BFKL}, which underlies all results in this section.

\medskip
For an $n^c$-size circuit $C$ with $n$ inputs define a generator $$G_C:\{0,1\}^{mn}\mapsto\{0,1\}^{mn+m}$$ which maps $m$ $n$-bit strings $x_1,\dots,x_m$ to $x_1,C(x_1),\dots,x_m,C(x_m)$. 

\begin{lemma}[from \cite{BFKL}]\label{l:gen}
There is a randomized p-time function $L$ such that for every $n^c$-size circuit $C$, % and generator $G_C$ with $m=poly(n,s)$, 
if an $s$-size circuit $D$ satisfies $$\Pr[D(x)=1]-\Pr[D(G_C(x))=1]\ge 1/s,$$ then the circuit $C$ is learnable by $L(D)$ over the uniform distribution with random examples, confidence $1/2m^2s$, up to error $1/2-1/2ms$.
\end{lemma}

\proof Given $D$, $L(D)$ chooses a random $i\in [m]$, random bits $r_{i},\dots,r_m$, random $n$-bit strings $x_1,\dots,x_n$ except $x_i$ and queries the bits $C(x_1),\dots, C(x_{i-1})$. For $x_i\in\{0,1\}^n$, let $p_i:=D(x_1,C(x_1),\dots,x_{i-1},C(x_{i-1}),x_i,r_i,\dots,x_m,r_m)$. Then $L(D)$ on $x_i$ predicts the value $C(x_i)$ by outputting $\neg r_i$ if $p_i=1$ and $r_i$ otherwise. By triangle inequality, random $i\in [m]$ satisfies $$\Pr[p_{i}=1]-\Pr[p_{i+1}=1]\ge 1/ms$$ with probability $1/m$. Since the probability over $r_i\dots,r_m,x_1,\dots,x_m$ that $L(D)$ predicts $C(x_i)$ correctly is $$\frac{1}{2}\Pr[p_i=1\mid r_i\ne C(x_i)]+\frac{1}{2}(1-\Pr[p_i=1\mid r_i= C(x_i)]),$$ and $\Pr[p_i=1]=\frac{1}{2} \Pr[p_i=1\mid r_i=C(x_i)]+\frac{1}{2}\Pr[p_i=1\mid r_i\ne C(x_i)],$ it follows that $$\Pr_{x_i}[L(D)(x_{i})=C(x_i)]\ge 1/2+1/2ms$$ with probability $1/2m^2s$ over the internal randomness of $L(D)$. \qed
\medskip

%Using standard methods, the confidence parameter $1/poly(n,s)$ in Lemma \ref{l:gen} can be boosted to $1-1/n$ with just a polynomial increase in the size of learner. (Similarly, the error parameter $1/2-1/poly(n)$ can be reduced to $1/n$ by applying a hardness amplification on $C$.)%rewrite%inthispara

%Note that Lemma \ref{l:gen} does not answer Question \ref{q:ach} because a part of the success of $F(D)$ in predicting $C(x_i)$ stems from the possibility that the partial function given to $D$ is not hard.
\bigskip

%\noindent {\bf Learning vs pseudorandom generators.} 
The proof of Lemma \ref{l:gen} implies that learning on average follows from breaking pseudorandom generators. Specifically, let $R$ be a p-size circuit which given $r$ bits outputs an $n^c$-size circuit $C$ and consider a generator $G:\{0,1\}^{mn+r}\mapsto \{0,1\}^{mn+m}$ which applies $R$ on its first $r$ input bits in order to output a circuit $C$ and then computes as a generator $G_C$ on the remaining $mn$ inputs. Breaking $G$ implies that we can break $G_C$ with significant probability over $C$ drawn from the distribution induced by $R$. Consequently, breaking $G$ means that we can learn a big fraction of $n^c$-size circuits w.r.t. $R$. Can we improve this average-case learning into a worst-case learning which works for all $n^c$-size circuits? Since efficient learning algorithms for p-size circuits yield natural properties useful against p-size circuits, which by \cite{RR} break pseudorandom generators, a positive answer would present an important dichotomy: cryptographic pseudorandom generators do not exist if and only if there are efficient learning algorithms for small circuits (with suitable parameters). This possibility has been explored by Oliveira-Santhanam \cite{OS} and Santhanam \cite{S19}, cf. Section \ref{s:prfs}.

\begin{question}[Dichotomy]\label{q:dichotomy} Assume that for each $\epsilon<1$ there is no pseudorandom generator $g:\{0,1\}^n\mapsto \{0,1\}^{n+1}$ computable in \Ppoly and safe against circuits of size $2^{n^{\epsilon}}$ for infinitely many $n$. Does it follow that p-size circuits are learnable by circuits of size $2^{O(n^\delta)}$, for some $\delta<1$, with confidence $1/n$, up to error $1/2-1/2^{O(n^\delta)}$?
\end{question}

\subsection{Worst-case learning from strong lower bound methods} 

The proof of Lemma \ref{l:gen} shows also that we can construct a worst-case learning algorithm assuming that given an oracle access to a pseudorandom generator we can efficiently produce its distinguisher. In particular, a single method breaking all pseudorandom generators would suffice.

\begin{definition} The circuit size problem $\GCSP[s,k]$ is the problem to decide whether for a given list of $k$ samples $(y_i,b_i)$, $y_i\in\{0,1\}^n, b_i\in\{0,1\}$, there exists a circuit $C$ of size $s$ computing the partial function defined by samples $(y_i,b_i)$, i.e. $C(y_i)=b_i$ for the given $k$ samples $(y_i,b_i)$. The parameterized minimum circuit size problem $\MCSP[s]$ stands for $\GCSP[s,2^n]$ where the list of $2^n$ samples defines the whole truth-table of a Boolean function.
\end{definition}

If we were extraordinary in proving circuit lower bounds, we could solve \GCSP efficiently. Note that $\MCSP[n^{O(1)}]\in\Ppoly$ is stronger assumption than the existence of \Ppoly-natural property useful against \Ppoly, which breaks pseudorandom generators.

The following theorem appeared (in different terminology) in Vadhan \cite{LR}, see also~\cite{ILO}.
%\begin{theorem}[Learning from succinct natural proofs]\label{t:gcsplear} Assume $\GCSP[n^c,n^d]\in\Ppoly$ for constants $d>c+1$. Then, $\Circuit[n^c]$ is learnable by \Ppoly with random examples, confidence $1-1/n$, up to error $1/n$.
%\end{theorem}

%\proof By Proposition \ref{p:randomsize}, with probability $\ge 1-1/n$, $n^d$ random queries to $f\in\Circuit[n^c]$ force each $n^c$-size circuit consistent with the queries to coincide with $f(y)$ on $\ge 1-1/n$ fraction of inputs. Hence, we can use $\GCSP[n^c,n^d]\in\Ppoly$ to compute $f(y)$ with probability $\ge 1-1/n$ as in box A. \qed
%\bigskip
%rewriterestofsubsection

\begin{theorem}[Learning from succinct natural proofs]\label{t:gcsplear}
Assume $\GCSP[n^c,n^{d}]\in\Ppoly$ for constants $d>c+1$. Then, $\Circuit[n^c]$ is learnable by \Ppoly over the uniform distribution with random examples, confidence $1/poly(n)$, up to error $1/2-1/poly(n)$.
\end{theorem}

\proof %For an $n^c$-size circuit $C$ define a generator $$G_C:\{0,1\}^{mn}\mapsto\{0,1\}^{mn+m}$$ which maps $m$ $n$-bit strings $x_1,\dots,x_m$ to $x_1,C(x_1),\dots,x_m,C(x_m)$. 
As the number of partial Boolean functions on a given set of $m$ inputs is $2^m$ and the number of $n^c$-size circuits is bouded by $2^{n^{c+1}}$, %the assumption 
$\GCSP[n^c,n^{d}]\in\Ppoly$ implies that for $m=n^{d}$ there are p-size circuits $D$ such that for each $n^c$-size circuit $C$, $$\Pr[D(x)=1]-\Pr[D(G_C(x))=1]\ge 1/2.$$ Now, it suffices to apply Lemma \ref{l:gen}. \qed %Therefore, there exists $i<m$ and p-size circuits $D'$ such that for each $n^c$-size circuit $C$, $$\Pr[D'(x_1,C(x_1),\dots,x_i,C(x_i),x_{i+1})=C(x_{i+1})]\ge 1/2+1/poly(n).$$\qed

\def\susanna{
\bigskip

An instance-specific version of Theorem \ref{t:gcsplear} in which the prediction of $C(x)$ on a single input $x$ can be obtained from proving a single circuit lower bound holds by definition if we want to predict the value of $C(x)$ only if it is determined by the queried/given samples as in box A.\footnote{The proof of Theorem \ref{t:gcsplear} does not give us an instance-specific learning - it exploits the fact that suitable circuit lower bounds yield more often a correct prediction than an incorrect one.} We conclude the section by pointing out that this could be strengthened to an instance-specific construction of a learning algorithm which predicts $C(x)$ on many inputs assuming an error concentration hypthesis.

Suppose the size of circuit $C$ is $s=n^k$. A simple counting argument shows that if we chose randomly $y_1,\dots, y_{n^{k_1}}$ for a sufficiently big $k_1$, then with high probability each circuit of size $n^{k}$ which coincides with $C$ on $y_1,\dots y_{n^{k_1}}$ differs from $C$ only on a $\le \frac{1}{n}$-fraction of inputs. However, it is possible that each circuit of size $n^{k}$ errs on a different set of inputs and that their collective set of errors is too big: if the function computed by $C$ can be actually computed by a circuit of size $s-O(n)$, then for each $x$, we can construct a circuit $C'$ such that $C'(x)\ne C(x)$ but still $C'(y_i)=C(y_i)$ for $i=1,\dots,n^{k_1}$. Is this the case also if the size $s$ is guaranteed to be the size of a minimal circuit computing the partial function $y_1,C(y_1),\dots,y_{n^{k_1}},C(y_{n^{k_1}})$?

\begin{question}[Error concentration]\label{q:ach} Let $s:\mathbb{N}\mapsto\mathbb{N}$ be a function. Is it true that for every Boolean function $f$ with (minimal) circuit complexity $s$, for each $n$, there is $m=poly(s,n)$ such that randomly chosen $n$-bit strings $y_1,\dots,y_m$ force the set of $x\in\{0,1\}^n$ satisfying $C(x)\ne f(x)$ for some circuit $C$ of size $s$ which coincides with $f$ on $y_1,\dots,y_m$ to be small with high probability? More precisely, does the following hold? $$\Pr_{y_1,\dots,y_m}[|\{x\in\{0,1\}^n \mid \exists\ s\text{-size }C\text{ s.t.}\bigwedge_{y_1,\dots,y_m} C(y_i)=f(y_i)\wedge C(x)\ne f(x)\}|\le \frac{2^n}{n}]\ge 1-\frac{1}{n}$$
\end{question}
%potential counter-example for the modified question where s is min-size of circuit approximating f: f(i,r1,...,r_{n/clogn})=1 iff h(r_i)=1 where h is a hard function on clogn bits, each ri has clogn bits. there are (n/clogn n/polylogn) many circuits erring on different inputs when computing f. f is in C[n^c+1]\C[n^c/2]. however, unclear if these circuits are optimal approximators. also not enough of them

Since the size $s$ of the circuit $C$ can be provided as a nonuniform advice, a positive answer to Question \ref{q:ach} would mean that for every Boolean function $f$ whose minimal circuit has size $s$, for many inputs $x$, we can predict $f(x)$ by chosing random strings $y_1,\dots,y_m$ and proving a single lower bound as in box A. %Similarly, we could obtain an instance-specific version of a result of Carmosino, Imagliazzo, Kabanets and Kolokolova \cite{CIKK} presented in the following section, if the error concentration was guaranteed for inputs $y_1,\dots,y_m$ specified by the used Nisan-Wigderson generator, cf. Section \ref{s:natcikk}. 
\textcolor{red}{To be erased: instance-specific part. Question 3 resolved by a simple counter-example of Erfan and Susanna. Unclear what would be the right formulation. Not so important anyway.}}

\subsection{Worst-case learning from natural proofs}\label{s:natcikk}

In Theorem \ref{t:gcsplear}, we can learn $f\in\Circuit[n^c]$ even if the algorithm for \GCSP works just for a significant fraction of partial truth-tables $(y_1,b_1),\dots,(y_{n^d},b_{n^d})$ with zero-error on easy partial truth-tables.  %rewrite'ofpartial truth-tables.. easy partial truth-tables.'
Carmosino, Impagliazzo, Kabanets and Kolokolova \cite{CIKK} proved that the assumption of Theorem \ref{t:gcsplear} can be weakened to the existence of a standard natural property. The price for this is that the resulting learning uses membership queries instead of random examples. 
%\medskip
The crucial idea is similar to the proof of Theorem \ref{t:main}: apply the natural property (as an algorithm for suitable \GCSP) on a Nisan-Wigderson generator $NW_f$ based on the function $f$, which we want to learn.

\begin{theorem}[Learning from natural proofs \cite{CIKK}]\label{t:cikk}
Let $R$ be a $\Ppoly$-natural property useful against $\Circuit[n^d]$ for some $d \geq 1$. Then, for each $\gamma\in (0,1)$, $\Circuit[n^k]$ is learnable by $\Circuit[2^{O(n^{\gamma})}]$ over the uniform distribution with non-adaptive membership queries, confidence 1, up to error $\frac{1}{n^k}$, where $k = \frac{d\gamma}{a}$ and $a$ is an absolute constant.%rewrite' over the uniform distribution'
\end{theorem}

%We could again consider an instance-specific version of Theorem \ref{t:cikk}. Given samples $(y_i,f(y_i))$, $i=1,\dots,k$, for $y_i\in\{0,1\}^m, f(y_i)\in\{0,1\}$, a function $f$ with circuit complexity $m^{k'}$ and $y_{k+1}\in\{0,1\}^m$, we can predict $f(y_{k+1})$ as in box A by solving $\GCSP[m^{k'},k+1]$. If $y_1,\dots,y_k$ satisfy $y_i=w|J_i(A)$ for a string $w\in\{0,1\}^{n^{2d}}$ and an $(n,n^{d})$-design $A$ as above (so $m=n^{d}$), then $f(y_{k+1})$ could be predicted by solving $\GCSP[n^{10k'd},k+1]$. However, it is not guaranteed that if the value $f(y_{k+1})$ is determined then the corresponding value w.r.t. to $n^{10k'd}$-size circuits will be determined as well. \textcolor{red}{Erase this paragraph.}
%rewriteabove'n^2d'2x'n^d'2x'n^10k'd'amedzerabelow

\subsection{Learning from breaking pseudorandom function families}\label{s:prfs}

Oliveira and Santhanam \cite{OS} showed that the assumption of the existence of natural proofs from Theorem \ref{t:cikk} can be further weakened to the existence of a distinguisher breaking non-uniform pseudorandom function families. Their result follows from a combination of Theorem \ref{t:cikk} and the Min-Max Theorem. Using their strategy but combining the Min-Max Theorem with Theorem \ref{t:gcsplear}, learning algorithms with random examples can be obtained from distinguishers breaking succinct non-uniform pseudorandom function families
\bigskip

A {\em two-player zero-sum game} is specified by an $r\times c$ matrix $M$ and is played as follows. MIN, the row player, chooses a probability distribution $p$ over the rows. MAX, the column player, chooses a probability distribution $q$ over the columns. A row $i$ and a column $j$ are drawn randomly from $p$ and $q$, and MIN pays $M_{i,j}$ to MAX. MIN plays to minimize the expected payment, MAX plays to maximize it. The rows and columns are called the {\em pure strategies} available to MIN and MAX, respectively, while the possible choices of $p$ and $q$ are called {\em mixed strategies}. The Min-Max theorem states that playing first and revealing one's mixed strategy is not a disadvantage: 
$$min_{p} max_j \sum_i p(i)M_{i,j}=max_q min_i \sum_j q(j) M_{i,j}.$$ Note that the second player need not play a mixed strategy - once the first player's strategy is fixed, the expected payoff is optimized for the second player by playing some pure strategy. The expected payoff when both players play optimally is called the {\em value} of the game. We denote it $v(M)$.

A mixed strategy is {\em $k$-uniform} if it chooses uniformly from a multiset of $k$ pure strategies. Let $M_{min}=min_{i,j} M_{i,j}$ and $M_{max}=max_{i,j} M_{i,j}$. Newman \cite{Nm}, Alth\"ofer \cite{Alt} and Lipton-Young \cite{LY} showed that each player has a near-optimal $k$-uniform strategy for $k$ proportional to the logarithm of the number of pure strategies available to the opponent.

\begin{theorem}[\cite{Nm, Alt, LY}]\label{thm:minmax} For each $\epsilon>0$ and $k\ge \ln(c)/2\epsilon^2$, $$min_{p\in P_k} max_j \sum_{i} p(i) M_{i,j}\le v(M)+\epsilon(M_{max}-M_{min}),$$ where $P_k$ denotes the $k$-uniform strategies for MIN. The symmetric result holds for MAX.
\end{theorem}

\begin{definition}[Succinct non-uniform PRF] An $(m,m')$-succinct non-uniform pseudorandom function family from circuit class $\mathcal{C}$ safe against circuits of size $s$ is a set $S$ of partial truth-tables $\langle (x_1,b_1),\dots, (x_m,b_m)\rangle$ where each $x_i$ is an $n$-bit string and $b_i\in\{0,1\}$ such that each partial truth-table from $S$ is computable by one of $m'$ circuits from $\mathcal{C}$ and for every circuit $D$ of size $s$, $$\Pr_{x}[D(x)=1]-\Pr_{x\in S}[D(x)=1]<1/s$$ where the first probability is taken over $x\in\{0,1\}^{m(n+1)}$ chosen uniformly at random and the second probability over partial truth-tables chosen uniformly at random from $S$.
\end{definition}

\begin{theorem}[Learning or succinct non-uniform PRF]\label{t:towardscore}
Let $c\ge 1$ and $s>n,m\ge 1$. There is an $(m,8s^4)$-succinct non-uniform PRF in $\Circuit[n^c]$ safe against $\Circuit[s]$ or there are circuits of size $poly(s)$ learning $\Circuit[n^c]$ over the uniform distribution with random examples, confidence $1/poly(s)$, up to error $1/2-1/poly(s)$.%rewrite' over the uniform distribution'
\end{theorem}

\proof %Assume there is no pseudorandom generator safe against circuits of size $s$. In particular, for every $n^c$-size circuit $C$ and generator $G_C$ from the proof of Lemma \ref{gcsp} %we can break generator $$G_C:\{0,1\}^{mn+n^{c+1}}\mapsto\{0,1\}^{mn+m}$$ which takes as input a string of length $mn+n^{c+1}$ encoding a circuit $C$ of size $n^c$ together with $m$ $n$-bit strings $x_1,\dots,x_m$ and outputs $x_1,C(x_1),\dots,x_m,C(x_m)$. That is, for each $C$ there is an $s$-size circuit $D$ such that $$|\Pr[D(x)=1]-\Pr[D(G_C(x))=1]|\ge 1/s.$$ 

%By OS, there is a circuit $L$ of size $poly(s)$ solving \Succinct. By Lemma \ref{gcsp} this implies circuits learning $\Circuit[n^d]$ with random examples.
Consider a two-player zero-sum game specified by a matrix $M$ with rows indexed by $n^c$-size circuits with $n$ inputs and columns indexed by $s$-size circuits with $m(n+1)$ inputs. Define the entry $M_{C,D}$ of $M$ corresponding to a row circuit $C$ and a column circuit $D$ as $$M_{C,D}:=|\Pr_x[D(x)=1]-\Pr_x[D(G_C(x))=1]|$$ for the generator $G_C$ from the proof of Lemma \ref{l:gen}. Hence $M_{max}-M_{min}\le 1$. 

If $v(M)\ge 1/4s$, then by Theorem \ref{thm:minmax} (with $\epsilon=1/8s$), there exist a multiset of $k\le 32n^{c+1}s^2$ $s$-size circuits $D^1,\dots, D^{k}$ such that for every $n^c$-size circuit $C$, a random $D$ from $D^1,\dots, D^{k}$ satisfies $$\text{E}[|\Pr[D(x)=1]-\Pr[D(G_C(x))=1]|]\ge 1/8s.$$ 

By Lemma \ref{l:gen}, for every $n^c$-size circuit $C$, one of the circuits $D^1,\dots, D^k$ (or their negations) can be used to learn $C$ with confidence $1/poly(s)$, up to error $1/2-1/poly(s)$. %Since we can recognize with high probability if a given circuit $(1/2+1/poly(n))$-approximates $C$, 
A $poly(s)$-size circuit using a random $D^i$ from $D^1,\dots, D^k$ or its negation thus learns $\Circuit[n^c]$ with random examples, confidence $1/poly(s)$, up to error $1/2-1/poly(s)$.
\medskip

If $v(M)<1/4s$, then by Theorem \ref{thm:minmax} (with $\epsilon=1/4s$), there exists a multiset of $k\le 8s^4$ $n^c$-size circuits $C^1,\dots, C^k$ such that for every $s$-size circuit $D$, a random $C$ from $C^1,\dots, C^k$ satisfies $$\text{E}[|\Pr[D(x)=1]-\Pr[D(G_C(x))=1]|]\le 1/2s.$$ Since $\text{E}[|\Pr[D(x)=1]-\Pr[D(G_C(x))=1]|]\ge|\Pr[D(x)=1]-\text{E}[\Pr[D(G_C(x))=1]]|$ a generator $$G:\{0,1\}^{mn+\lceil\log k\rceil}\mapsto \{0,1\}^{mn+m}$$ which takes as input a string of length $mn+\lceil\log k\rceil$ encoding (an index of) a circuit $C$ from $C^1,\dots, C^k$ together with $m$ $n$-bit strings $x_1,\dots,x_m$ and outputs $x_1,C(x_1),\dots, x_m, C(x_m)$ is safe against circuits of size $s$. The range of $G$ defines an $(m,8s^4)$-succinct non-uniform PRF in $\Circuit[n^c]$ safe against $\Circuit[s]$.
\qed

\bigskip%rewritebelow:' of small complexity'
Note that the existence of a generator $G$ from the proof of Theorem \ref{t:towardscore} follows directy from a counting argument if we do not require that $G$ defines a PRF of small complexity: a random set of $poly(s,n)$ strings (yielding a non-uniform pseudorandom generator mapping $\{0,1\}^{O(\log s)}$ to $\{0,1\}^n$) fools circuits of size $s$. %Ideally, we would like to replace succinct non-uniform PRFs in Theorem \ref{t:towardscore} by a uniform pseudorandom generator, thus answering Question \ref{q:dichotomy}. %The pseudorandom generator we want is cryptographic, i.e. safe against circuits of size much bigger than the running time of the generator. For complexity-theoretic generators the implication is known: if $\mathsf{E}$ requires circuits of subexponential size (e.g. because $\GCSP[n^{O(1)},n^{O(1)}]$ is hard for circuits of size $2^{O(n)}$ or learning p-size circuits is hard for subexponential-size circuits), then there are NW-generators mapping $n^{O(1)}$ bits to $2^n$ bits computable in time $2^{n^{O(1)}}$ and safe against circuits of size $2^{O(n)}$. %Nevertheless, we can obtain an alternative construction of complexity-theoretic generators if we assume a uniform version of the Min-Max theorem.

\def\unimin{
\bigskip
\noindent {\bf Lowering the bar to complexity-theoretic PRGs.} Let $M^{n,m}$ be a sequence of $2^n\times 2^s$ matrices specifying two-player zero-sum games. We say that a $poly(s)$-time algorithm $A$ is a sampling algorithm of a mixed strategy $q$ of the column player if $\forall j\in [2^s]$, $\Pr_{x\in\{0,1\}^{n}}[A(x)=j]=q(j)$.

\begin{hypothesis}[Uniform Min-Max]\label{h:uniminmax} Let $M^{n,s}$ be a sequence of $2^n\times 2^s$ matrices specifying two-player zero-sum games. Assume there is a p-time algorithm which given a sampling algorithm of a mixed strategy $q$ of the column player outputs a pure strategy $i$ of the row player such that $\sum_j q(j)M_{i,j}\le v(M)$. Then there is a p-time algorithm which given $1^s$ outputs a sampling algorithm of a mixed strategy $q'$ of the column player such that $min_i \sum_j q'(j)M_{i,j}\ge v(M)-1/2^s$.
\end{hypothesis}

A uniform version of the Min-Max theorem has been proved by Vadhan-Zheng \cite{VZ}. However, their algorithm does not seem to generate sampling algorithms of mixed strategies - just an implicit description of such strategies.

\begin{theorem}[Non-refutation of Learning or PRGs assuming uniform Min-Max]\label{t:almostcore}
Assume the uniform Min-Max hypothesis. Then, there is a p-time computable pseudorandom generator safe against $\Circuit[s]$, where $s\ge poly(n)$, or for every constant $c$, there is no p-time algorithm which given a circuit $D$ of size $poly(s,n)$ outputs a circuit $C$ of size $n^c$ such that $D$ does not learn $C$ with random examples, confidence $1/n$, up to error $1/2-1/poly(s,n)$. 
\end{theorem}

\proof Proceed as in the proof of Theorem \ref{t:towardscore} but distinguish the following two cases. If for infinitely many $n$, $v(M)\ge 1/4s(n)$, then for infinitely many $n$, there are circuits of size $poly(s,n)$ learning circuits of size $n^c$ with random examples, confidence $1/n$, up to error $1/2-1/poly(s,n)$. If for all sufficiently big $n$, $v(M)<1/4s(n)$, then apply uniform Min-Max hypothesis to conclude that for all sufficiently big $n$, for every circuit $D$ of size $s(n)$, the generator $$G:\{0,1\}^{mn+poly(s)}\mapsto \{0,1\}^{mn+m}$$ which takes as input a string of length $mn+poly(m)$ encoding an input of the sampling algorithm from the uniform Min-Max hypothesis together with $m$ $n$-bit strings $x_1,\dots,x_m$ and outputs $x_1,C(x_1),\dots, x_m, C(x_m)$ is safe against circuits of size $s$. \qed

\bigskip

Since the existence of an efficient algorithm generating strategies for the second Player follows from the provability of a suitable statement in intuitionistic \SB we can instead assume that the non-existence of an efficient learner is also feasibly provable.

\begin{corollary}[Consistency of Learning or PRGs assuming uniform Min-Max]\label{c:conscore}
Assume uniform Min-Max hypothesis. Then, there is a p-time computable pseudorandom generator safe against $\Circuit[s]$, where $s\ge poly(n)$, or for every constant $c$, it is consistent with intuitionistic $\SB$ that there are circuits of size $poly(s,n)$ learning $\Circuit[n^c]$ with random examples, confidence $1/n$, up to error $1/2-1/poly(s,n)$. 
\end{corollary}

\proof[Proof sketch.] Proceed as in the proof of Theorem \ref{t:towardscore} but in the case $v(M)<1/4s$ apply witnessing theorem. \qed
}

\subsection{Superbits vs demibits}\label{s:rudich}

Rudich \cite{Rs} proposed a conjecture about the existence of superbits, a version of pseudorandom generators safe against nondeterministic circuits, and showed that it rules out the existence of \NP-natural properties against \Ppoly. He then asked whether the existence of superbits follows from a seemingly weaker assumption of the existence of so called demibits. We note that an affirmative answer to his question would resolve Question \ref{q:dichotomy} in nondeterministic setting.

\begin{definition}[Superbit] A function $g:\{0,1\}^n\mapsto\{0,1\}^{n+1}$ computable by p-size circuits is a superbit if there is $\epsilon<1$ such that for infinitely many input lengths $n$, for all nondeterministic circuits $C$ of size $|C|\le 2^{n^{\epsilon}}$, $$\Pr_{x\in\{0,1\}^{n+1}}[C(x)=1]-\Pr_{x\in\{0,1\}^n}[C(g(x))=1]<1/|C|.$$
\end{definition}

\begin{definition}[Demibit] A function $g:\{0,1\}^n\mapsto\{0,1\}^{n+1}$ computable by p-size circuits is a demibit if there is $\epsilon<1$ such that for infinitely many input lengths $n$, no nondeterministic circuit $C$ of size $|C|\le 2^{n^{\epsilon}}$ satisfies $$\Pr_{x\in\{0,1\}^{n+1}}[C(x)=1]\ge 1/|C|\ \ \ \text{ and }\ \ \ \Pr_{x\in\{0,1\}^n}[C(g(x))=1]=0.$$
\end{definition}

\begin{proposition}[Question \ref{q:dichotomy} vs Rudich's problem] Assume the existence of demibits implies the existence of superbits. Then, either superbits exist or for each $c\ge 1$, for each $\epsilon<1$, $\Circuit[n^c]$ is learnable by $\Circuit[2^{O(n^{\epsilon})}]$ over the uniform distribution with random examples, confidence $1/2^{O(n^{\epsilon})}$ up to error $1/2-1/2^{O(n^{\epsilon})}$, where the learner is allowed to generate a nondeterministic or co-nondeterministic circuit approximating the target function.%rewrite' over the uniform distribution'
\end{proposition}

\proof Assume superbits do not exist and their non-existence implies the non-existence of demibits. Consider a generator $G:\{0,1\}^{mn+n^{c+1}}\mapsto\{0,1\}^{mn+m}$, with $m=n^{c+1}+1$, which interprets the first $n^{c+1}$ bits of its input as a description of an $n^c$-size circuit $C$ and then computes on the remaining $mn$ inputs as generator $G_C$ from Lemma \ref{l:gen}. Since $G$ is not a demibit, for each $\epsilon<1$ there are nondeterministic circuits $D$ of size $2^{(mn+m-1)^{\epsilon}}$, such that for each $n^c$-size circuit $C$, $$\Pr[D(x)=1]-\Pr[D(G_C(x))=1]\ge 1/|D|.$$ By the proof of Lemma \ref{l:gen}, this means that $n^c$-size circuits are learnable by circuits of size $poly(|D|)$ with confidence $1/poly(|D|)$ up to error $1/2-1/poly(|D|)$, except that the learner might generate nondeterministic (if $r_i=0$) or co-nondeterminitic (if $r_i=1$) circuit approximating the target function. \qed

}
%Another form of automatability speedup: P>lb > P>tt in lin size whp.

%\section{Concluding remarks and open problems}\label{s:concluding}

\section*{Acknowledgements}%rewriteack

We would like to thank Jan Kraj\'{i}\v{c}ek and Iddo Tzameret for comments on a draft of the paper.  %rewriteoxpredchvetaanaslmedzera
This project has received funding from the European Union's Horizon 2020 research and innovation programme under the Marie Sk\l{}odovska-Curie grant agreement No 890220.
\medskip

\noindent\fbox{\includegraphics[width=48pt,height=31pt]{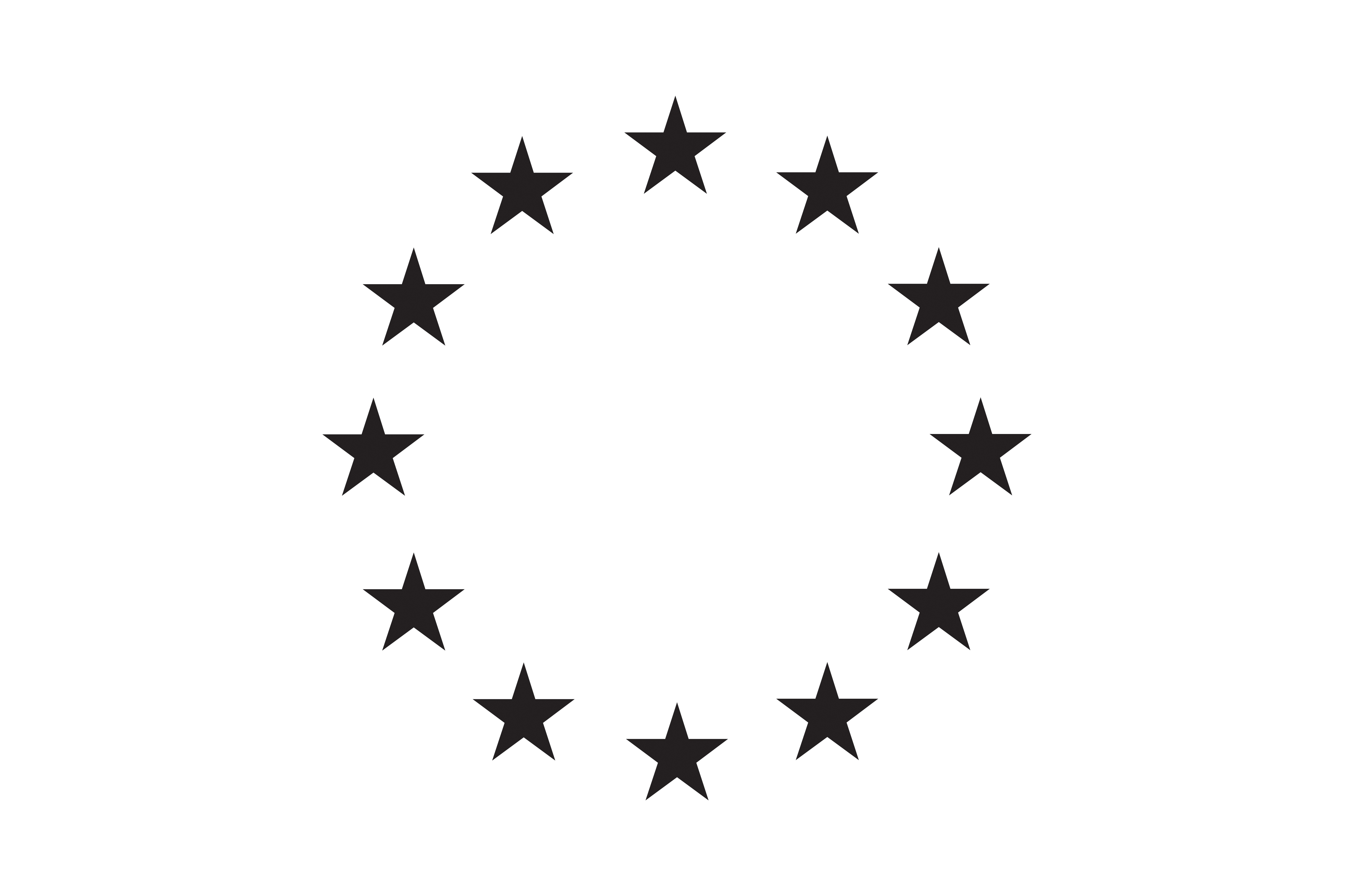}}

\end{document}